%% file: paper.tex
\documentclass[a4paper,fontsize=10pt,listof=totoc,numbers=noenddot]{scrartcl}

\usepackage[a4paper,top=3cm,bottom=3cm,left=2cm,right=2cm]{geometry}
\usepackage[onehalfspacing]{setspace}
\usepackage{csquotes}
\usepackage{scrlayer-scrpage}
\usepackage{subfiles} 
\usepackage{microtype}
\usepackage{graphicx}
\usepackage{subcaption}
\usepackage[hidelinks]{hyperref}
\usepackage{geometry}
\usepackage[style=numeric,sorting=none]{biblatex}
\usepackage{amsmath}
\usepackage{amssymb}
\usepackage[nameinlink]{cleveref}
\usepackage{csvsimple}
\usepackage{tikz}
\usepackage{enumitem}
\usepackage{listings}
\usepackage{longtable}
\usepackage{multirow}
\usepackage[T1]{fontenc}
\usepackage{lmodern}
\usepackage{svg}
\usepackage{draftwatermark}
\usepackage{lastpage}
\usepackage{todonotes}
\usepackage{censor}
\usepackage{array}

\SetWatermarkText{}
\SetWatermarkScale{2}

\clearpairofpagestyles
\ihead{Stoll et al.}
\crefname{figure}{figure}{figures}
\Crefname{figure}{Figure}{Figures}
\crefname{table}{table}{tables}
\Crefname{table}{Table}{Tables}

\graphicspath{./}

\usetikzlibrary{calc}

\AtBeginDocument{
  \setlength{\abovedisplayskip}{3pt}
  \setlength{\belowdisplayskip}{3pt}
  \setlength{\abovedisplayshortskip}{0pt}
  \setlength{\belowdisplayshortskip}{3pt}
}

\begin{document}

\begin{titlepage}
    \newcommand{\HRule}{\rule{\linewidth}{0.5mm}}
    \center
    \HRule\\[0.2cm]
    
    {\huge\bfseries Delay Attacks on the German Smart Metering Infrastructure:}\\[0.2cm]
    {\huge\bfseries A Security Analysis of CLS Channel Timing Constraints}\\[0.2cm]

    \HRule\\[1cm]
    Corresponding Author \\ [0.2cm]
    {\large\bfseries Fabio Stoll (stollf@hs-albsig.de)$^{1}$} \\
    Authors \\ [0.2cm]
    {\large\bfseries Benjamin Pottkamp$^{1}$, Heiko Lorenz$^{2}$, Shalaka Kale$^{2}$, \\ Jessica Rövekamp$^{1}$, Joachim Gerlach$^{1}$} \\ [0.2cm]
    {\large\bfseries $^{1}$Albstadt-Sigmaringen University\\ $^{2}$ Ulm University of Applied Sciences} \\ [1cm]
    {\large\bfseries Abstract}
    \begin{abstract}
        \subfile{abstract.tex}
    \end{abstract}
\end{titlepage}
\pagenumbering{Roman}
\tableofcontents
\pagebreak
\listoffigures
\listoftables
\pagebreak
\pagenumbering{arabic}

\section{Introduction} \label{sec:intro}
\subsection{Motivation} \label{sec:intro:motivation}
\subfile{intro/motivation.tex}
\subsection{Related Work} \label{sec:intro:related}
\subfile{intro/related.tex}
\subsection{Contents of this Work} \label{sec:intro:contents}
\subfile{intro/contents.tex}
\subsection{Contribution of this Work} \label{sec:intro:contribution}
\subfile{intro/contribution.tex}

\section{Theory} \label{sec:theory}
\subsection{System Architecture and Technical Assumptions} \label{sec:theory:scenario}
\subfile{theory/scenario.tex}
\subsection{Threat Model} \label{sec:theory:capabilities}
\subfile{theory/capa.tex}
\subsection{Delay Attacks} \label{sec:theory:delay}
\subfile{theory/delay.tex}
\subsection{Attacking DAP} \label{sec:theory:dap}
\subfile{theory/dap.tex}
\subsection{Attacking CLS.EEDI} \label{sec:theory:clseedi}
\subfile{theory/clseedi.tex}
\subsection{Attacking IEC 61850} \label{sec:theory:iec61850}
\subfile{theory/iec61850.tex}

\section{Evaluation} \label{sec:evaluation}
\subsection{Laboratory Setup} \label{sec:evaluation:lab}
\subfile{evaluation/lab.tex}
\subsection{Methodology} \label{sec:evaluation:methodology}
\subfile{evaluation/meth.tex}
\subsection{Results} \label{sec:evaluation:results}
\subfile{evaluation/res.tex}

\section{Impact} \label{sec:impact}
\subsection{Use Cases} \label{sec:impact:cases}
\subfile{impact/usecases.tex}
\subsection{Static Load-Altering Attack} \label{sec:impact:laa}
\subfile{impact/laa.tex}
\subsection{Coordinated Static Load-Altering Attack} \label{sec:impact:CoLAA}
\subfile{impact/CoLAA.tex}

\section{Countermeasures} \label{sec:countermeasure}
\subsection{Enforcing Secure Configurations} \label{sec:countermeasure:short}
\subfile{countermeasure/shortTem.tex}
\subsection{Securing Application-Layer Implementations} \label{sec:countermeasure:mid}
\subfile{countermeasure/midTerm.tex}
\subsection{Securing Protocols} \label{sec:countermeasure:long}
\subfile{countermeasure/longTerm.tex}

\section{Discussion} \label{sec:discussion}
\subsection{Generalization} \label{sec:discussion:gen}
\subfile{discussion/gen.tex} 
\subsection{Limitations} \label{sec:discussion:limit}
\subfile{discussion/limit.tex}
\subsection{Practical Impact} \label{sec:discussion:impa}
\subfile{discussion/impa.tex}

\section{Conclusion} \label{sec:conclusion}
\subsection{Summary} \label{sec:conclusion:summary}
\subfile{conclusion/summary.tex}
\subsection{Future Work} \label{sec:conclusion:future}
\subfile{conclusion/future.tex}
\pagebreak

\appendix

\section*{CRediT Statement}
\subfile{appendix/contri.tex}
\section*{Acknowledgments}
\subfile{appendix/ack.tex}
\section*{AI Disclaimer}
\subfile{appendix/ai.tex}
\section*{Conflicts of Interest}
\subfile{appendix/conf.tex}
\section*{Responsible Disclosure}
\subfile{appendix/disc.tex}
\printbibliography[heading=bibintoc]

\end{document}

%% file: abstract.tex
This work analyzes the feasibility of delay attacks on control signals transmitted via the Controllable Local System (CLS) channel of the German Smart Metering Infrastructure (SMI). It combines theoretical analysis with experimental validation under a threat model aligned to the Common Criteria Protection Profile for the Smart Meter Gateway (SMGW) \cite{tech:bsi:pp-smgw} and assesses the potential impact on the power grid if the identified attack vector is exploited across multiple CLS channels simultaneously. We also outline mitigation strategies, including SMGW configuration restrictions, implementation-level changes, and protocol extensions.

Our results show that an on-path attacker in the Wide Area Network (WAN) with sufficient contextual knowledge can feasibly execute delay attacks, with a theoretical upper bound of roughly 48 hours for some deployed protocol configurations. Projecting from a single CLS to several hundred thousand CLS devices indicates such an adversary could cause a significant frequency deviation potentially resulting in load shedding. Scaling the attack requires contextual knowledge for each targeted implementation and configuration; whether this knowledge can be broadly reused across CLS channels is uncertain but may become easier to obtain as standardization progresses.

Time restricted transmissions in the FNN Steuerbox \cite{tech:FNN:STB} and in applications using CLS.EEDI \cite{tech:standard:cls-eedi} are implementation-specific and can therefore be addressed by manufacturers. By contrast, ensuring application-data time limitations in TLS~1.3 requires protocol-level extensions. The TLS extensions proposed here offer a sustainable mitigation while preserving backward compatibility.

Other communication channels outside the SMI (for example, proprietary remote terminal units used to control a CLS) are outside this work's scope and may exhibit similar or worse vulnerabilities.

%% file: intro/motivation.tex
The German Smart Metering Infrastructure (SMI) is a highly regulated infrastructure that serves as a secure communication platform within the German energy sector \cite{tech:bsi:tr-03109}. The scope, features and functions of the Smart Meter Gateway (SMGW) are outlined by the German Federal Office for Information Security (German: Bundesamt für Sicherheit in der Informationstechnik, BSI). Devices used in production must obtain certification from the BSI to demonstrate compliance \cite{law:deu:msbg,tech:bsi:tr-03109-1}.

The SMGW is primarily used to read and process metering data about power flows at the grid connection point. In addition it can act as a TLS proxy to enable a secure communication channel - referred to as the CLS channel - between an active External Market Participant (aEMT) (e.g., a Distribution System Operator, DSO) and a Controllable Local System (CLS) (e.g., a photovoltaic inverter) located at a customer's premises \cite{tech:bsi:tr-03109-1,tech:bsi:pp-smgw}. A CLS is connected to the SMGW using a CLS Adapter (CLS-A) \cite{tech:bsi:tr-03109-5,tech:bsi:tr-03109-1}. Since we focus on the CLS channel, the SMGW is primarily considered in its role as a TLS proxy.

For several use cases defined in §34 of the Act on Metering Point Operations and Data Communication in Smart Energy Networks (MsbG), German law mandates the use of the CLS channel provided by the SMGW to control a CLS installed at the customer premises \cite{law:deu:msbg}. Consequently, the SMGW must implement appropriate security mechanisms to comply with the requirements outlined in Chapter 3 of the MsbG (§19-§28) \cite{law:deu:msbg} and further substantiated in the BSI Common Criteria Protection Profile \cites{tech:bsi:pp-smgw,tech:bsi:pp-hsm}.

Given the planned large-scale rollout of smart meters defined in §45 MsbG \cite{law:deu:msbg}, the additional use-cases planned for the future described in §34 MsbG and the expected increase in grid-connected CLS devices \cite{law:bentza:netzentwicklungsplanSzenarien}, vulnerabilities in the SMI specification could have significant consequences for the stability and security of the German power grid if exploited successfully on a large-scale.

Previous work by Fu et al. has shown that in Internet-of-Things (IoT) environments the usage of TLS alone does not necessarily mitigate delay attacks \cite{Paper:Delay:phantomDelay}. Such attacks may allow an adversary to manipulate system behavior by delaying control signals without modifying their content \cite{Paper:Delay:phantomDelay,Paper:Delay:delayBasedAutomationInterference}.

In this work, we investigate whether this observation also applies to the CLS channel provided by the SMGW. Specifically, this work addresses the following research questions:

\begin{enumerate}
    \item What is the theoretical maximum delay that an attacker could introduce for a control signal between an aEMT and a CLS while keeping the control signal successfully implemented?
    \item Do application-layer protocols such as CLS.EEDI or IEC 61850, when used within the CLS channel, sufficiently mitigate delay attack vulnerabilities?
    \item What impact could delayed control signals have on the German power grid?
    \item Which countermeasures can be implemented at the SMGW, CLS-A or aEMT to mitigate delay attacks?
\end{enumerate}

%% file: intro/related.tex
Delay attacks in the context of time synchronization have been discussed in \cite{Paper:Delay:DelayAttackPTP}. In these attacks, an adversary introduces asymmetric delays in protocols such as NTP or PTP, violating the assumption of symmetric network latency. As a result, the time values are computed incorrectly. 

Fu et al. extended the concept of delay attacks to IoT devices and showed that TLS alone is insufficient to mitigate such threats \cite{Paper:Delay:phantomDelay}. They proposed a taxonomy that classifies delay attacks along two dimensions: impact and message content. Based on impact, delay attacks are divided into state-update delay attacks, action delay attacks, and erroneous execution attacks, which include the subtypes spurious execution and disabled execution. Based on message content, they distinguish between event message delay attacks and command message delay attacks. Mattsson et al. analyze delay attacks on the protocol CoAP \cite{ietf-core-attacks-on-coap-06}.

Chi et al. study various impacts of delay attacks on smart home automation systems and show that such attacks can trigger cross-rule interference \cite{Paper:Delay:delayBasedAutomationInterference}. They further show that many of these attacks could be realized through Wi-Fi jamming, effectively reducing them to DoS attacks \cite{Paper:Delay:delayBasedAutomationInterference}. 

Delay attacks in the context of smart grids have been studied in \cite{Paper:Delay:SmartGridAGC,Paper:Delay:PowerGridControlDelay}. However, none of these works consider delay attacks in the context of TLS-secured communication. They nevertheless demonstrated the potential impact, including violations of the grid stability criteria, in small and nano grids.

Several works concluded that countermeasures are necessary and should be implemented either on the application or transport layer \cite{Paper:Delay:phantomDelay,ietf-core-attacks-on-coap-06}. 

Britz et al. analyzed the residual risk of the SMI \cite{Paper:SMI:STRIDE}. However, their work does not consider Delay Attacks. 

Despite these contributions, existing work does not analyze delay attacks in smart metering infrastructures such as the strictly regulated German SMI. The resilience of protocols used within the German SMI, such as CLS.EEDI \cite{tech:standard:cls-eedi} and IEC 61850 as implemented by the FNN Steuerbox (FNN-STB)\cite{tech:FNN:STB}, against delay attacks has not been analyzed yet. The impact of large-scale delay attacks targeting several thousand consumers or producers has also not been analyzed. These gaps motivate the research questions addressed in this work.

%% file: intro/contents.tex
\Cref{sec:theory} aims to answer the first and second research questions on a purely theoretical basis. First, we define the system architecture in \cref{sec:theory:scenario} and two threat models in \cref{sec:theory:capabilities}. Based on these definitions, we analyze the vulnerability of the system to delay attacks in \cref{sec:theory:delay}. We then demonstrate the general vulnerability of the CLS channel using an intentionally insecure protocol in \cref{sec:theory:dap}. Subsequently, we analyze the application-layer protocols CLS.EEDI and IEC 61850 within the CLS channel and show that the vulnerability to delay attacks persists in \cref{sec:theory:clseedi} and \cref{sec:theory:iec61850}, respectively.

\Cref{sec:evaluation} validates the theoretical results using a hardware-in-the-loop laboratory setup and provides empirical results addressing the first and second research questions. First, we outline the laboratory setup in \cref{sec:evaluation:lab}. Second, we describe the methodology used to evaluate the protocol implementations in \cref{sec:evaluation:methodology}. Finally, we present and discuss the results in \cref{sec:evaluation:results}.

\Cref{sec:impact} discusses the potential large-scale impact of the proposed attack and thereby addresses the fourth research question. \Cref{sec:impact:cases} outlines the use cases where the SMI is relevant, while \cref{sec:impact:laa} analyses the impact a single CLS can have on the grid and \cref{sec:impact:CoLAA} projects this impact to several hundred thousand simultaneously attacked CLS.

In \cref{sec:countermeasure}, we address the third research question by proposing mitigation strategies, including configuration changes in \cref{sec:countermeasure:short}, implementation adjustments in \cref{sec:countermeasure:mid}, and protocol extensions in \cref{sec:countermeasure:long}.

\Cref{sec:discussion} critically discusses the results. \Cref{sec:discussion:gen} examines the generalizability of our findings, while \cref{sec:discussion:limit} addresses their limitations. In \cref{sec:discussion:impa}, we discuss the feasibility of large-scale attacks.

Finally, we conclude this work in \cref{sec:conclusion} by summarizing the findings in \cref{sec:conclusion:summary} and outlining future research directions in \cref{sec:conclusion:future}.

%% file: intro/contribution.tex
This work contributes as following to answer the identified research gap:
\begin{itemize}
    \item It proposes a formal model to analyze delay attacks targeting control signals within the German SMI.
    \item Based on this model, it proposes an attack that enables an adversary to delay the implementation of control signals for up to 48 hours.
    \item It performs a detailed security analysis of the protocols CLS.EEDI and IEC 61850 with respect to their ability to verify the age of control signals.
    \item It validates our theoretical findings in a hardware-in-the-loop laboratory setup and provide empirical results.
    \item It assesses the potential large-scale impact of delayed control signals on power grid stability.
    \item It proposes mitigation strategies, including configuration changes, implementation adjustments, and protocol extensions.
\end{itemize}

%% file: theory/scenario.tex
\begin{figure}[h]
    \centering
    \includegraphics[width=\textwidth]{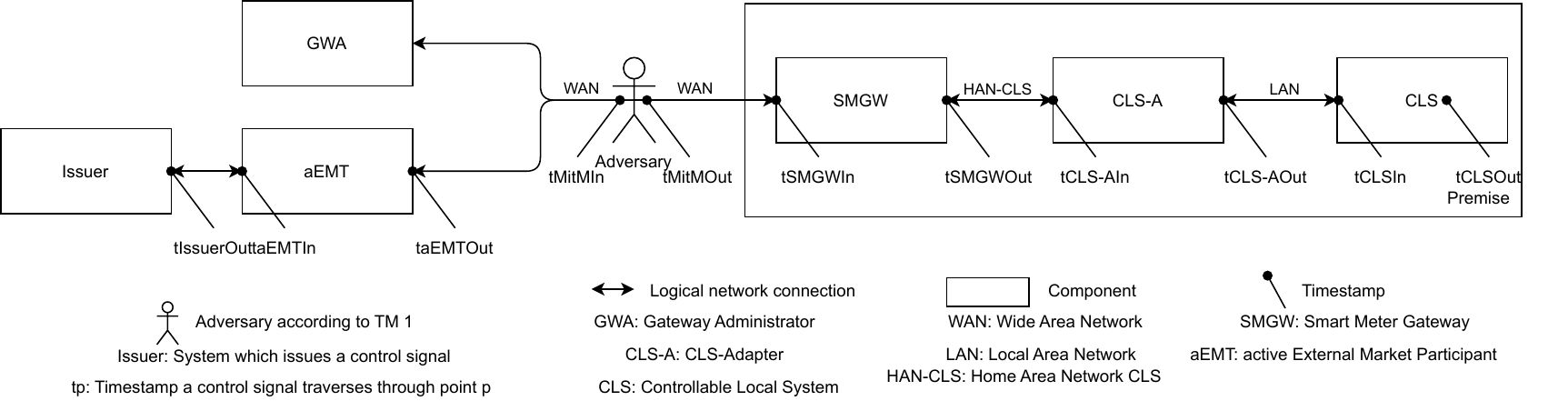}
    \caption{Overview of the system architecture}
    \label{fig:theory:scenario}
\end{figure}

The system architecture, illustrated in \cref{fig:theory:scenario} and derived from \cite{tech:bsi:tr-03109-1}, consists of an aEMT that is connected, via an SMGW, to a CLS located at the customer premises using a CLS-A and the CLS channel. A CLS-A is a device which connects a CLS to the SMGW and is specified in \cite{tech:bsi:tr-03109-5}. A CLS may represent a consumer (e.g., a heat pump), a producer (e.g., a PV inverter), both (e.g., a battery storage system), or an energy management system (EMS) that processes control signals to coordinate multiple connected CLS devices. We assume that the CLS-A does not directly forward the raw information contained in the control signal. It rather forwards the intended control action derived from that information. For the purposes of this work, these different CLS types are not treated separately. Instead, the CLS is modeled abstractly as a system that influences the active power flow at the grid connection point. 

The adversary shown in \cref{fig:theory:scenario} is meant to illustrate an abstract adversary with capabilities on one or more communication paths in the WAN. The adversary's exact capabilities will be defined in detail in \cref{sec:theory:capabilities}.

The attack proposed in this work can target several types of control signals. These may directly limit the active power flow, such as the use cases Limit of Active Power Consumption (LPC) or Limit of Active Power Production (LPP) \cite{tech:VDE:ARE_2829_6_1}, or indirectly influence it, for example through pricing signals \cite{Paper:SmartMetersCyberAttacks}. The purpose of such control signals is either to maintain grid stability or to optimize power flow according to economic objectives. However, for the purpose of this analysis, these types can all be abstracted as specifying a positive or negative target value for the active power flow either directly or indirectly. This abstraction simplifies the analysis without loss of generality.

To formally describe the message flow within the system architecture, we define a set of timestamps $T = \{t_p : p \in P\}$ based on the points in time at which a message traverses specific points $p \in P$ within the system architecture, with $P$ as the set of all points. The location of each timestamp is shown in \cref{fig:theory:scenario}. The precise definitions of the individual timestamps are provided in \cref{tab:theory:delay:timestamps}.

\begin{table}
\renewcommand{\arraystretch}{1.3}
\caption{Definition of timestamps}
\label{tab:theory:delay:timestamps}
\centering
\begin{tabular}{|c|l|}\hline
Timestamp & Definition \\\hline
$t_{\text{IssuerOut}}$ & Control signal issued by Issuer\\
$t_{\text{aEMTIn}}$ & Control signal received by aEMT\\
$t_{\text{aEMTOut}}$ & Control signal forwarded to the CLS channel by aEMT\\
$t_{\text{MitMIn}}$ & Control signal interception by the adversary\\
$t_{\text{MitMOut}}$ & Control signal released to the SMGW by the adversary \\
$t_{\text{SMGWIn}}$ & Control signal received by the SMGW \\
$t_{\text{SMGWOut}}$ & Control signal forwarded to the CLS-A by the SMGW\\
$t_{\text{CLS-AIn}}$ & Control signal received by the CLS-A \\
$t_{\text{CLS-AOut}}$ & Control signal forwarded to the CLS by the CLS-A\\
$t_{\text{CLSIn}}$ & Control signal received by the CLS\\
$t_{\text{CLSOut}}$ & Control signal implemented by the CLS\\\hline
\end{tabular}
\end{table}

$\Delta_{p_x\rightarrow p_y}=t_y - t_x$ defines the time it takes for a message to travel between the points $p_x$ and $p_y$ according to \cref{tab:theory:delay:timestamps} in the system architecture with $t_x$ and $t_y$ as the timestamps at which the message traversed $p_x$ and $p_y$, respectively. 

Using these definitions the total delay between issuing a control signal and its implementation ($\Delta_{\text{IssuerOut}\rightarrow\text{CLSOut}}$) can be expressed as $\Delta_{\text{IssuerOut}\rightarrow\text{CLSOut}} = \Delta_{\text{IssuerOut}\rightarrow\text{CLSOut},n} + \Delta_{\text{IssuerOut}\rightarrow\text{CLSOut},a} = t_{\text{CLSOut}} - t_{\text{IssuerOut}}$, where ($\Delta_{\text{IssuerOut}\rightarrow\text{CLSOut},n}$) represents the non-artificial (natural) delay, e.g. caused by network transmission and component processing times, and ($\Delta_{\text{IssuerOut}\rightarrow\text{CLSOut},a} = t_{\text{MitMOut}} - t_{\text{MitMIn}}$) represents artificial delay caused by the adversary. 

For the theoretical part of this work, we assume an idealized setting in which the non-adversarial delay is negligible, i.e., $\Delta_{\text{IssuerOut}\rightarrow\text{CLSOut},n} = 0$. Under this assumption, the total delay is entirely determined by the adversarial contribution, yielding $\Delta_{\text{IssuerOut}\rightarrow\text{CLSOut}} = \Delta_{\text{IssuerOut}\rightarrow\text{CLSOut},a} = t_{\text{CLSOut}} - t_{\text{IssuerOut}} = t_{\text{MitMOut}} - t_{\text{MitMIn}}$. This assumption isolates the adversarial effect. We define $\Delta_{\max}$ as the maximum delay an adversary can successfully impose on the control signal under worst case assumptions.

We assume that a TLS session between the aEMT and the SMGW exists for the entire duration of the attack and is used exclusively to transport control signals for the CLS, as well as any additional information required to ensure their validity. Additional TLS sessions may exist but are assumed to have no immediate relevance to the control signals under consideration. Furthermore, we assume that the customer premises contain only a single CLS-A which provides connectivity for a single CLS to the aEMT. This assumption does not imply that scenarios involving multiple CLS-A devices or multiple CLS devices are not vulnerable; rather, such scenarios may increase the complexity of identifying and targeting relevant control signal packets.

Regarding the WAN configuration profile of the CLS channel, we assume that the TLS session is configured to remain permanently active and is renewed every 48 hours, in accordance with the maximum permitted session duration defined in \cite{tech:bsi:tr-03109-1}. Additionally, we assume that no idle timeout is configured. Consequently, the TLS session under attack may remain active for up to 48 hours and is automatically re-established thereafter. These assumptions are reasonable, as such configurations are not prohibited by the SMI specifications \cite{tech:bsi:tr-03109-1,tech:bsi:tr-03116-3,tech:bsi:pp-smgw}. From a technical perspective such long-standing connections reduce the overhead by reducing the number of handshakes and therefore the traffic. No further assumptions are made about the TLS session beyond general compliance with the requirements specified in \cite{tech:bsi:pp-smgw,tech:bsi:tr-03109-1,tech:bsi:tr-03116-3}. Therefore, either TLS 1.2 (according to \cite{rfc5246}, with constraints from \cite{rfc8446} and \cite{tech:bsi:tr-03116-3}) or TLS 1.3 (according to \cite{rfc8446}, with constraints from \cite{tech:bsi:tr-03116-3}) may be used \cite{tech:bsi:tr-03109-1,tech:bsi:tr-03116-3,tech:bsi:pp-smgw}.

We further assume that the CLS channel is configured according to the communication scenarios HKS.TLSPROXY.SOCKS or HKS.TLSPROXY.SRV (formerly HKS3) \cite{tech:bsi:tr-03109-1}. In these scenarios, the CLS-A initiates a TLS session to the SMGW, which in turn establishes a TLS session with the aEMT \cite{tech:bsi:tr-03109-1}. As a result, the lifetimes of both TLS sessions are coupled: termination of the TLS session between the SMGW and CLS-A also terminates the TLS session between the SMGW and the aEMT, and vice versa \cite{tech:bsi:tr-03109-1,tech:bsi:tr-03109-5}. The CLS-A and aEMT do not have strict requirements to close the CLS channel after a certain amount of time \cite{tech:bsi:tr-03109-5,tech:bsi:tr-03109-6}.

Between the completion of the TLS handshake and the transmission of a control signal, a time interval of up to 48 hours may elapse, corresponding to the maximum permitted session duration \cite{tech:bsi:tr-03109-1}. However, if the session duration is calculated based on an earlier event (e.g., the first message of the TLS or TCP handshake), the effective time window for data transmission may be slightly shorter. Notably, neither the protection profiles \cite{tech:bsi:pp-smgw,tech:bsi:pp-hsm} nor the SMGW specifications \cite{tech:bsi:tr-03109-1-detail,tech:bsi:tr-03109-1}, including the associated test specifications \cite{tech:bsi:tr-03109-1-test}, clearly define which event serves as the reference point for session lifetime calculation. If the completion of the TLS handshake is used as the reference, it may theoretically be possible to manipulate the handshake timing to extend the effective session duration beyond 48 hours. Such an attack is not considered in detail in this work. Instead, we assume that the session lifetime is calculated based on the timestamp of TLS handshake completion. Additionally, REQ.HAN.Kommunikationsszenarien.61 recommends that no more than 10 seconds elapse between the initiation and completion of the TLS handshake \cite{tech:bsi:tr-03109-1}.

%% file: theory/capa.tex
To analyze potential delay attacks, based on the previously specified system architecture, we define a threat model that specifies the capabilities available to the adversary.

First, the adversary is assumed to be capable of establishing an active on-path Machine-in-the-Middle (MitM) position on the following communication paths within the Wide Area Network (WAN), without having access to the GWA, the aEMT, or the SMGW:

\begin{itemize}
\item GWA $\rightleftharpoons$ SMGW
\item aEMT $\rightleftharpoons$ SMGW
\end{itemize}

If the attack can be executed in a manner that prevents detection by the aEMT, a simplified threat model can be considered in which the adversary only requires a MitM position on the communication path aEMT $\rightleftharpoons$ SMGW. In this work, we refer to this as Threat Model 2 (TM 2), whereas the threat model that additionally includes capabilities on the communication path GWA $\rightleftharpoons$ SMGW is referred to as Threat Model 1 (TM 1).

Within this position, the adversary is capable of capturing, dropping, delaying, and injecting packets, and of compromising any protocol layer below the TLS layer. However, the adversary is cryptographically bounded and therefore cannot read or modify the application-layer payload (e.g., control signals) protected by TLS.

Second, the adversary is assumed to possess contextual knowledge that enables the identification of relevant packets. In particular, the adversary must be able to determine whether a packet exchanged between the aEMT and the SMGW contains a control signal relevant to the attack objective. Additionally, the adversary must be able to identify whether packets on the communication path GWA $\rightleftharpoons$ SMGW contain requests to reset the CLS channel. Under the technical assumptions made in \cref{sec:theory:scenario}, the adversary's assumed ability to infer the relevance of TLS application data is further justified. Specifically, after completion of the TLS handshake, exchanged TLS application data is assumed to consist exclusively of control signals or TLS control messages. For the latter, we assume that no such messages are exchanged during the period relevant to the attack.

Importantly, the adversary does not need to have access to the plaintext of the encrypted payload. Instead, such knowledge may be inferred from traffic analysis techniques, such as packet size analysis, timing characteristics, communication frequency patterns, and correlations between multiple TLS streams. TLS 1.3 explicitly acknowledges traffic analysis as a residual risk and states that, in general, no complete mitigation covering all side channels is known \cite{rfc8446}, which supports the assumptions made in this work. Prior research has demonstrated the practical feasibility and impact of such techniques \cite{paper:WhyYouWentToClient,paper:HTTPSFingerprinting}. However, TLS 1.3 specifies an optional padding functionality, which hides the actual packet size on a per-TLS-record basis \cite{rfc8446}.

These assumptions extend the capabilities of an active attacker, as defined in the threat model for TLS \cite{rfc3552}, and those of a remote attacker, as defined in the threat model for the SMGW \cite{tech:bsi:pp-smgw}, with the possession of contextual knowledge. However, the SMI specification explicitly assumes an adversary capable of manipulating WAN communication; consequently, TLS, in combination with the Smart Metering Public Key Infrastructure (SM-PKI) \cite{tech:bsi:tr-03109-4,tech:bsi:certPolicy}, is mandated to fulfill the security objectives defined in \cite{law:deu:msbg} and further specified in \cite{tech:bsi:pp-smgw}. In particular, the threat T.DataModificationWAN explicitly describes a remote attacker's objective to influence the behavior of a CLS by modifying WAN communication \cite{tech:bsi:pp-smgw}.

Within this work, the adversary's objective is to delay the execution of control signals transmitted via the CLS channel, i.e., to shift their physical impact on the power grid by an attacker-controlled duration. The potential consequences of such delays are discussed in detail in \cref{sec:impact}.

The operational structure of the German energy market relies heavily on 15-minute intervals for billing and forecasting. From this perspective, a delay exceeding 15 minutes can be considered a clear disruption and, thus, a successful attack. Consequently, we define a critical threshold $\Delta_{\text{crit}}=15\,\text{minutes}$ and an attack as successful if $\Delta_{\text{IssuerOut}\rightarrow\text{CLSOut}}>15\,\text{minutes}$ under either threat model. Nevertheless, it is important to note that even shorter delays may still adversely affect the power grid's performance and stability \cite{Paper:Delay:SmartGridAGC,tech:uenb:pq-bedingungen,tech:uenb:rr-itAnf,tech:bnetza:gpke}.

%% file: theory/delay.tex
\subsubsection*{Foundation}
A delay attack is a network-based attack in which an adversary records and drops parts of a communication stream or individual packets at a given point in time and subsequently forwards or injects the captured stream or packets at a later time \cite{Paper:Delay:phantomDelay,Paper:Delay:DelayAttackPTP}. In contrast to replay and denial-of-service (DoS) attacks, the captured stream or packets are delivered exactly once — neither multiple times (as in replay attacks) nor not at all (as in DoS attacks). Furthermore, the transmitted application-layer data is not modified.

Such a distinction between replay and delay attacks is not always made in the literature. Consequently, delay attacks are sometimes considered a subset or trivial variant of replay attacks \cite{tech:bsi:pp-smgw,tech:iso:15408_2,paper:replay:noneLinarNetworkedSystem}. However, we explicitly adopt this distinction here, as it aligns with the definition used in the informal threat model for TLS~1.3 \cite{rfc3552,rfc8446}.

To formally describe the delay attack, we consider the transmission of control signal message $m$ between two points $p_x$ and $p_y$. For a successful delay attack we require that $\Delta_{\text{crit}} \le \Delta_{p_x\rightarrow p_y} < \infty$. For our system architecture $p_x$ corresponds to $\text{IssuerOut}$ and $p_y$ corresponds to $\text{CLSOut}$.

Using our formal description we describe the delay of the original message $m$ with $\Delta_{p_x\rightarrow p_y}=\Delta_{p_x\rightarrow p_y,n}$ and the delay of the replayed message $m'$ with $\Delta_{p_x\rightarrow p_y}=\Delta_{p_x\rightarrow p_y,n}+\Delta_{p_x\rightarrow p_y,a}$ in case of a replay attack. However, our threat model does not allow $m'$ because the adversary is cryptographically bounded and TLS guarantees freshness for 1-RTT application-data. For a DoS attack we can describe the adversarial delay as $\Delta_{p_x\rightarrow p_y,a}=\infty$, which results in $\Delta_{p_x\rightarrow p_y}=\infty$. An unsuccessful delay attack ($\Delta_{p_x\rightarrow p_y} > \Delta_{\max}$) effectively becomes a DoS attack, since $m$ will no longer be processed by receiver $p_y$. Hence, our formal model distinguishes between replay, delay, and DoS attacks.

A possible mitigation mechanism for detecting delay attacks is the use of timestamps transmitted within the same packet as the control signal in a manner that prevents forgery, for example as part of the TLS application data \cite{Paper:Delay:phantomDelay}. Another approach is the detection of statistical deviations or anomalies in the round-trip time (RTT) \cite{Paper:Delay:DelayAttackPTP,Paper:Delay:MPMediator}. The latter approach is mandated by the National Metrology Institute (German: Physikalisch-Technische Bundesanstalt, PTB) for time synchronization between the SMGW and GWA \cite{tech:ptb:timesynchSMGW}.

To fulfill the time synchronization requirement of the PTB, TR-03109-1 specifies that the maximum RTT between the SMGW and GWA must remain below the maximum allowed time deviation between the SMGW and the PTB reference time \cite{tech:bsi:tr-03109-1}. If this condition is violated, the received time value is considered invalid until a subsequent synchronization attempt satisfies this requirement, as defined in Section 3.10.5.1 of \cite{tech:bsi:tr-03109-1}. TR-03109-5 defines a similar mechanism for the CLS-A, with a fixed maximum RTT of 9 seconds, as specified in REQ.FA.DoTimeSync.40 of \cite{tech:bsi:tr-03109-5}. However, these mechanisms apply exclusively to time synchronization and do not extend to other communication aspects of the SMGW or CLS-A \cite{tech:bsi:tr-03109-1,tech:bsi:pp-smgw,tech:bsi:tr-03109-5}.

In addition to explicit detection mechanisms, the maximum achievable delay may be implicitly constrained by timeout mechanisms and maximum session-duration limits enforced by communication endpoints \cite{dis:ChenglongFu}. These include timeouts triggered by liveness checks or heartbeat mechanisms. For the SMGW, such a constraint is given by the maximum TLS session lifetime of 48 hours, as discussed earlier in \cref{sec:theory:scenario}. If a sender resends the same control signal using a new TLS session due to events controllable by the adversary, the maximum delay is not restricted by the TLS session lifetime. We restrict our analysis to attacks targeting a single TLS session.

The CLS channel is considered a trusted and secure channel within the SMI specification due to the use of TLS. Consequently, no explicit requirements are defined for proxied application data or the application-layer protocol between the aEMT and the CLS-A, as stated in Section 4.2.11.1 of \cite{tech:bsi:tr-03109-1} and Sections 4.4.4.4, 4.4.5.4, and 4.4.6.4 of \cite{tech:bsi:tr-03109-5}. This allows the use of a wide range of protocols within the CLS channel (e.g., CLS-EEDI or IEC 61850) between the aEMT and the CLS-A, including protocols that are intended to be vulnerable to delay attacks.

Fu et al. demonstrated that delay attacks may be feasible even when TLS is used. Furthermore, they showed that, under certain conditions, such attacks can remain stealthy, i.e., undetectable by both the client and the server \cite{Paper:Delay:phantomDelay}. 

If an aEMT detects suspicious behavior within the TLS session or the associated TCP session, it may request the GWA to terminate the affected session, for example by deleting the corresponding communication profile or rebooting the SMGW \cite{tech:bsi:tr-03109-1}. Such actions may be triggered by availability concerns on the side of the aEMT.

\subsubsection*{Formal Model of Vulnerability}
To be able to determine the vulnerability of the system architecture for delay attacks we need to formally define degrees of vulnerability.

Let $C=\{\text{Issuer},\text{aEMT},\text{SMGW},\text{CLS-A},\text{CLS}\}$ be the set of all components in the system architecture, $L_c$ the set of communication layer implementations of $c\in C$ which are ordered by the location of each layer in the network stack, $\mathcal{T} = \{\textsf{TM1}, \textsf{TM2}\}$ the set of threat models defined in \cref{sec:theory:capabilities} and $D = \{ \textsf{v}, \textsf{lv}, \textsf{i}\}$ the set of possible degrees of vulnerability where $\textsf{v}$ denotes \textit{vulnerable}, $\textsf{lv}$ denotes \textit{limited vulnerable} and $\textsf{i}$ denotes \textit{invulnerable}.

The vulnerability of an individual layer can be determined by the function $V_{\text{layer}} : L_c\times \mathcal{T} \rightarrow D$ defined as
\[
V_{\text{layer}}(l,t) =
\begin{cases}
\textsf{i}, & \text{if } l \text{ implemented by } c \text{ provides mandatory mechanisms } \\
            & \text{or optional mechanisms that are mandated to be used by other layer} \\
            & \text{to guarantee } \Delta_{\max} < \Delta_{\text{crit}} \text{ under } t \\

\textsf{lv}, & \text{if } l \text{ implemented by } c \text{ provides optional mechanisms } \\
             & \text{which are neither mandated nor prohibited by other layers } \\
             & \text{to guarantee } \Delta_{\max} < \Delta_{\text{crit}} \text{ under } t \\

\textsf{v}, & \text{if } l \text{ implemented by } c \text{ provides no mechanisms } \\
            & \text{or only optional mechanisms that are prohibited by other layers } \\
            & \text{to guarantee } \Delta_{\max} < \Delta_{\text{crit}} \text{ under } t \\
\end{cases}
.\]
Further we define an aggregation function $V_{\text{comp}}\colon C\times \mathcal{T} \rightarrow D$ to determine the degree of vulnerability of a component as follows:

\[V_{\text{comp}}(c,t) =
\begin{cases}
\textsf{i}, & \text{if } \exists l \in L_c : V_{\text{layer}}(l,t) = \textsf{i} \\
\textsf{lv}, & \text{if } \forall l \in L_c : V_{\text{layer}}(l,t) \neq \textsf{i} \; \land \; \exists l \in L_c : V_{\text{layer}}(l,t) = \textsf{lv} \\
\textsf{v}, & \text{if } \forall l \in L_c : V_{\text{layer}}(l,t) = \textsf{v} \\
\end{cases}\]
since no layers must guarantee $\Delta_{\max} < \Delta_{\text{crit}}  \text{ under } t$ for a Delay Attack to be successful.

The vulnerability of the overall system architecture is defined by an aggregation function $V_{\text{sys}} : \mathcal{P}(C)\setminus\emptyset \times T \rightarrow D$, defined by:
\[V_{\text{sys}}(C,t) =
\begin{cases}
\textsf{i}, & \text{if } \exists c \in C : V_{\text{comp}}(c,t) = \textsf{i}, \\
\textsf{lv}, & \text{if } \forall c \in C : V_{\text{comp}}(c,t) \neq \textsf{i} \; \land \; \exists c \in C : V_{\text{comp}}(c,t) = \textsf{lv}, \\
\textsf{v}, & \text{if } \forall c \in C : V_{\text{comp}}(c,t) = \textsf{v}.
\end{cases}\]
since no component must guarantee $\Delta_{\max} < \Delta_{\text{crit}}  \text{ under } t$ for a Delay Attack to be successful.

To calculate $\Delta_{\max}$ we define the function 
$\delta^{\max}_{\text{sys}} : \mathcal{P}(C)\setminus\emptyset\times T \rightarrow \mathbb{R}$, given by \[\delta^{\max}_{\text{sys}}(C,t) = \min(\{\delta^{\max}_{\text{comp}(c,t)} | c \in C\})\] 
with $\delta^{\max}_{\text{comp}} : C\times T \rightarrow \mathbb{R}$, given by \[\delta^{\max}_{\text{comp}}(c,t) = \min(\{\delta^{\max}_{\text{layer}(l,t)} | l \in L_c\})\] 
with $\delta^{\max}_{\text{layer}} : L_c\times T \rightarrow \mathbb{R}$, given by $\delta^{\max}_{\text{layer}}(l,t)$ with the maximum delay an adversary can enforce on  $l$ with $l$ configured and implemented according to all applicable specifications and subject to all constraints and requirements enforced by upper layers under $t$.

Finally, we assume that the SMGW is not capable of interpreting or interacting with the application-layer protocol encapsulated within the CLS channel. Therefore, the classification $V_{comp}(\text{SMGW},t)$ is independent of the protocol used within the CLS channel, which is consistent with the design assumptions of the CLS channel specification \cite{tech:bsi:tr-03109-5,tech:bsi:tr-03109-1}. 

\subsubsection*{Generalized Delay Attack}

To describe a delay attack on the system architecture with the intention of manipulating the behavior of a CLS, as outlined in \cref{fig:theory:scenario}, we first describe and analyze the intended interactions between the different components. Based on this, we then introduce the adversary with the capabilities outlined in \cref{sec:theory:capabilities} into the interaction. This enables a comparison between benign and compromised interactions.

\begin{figure}[h]
    \centering
    \begin{subfigure}{0.49\textwidth}
        \centering
        \includegraphics[width=0.99\textwidth]{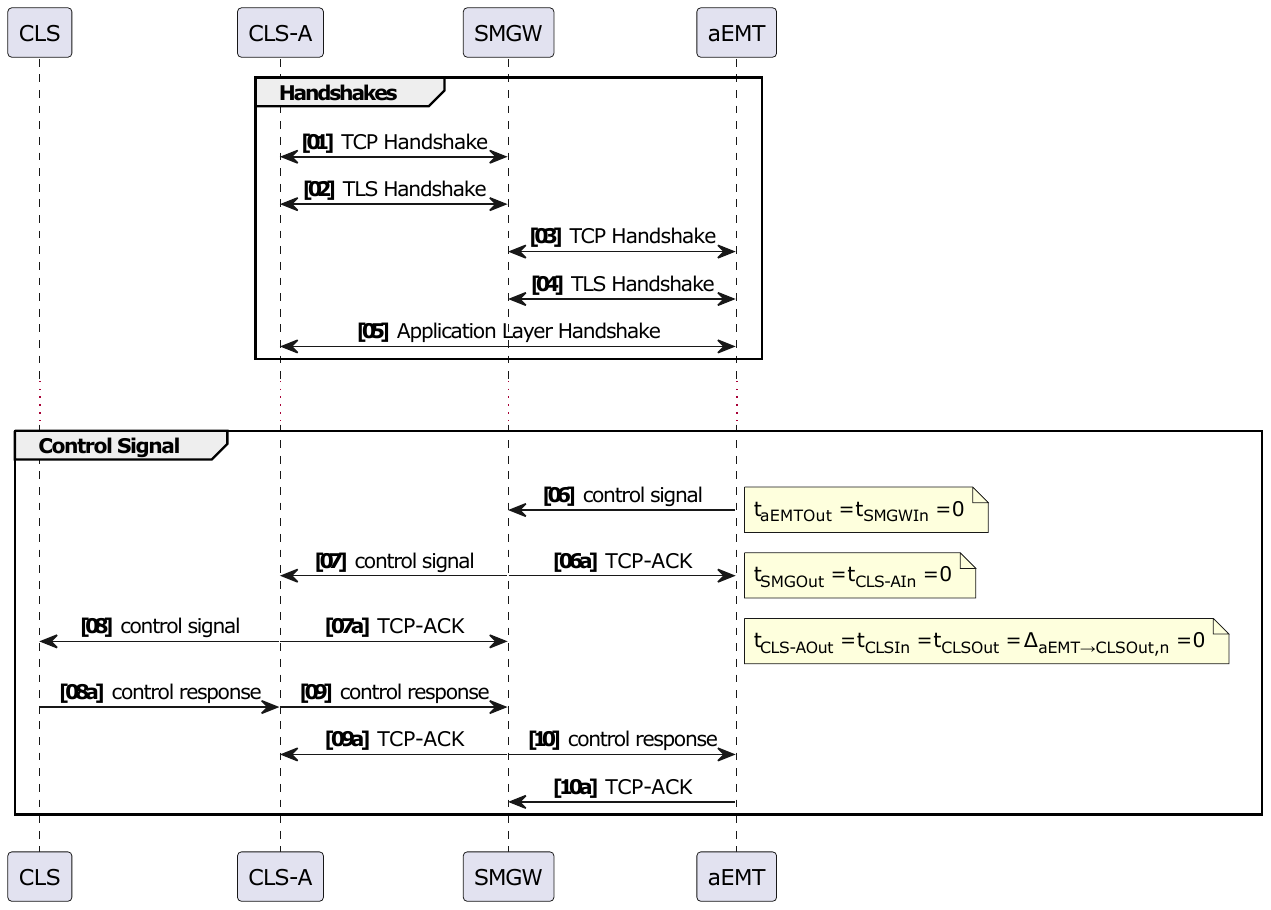}
        \caption{Intended}
        \label{fig:theory:delay:normal}
    \end{subfigure}
    \begin{subfigure}{0.49\textwidth}
        \centering
        \includegraphics[width=0.99\textwidth]{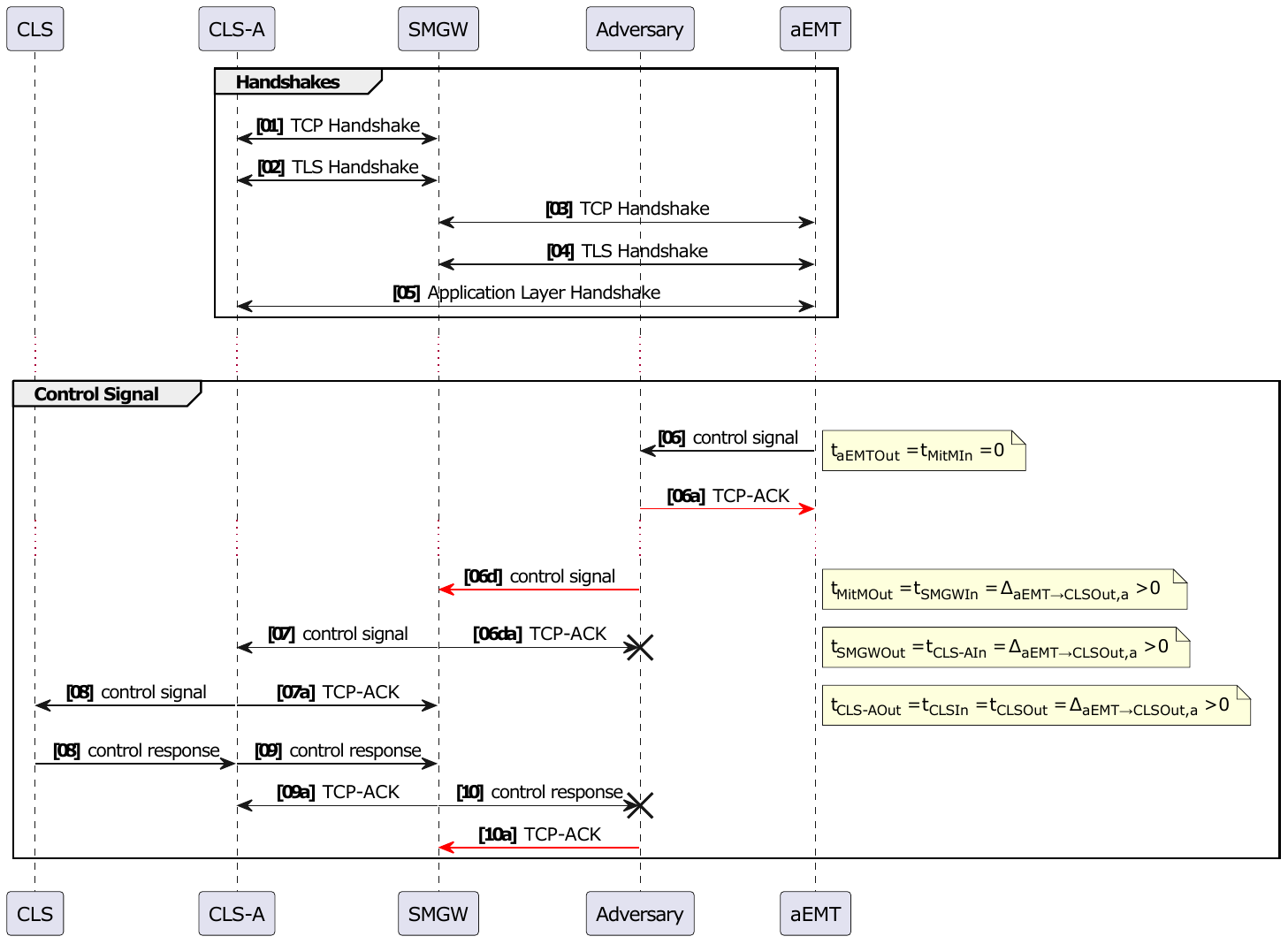}
        \caption{Attacked}
        \label{fig:theory:delay:attacked}
    \end{subfigure}
    \caption{Sequence diagram of a Generalized Delay Attack}
    \label{fig:theory:delay}
\end{figure}

\Cref{fig:theory:delay} illustrates an example of the interaction between the components of the system architecture for the purpose of issuing and transmitting a control signal, as well as the corresponding response. The Issuer is considered as part of the aEMT. The TCP, TLS, and application-layer handshakes (messages 01 to 05) are illustrated abstractly due to their irrelevance for the considered attack. The control signals transmitted via the SMGW's CLS channel are represented by messages 06 and 07. Messages 06a and 07a represent the TCP acknowledgments (ACKs) corresponding to messages 06 and 07, respectively. Message 08 represents the control signal forwarded to the CLS, which is directly connected to the CLS-A. This connection may use either digital or analog interfaces \cite{tech:bsi:tr-03109-5}. However, this is not relevant for the analysis, as neither the CLS-A, the CLS, nor the connection between CLS-A and CLS is directly targeted by the attack. After the control signal arrives at the CLS, we assume that the CLS responds after successfully implementing the control signal (message 08a). This response is transmitted back to the aEMT via messages 09 and 10. Messages 09a and 10a correspond to the TCP ACKs for messages 09 and 10, respectively. Note that such a response is not strictly required but is considered here for completeness.

By extending \cref{fig:theory:delay} with an adversary, it becomes possible to describe a delay attack on the control signal based on the previously identified weaknesses in the SMGW specifications. A corresponding sequence diagram is shown in \cref{fig:theory:delay}. As illustrated, the adversary forwards the TCP and TLS handshake messages (messages 03 and 04) without modification or delay. Notably, the adversary does not require specific capabilities at this stage of the communication and may therefore also target an already established TLS session.

When the control signal (message 06) is transmitted, the adversary captures the message and spoofs the corresponding TCP ACK (message 06a) by injecting a TCP packet, thereby preventing the aEMT from detecting any anomaly. Consequently, the aEMT does not retransmit the message, as the TCP layer indicates successful delivery. After a delay chosen by the adversary, it forwards the captured message to the SMGW (message 06d). The corresponding TCP ACK (message 06da) is dropped by the adversary to avoid revealing the attack to the aEMT.

Subsequent processing within the premises proceeds as if no delay had been introduced. The adversary further suppresses evidence of the attack by dropping the response message (message 10). To prevent retransmissions and avoid connection termination by the SMGW or CLS-A, the adversary spoofs the corresponding TCP ACK (message 10a), thereby preventing the CLS-A from detecting any irregularity.

Since our threat model allows an adversary to manipulate the layer TCP and below for the component aEMT and SMGW we do not need to analyze each layer individually and can subsum the layers TCP and beneath as the layer TCP. We consider the ability of the GWA to reset the TLS connection upon request of the aEMT as part of the aEMT. In case of the CLS-A the layers TLS and beneath are created freshly when the SMGW forwards data received on the TLS connection sent by the aEMT. Consequently, the layer sets are defined as $L_{\text{aEMT}}=APPS_{\text{aEMT}} \cup \{TLS_{\text{aEMT}},TCP_{\text{aEMT}}\}$, $L_{SMGW}=\{TLS_{\text{SMGW}},TCP_{\text{SMGW}}\}$ and $L_{\text{CLS-A}}=APPS_{\text{CLS-A}} \cup \{TLS_{\text{CLS-A}}\}$. $APPS_c \subset L_c$ defined as $APPS_c = \{APP^i_c|0 \leq i \leq n, n\in \mathbb{N} \}$ is a placeholder for all the application layer protocols $APP^i_c$ implemented by component $c \in C$ and will be specified later. 

For the aEMT we can reason the following:
\[\forall t \in T : (V_{\text{layer}}(TCP_{\text{aEMT}},t)= \textsf{v} \; \land \; \delta^{\max}_{layer}(TCP_{\text{aEMT}},t)=\infty)\]
since the adversary's capabilities allow full control over the TCP layer.
\[\forall t \in T : (V_{\text{layer}}(TLS_{\text{aEMT}},t)= \textsf{v} \; \land \; \delta^{\max}_{layer}(TLS_{\text{aEMT}},t)=\infty)\]
since TLS does not bound transmission delay for application data \cite{rfc8446,Paper:Delay:phantomDelay}.

For the aEMT's application-layer protocol implementation we assume
\[\forall APP^i_{\text{aEMT}} \in APPS_{\text{aEMT}} : (V_{\text{layer}}(APP^i_{\text{aEMT}},\textsf{TM1})= \textsf{v} \; \land \; \delta^{\max}_{layer}(APP^i_{\text{aEMT}},\textsf{TM1})=\infty)\]
and 
\[ \exists APP^i_{\text{aEMT}} \in APPS_{\text{aEMT}} : (V_{\text{layer}}(APP^i_{\text{aEMT}},\textsf{TM2})= \textsf{lv} \; \land \; \delta^{\max}_{layer}(APP^i_{\text{aEMT}},\textsf{TM2})=\infty)\]
since the aEMT may detect anomalies (e.g., missing control response) but any countermeasures such as connection resets can be prevented by the adversary under the TM 1 but not under TM 2.

Consequently, 
\[V_{\text{comp}}(\text{aEMT},\textsf{TM1})=\textsf{v} \; \land \; \delta^{\max}_{layer}(APP_{\text{aEMT}},\textsf{TM1})=\infty\]  
and
\[V_{\text{comp}}(\text{aEMT},\textsf{TM2})=\textsf{lv} \; \land \; \delta^{\max}_{layer}(APP_{\text{aEMT}},\textsf{TM2})=\infty.\] 

For the SMGW we can reason that 
\[\forall t \in T : (V_{\text{layer}}(TCP_{\text{SMGW}},t)= \textsf{v} \; \land \; \delta^{\max}_{layer}(TCP_{\text{SMGW}},t)=\infty)\]
since our threat models give the adversary full control over the TCP layer and below within the WAN.

Further,
\[\forall t \in T : (V_{\text{layer}}(TLS_{\text{SMGW}},t)= \textsf{v} \; \land \; \delta^{\max}_{layer}(TLS_{\text{SMGW}},t)=48\, \text{hours})\]
using the same argumentation as for $TLS \in L_{aEMT}$ and additionally taking limitation of the CLS channel's lifetime of 48 hours \cite{tech:bsi:tr-03109-1} into account.

Consequently, 
\[\forall t \in T : (V_{\text{comp}}(\text{SMGW},t)=\textsf{v} \; \land \; \delta^{\max}_{comp}(\text{SMGW},t)=48\, \text{hours})\]

For the CLS-A we can reason the following:
\[ \forall t \in T : (V_{\text{layer}}(TLS_{\text{CLS-A}},t)= \textsf{v} \; \land \; \delta^{\max}_{layer}(TLS_{\text{CLS-A}},t)=48\, \text{hours})\]
since packets on this connection are freshly created and the lifetime of the TLS session is restricted to 48 hours \cite{tech:bsi:tr-03109-1}. TLS closeNotify must be processed by the CLS-A \cite{tech:bsi:tr-03109-5} and cannot be prevented by our threat model as this connection resides within the HAN-CLS.

For the CLS-A's application-layer protocol implementation, due to missing specification \cite{tech:bsi:tr-03109-1,tech:bsi:tr-03109-5}, we assume no timing guarantees:
\[\forall APP^i_{\text{CLS-A}} \in APPS_{\text{CLS-A}}, \forall t \in T : (V_{\text{layer}}(APP^i_{\text{CLS-A}},t)= \textsf{v} \; \land \; \delta^{\max}_{layer}(APP^i_{\text{CLS-A}},t)=\infty)\]

Consequently,
\[\forall t \in T : (V_{\text{comp}}(\text{CLS-A},t)=\textsf{v} \; \land \; \delta^{\max}_{comp}(\text{CLS-A},t)=48\, \text{hours})\]

Finally, we can reason that:
\[V_{\text{sys}}(\{\text{aEMT},\text{SMGW},\text{CLS-A}\},\textsf{TM1}) = \textsf{v} \; \text{with} \; \delta^{\max}_{sys}(\{\text{aEMT},\text{SMGW},\text{CLS-A}\},\textsf{TM1})=48\,\text{hours}\]
and
\[V_{\text{sys}}(\{\text{aEMT},\text{SMGW},\text{CLS-A}\},\textsf{TM2}) = \textsf{lv} \; \text{with} \; \delta^{\max}_{sys}(\{\text{aEMT},\text{SMGW},\text{CLS-A}\},\textsf{TM2})=48\,\text{hours.}\]

\subsubsection*{Impact}
With respect to the use case of the information transported between two parties, a delay attack can be considered an attack on the availability property of the transmitted information, as the information is unavailable when required. If a system (e.g., a CLS) additionally requires the timely arrival of information to maintain integrity, delaying such information also constitutes a violation of the integrity property. In contrast to availability the integrity is within the scope of the SMGW's common criteria protection profile \cite{tech:bsi:pp-smgw}.

\begin{figure}[h]
    \centering
    \begin{subfigure}{0.24\textwidth}
        \centering
        \includegraphics[width=0.99\textwidth]{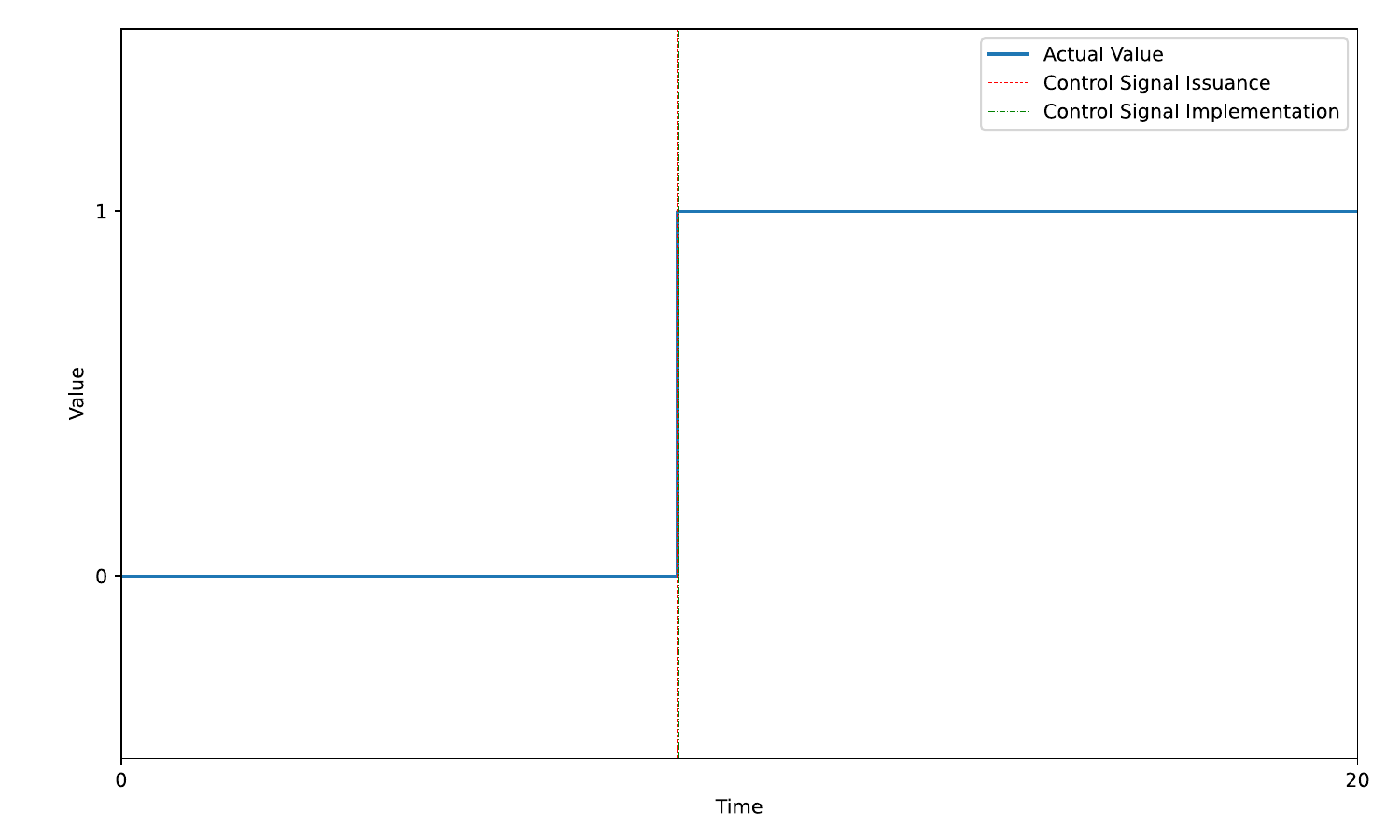}
        \caption{Intended}
        \label{fig:theory:delay:graph:Normal}
    \end{subfigure}
    \begin{subfigure}{0.24\textwidth}
        \centering
        \includegraphics[width=0.99\textwidth]{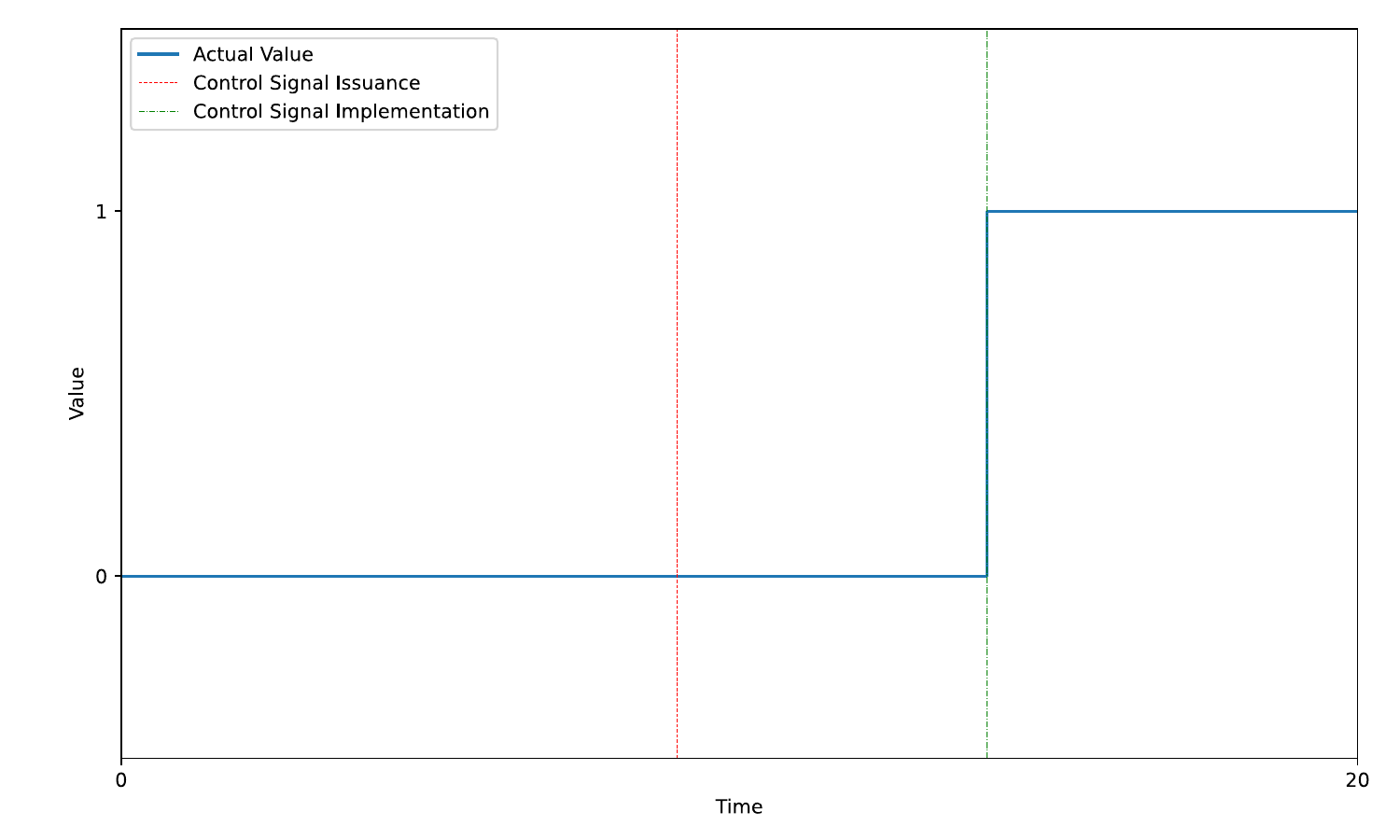}
        \caption{Delay-Attack}
        \label{fig:theory:delay:graph:delay}
    \end{subfigure}
    \begin{subfigure}{0.24\textwidth}
        \centering
        \includegraphics[width=0.99\textwidth]{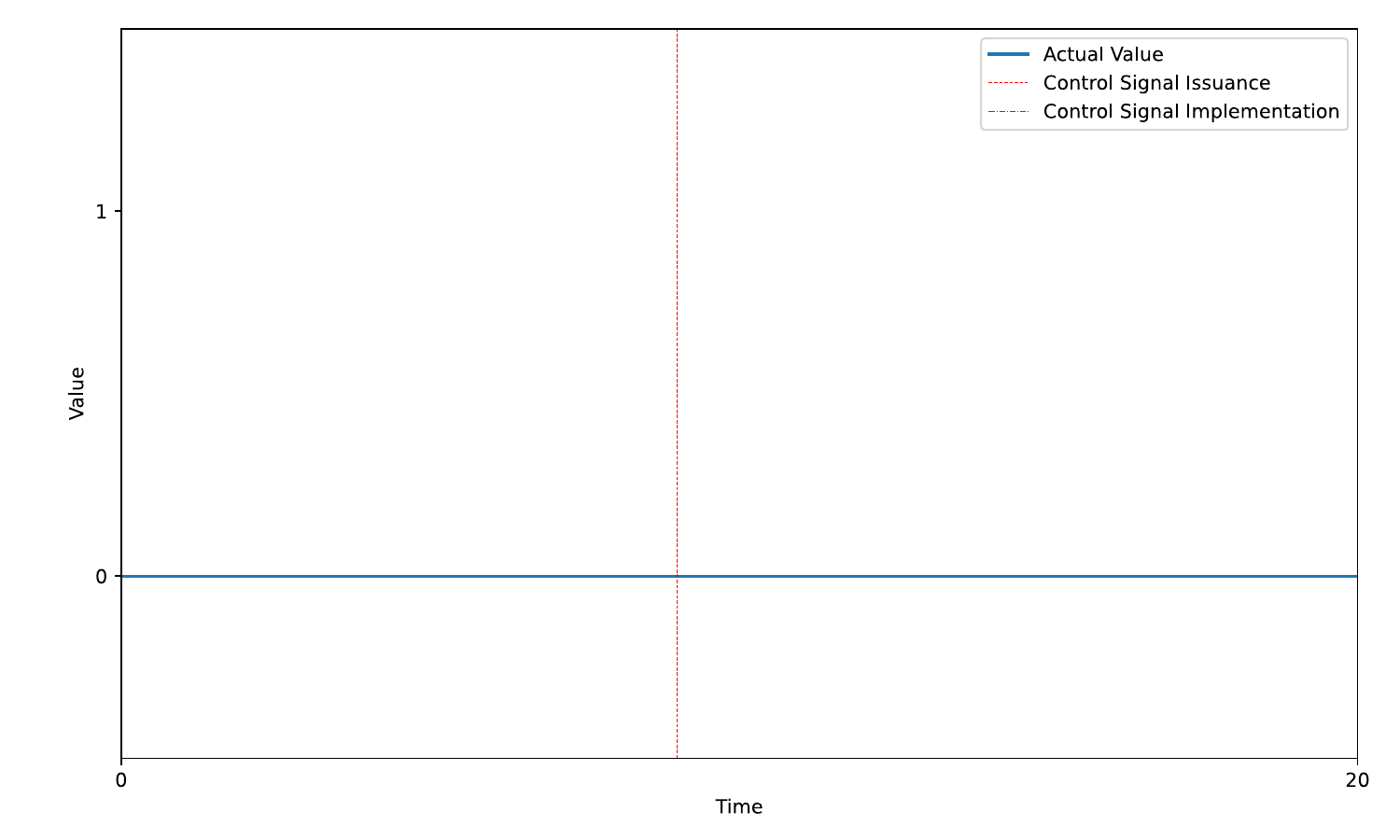}
        \caption{DoS-Attack}
        \label{fig:theory:delay:graph:dos}
    \end{subfigure}
    \begin{subfigure}{0.24\textwidth}
        \centering
        \includegraphics[width=0.99\textwidth]{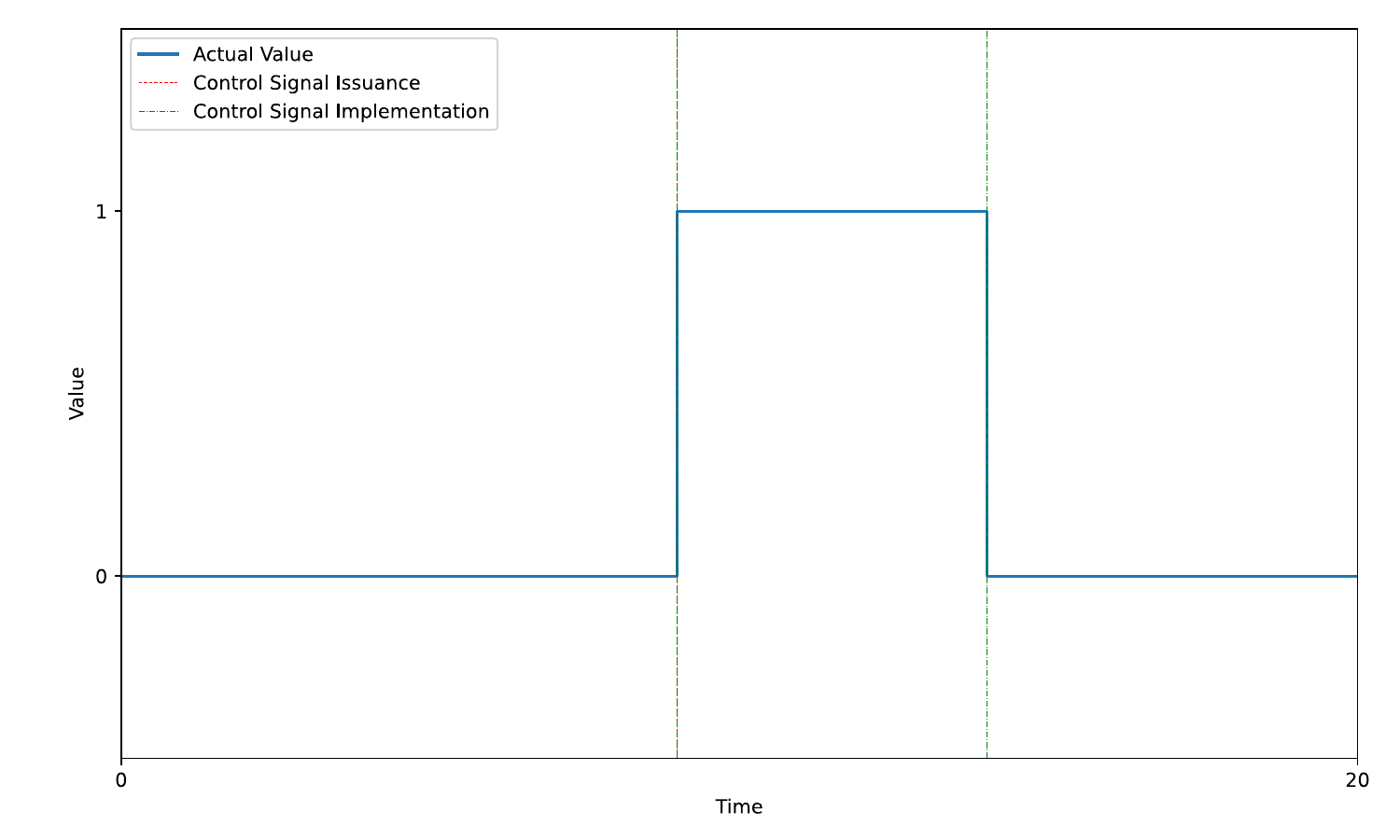}
        \caption{Replay-Attack}
        \label{fig:theory:delay:graph:replay}
    \end{subfigure}
    \caption{Impact of different attacks to an abstract binary state}
    \label{fig:theory:delay:graph}
\end{figure}

A more practical illustration of an abstract value transition at the CLS is provided in \cref{fig:theory:delay:graph}, where a delay attack (\cref{fig:theory:delay:graph:delay}) is contrasted with the intended behavior (\cref{fig:theory:delay:graph:Normal}), as well as with the behavior under a DoS attack (\cref{fig:theory:delay:graph:dos}) and a replay attack (\cref{fig:theory:delay:graph:replay}).

\Cref{fig:theory:delay:graph:delay} illustrates that, under a delay attack, the control signal arrives at a later point in time compared to the intended behavior shown in \cref{fig:theory:delay:graph:Normal}. In contrast, under a DoS attack, the control signal never arrives; consequently, no state transition occurs. As shown in \cref{fig:theory:delay:graph:replay}, replaying the control signal results in an additional state transition.

%% file: theory/dap.tex
Due to the absence of requirements for protocols encapsulated in the CLS channel, we define a simple plaintext protocol, referred to as the Dumbest Assumable Protocol (DAP). DAP is intentionally designed to be vulnerable to delay attacks under both threat models and serves as a reference protocol for vulnerability analysis. It consists of a single plaintext message (control signal) that allows an aEMT to configure the power-flow settings of a CLS. The aEMT does not expect any response from the CLS, and DAP does not define any response messages.

\begin{figure}[h]
    \centering
    \begin{subfigure}{0.49\textwidth}
        \centering
        \includegraphics[width=0.99\textwidth]{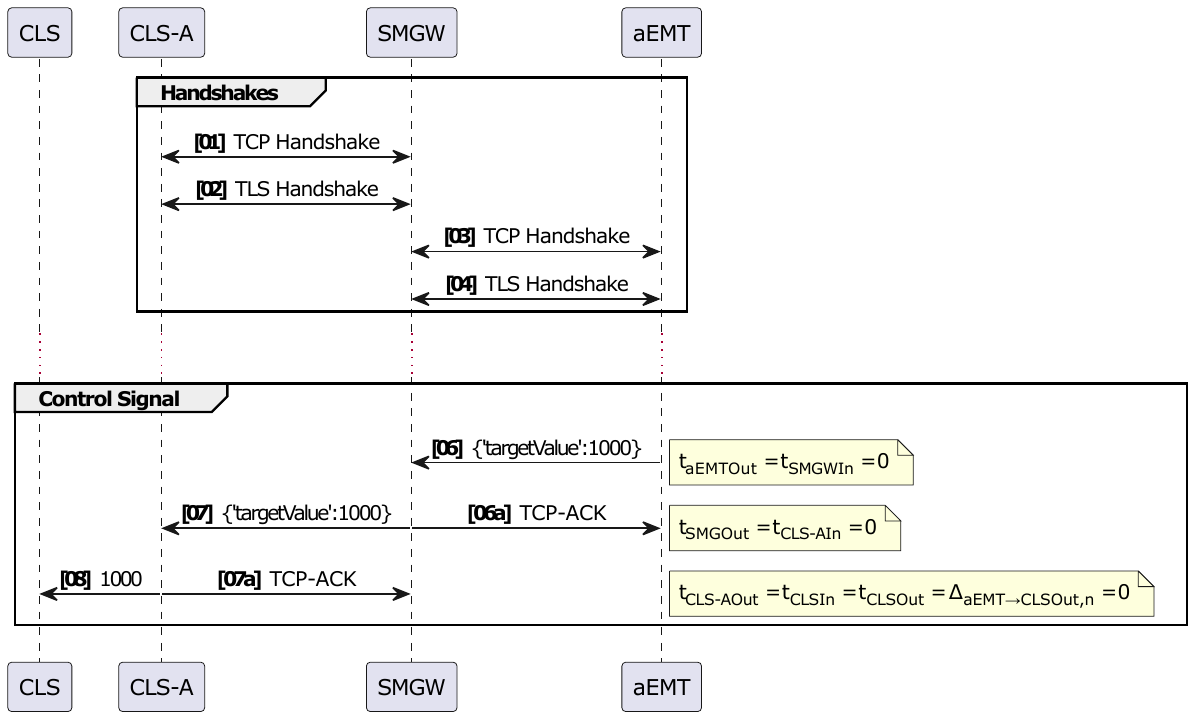}
        \caption{Sequence diagram of the DAP}
        \label{fig:theory:dap:seq}
    \end{subfigure}
    \begin{subfigure}{0.49\textwidth}
        \centering
        \includegraphics[width=0.99\textwidth]{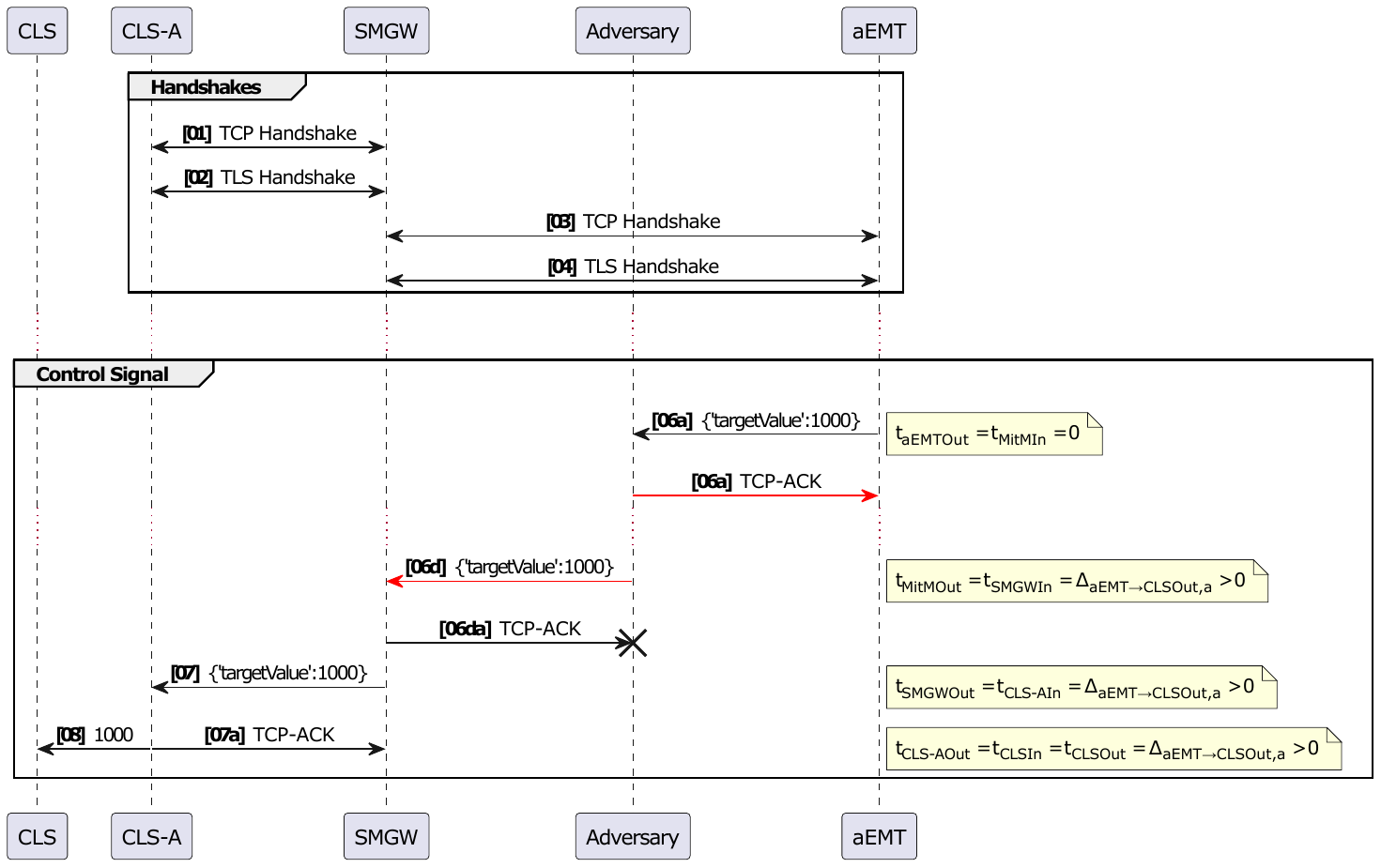}
        \caption{Sequence diagram of the DAP under attack}
        \label{fig:theory:dap:seqAttacked}
    \end{subfigure}
    \caption{Sequence diagrams of DAP}
    \label{fig:theory:dap}
\end{figure}

A sequence diagram of the intended behavior of DAP is shown in \cref{fig:theory:dap:seq}. Since DAP does not include an application-layer handshake, message 05 is omitted. Messages 06 and 07 refer to the plaintext application-layer message encapsulated within TLS ApplicationData records. The description of the remaining messages and packets is analogous to that used in \cref{fig:theory:delay:normal}.

\Cref{fig:theory:dap:seqAttacked} illustrates a delay attack on DAP. As shown, the attack can be carried out easily by spoofing the TCP ACK (message 06a) corresponding to the control signal (message 06) and delaying its transmission (message 06d). By dropping the TCP ACK from the SMGW (message 06da), the aEMT receives no indication that the control signal has been delayed. The description of the remaining messages and packets is analogous to that used in \cref{fig:theory:delay:attacked}.

$L_{\text{SMGW}}$ is not modified by DAP. $APPS_{c}$ consists of the single layer $DAP_c$, leading to
\[L_{\text{aEMT}}=\{DAP_{\text{aEMT}},TCP_{\text{aEMT}},TLS_{\text{aEMT}}\} \; \text{and} \; L_{\text{CLS-A}}=\{DAP_{\text{CLS-A}},TCP_{\text{CLS-A}},TLS_{\text{CLS-A}}\}.\] 

Because DAP does not specify any requirements for $TCP_{\text{aEMT}}$, $TLS_{\text{aEMT}}$, $TCP_{\text{CLS-A}}$ and $TLS_{\text{CLS-A}}$, we can reuse the analysis of the generalized delay attack for these layers.

For the application-layer implementation of the aEMT we can reason that:
\[\forall t \in T : (V_{\text{layer}}(DAP_{\text{aEMT}},t)= \textsf{v} \; \land \; \delta^{\max}_{layer}(DAP_{\text{aEMT}},t)=\infty)\]

Consequently,
\[\forall t \in T : (V_{\text{comp}}(\text{aEMT},t)=\textsf{v} \; \land \; \delta^{\max}_{comp}(\text{aEMT},t)=\infty)\]

For the application-layer implementation of the CLS-A we can reason that:
\[\forall t \in T : (V_{\text{layer}}(DAP_{\text{CLS-A}},t)= \textsf{v} \; \land \; \delta^{\max}_{layer}(DAP_{\text{CLS-A}},t)=\infty)\]
since the data transmitted using DAP does not provide indication to determine its age.

Consequently,
\[\forall t \in T : (V_{\text{comp}}(\text{CLS-A},t)=\textsf{v} \; \land \; \delta^{\max}_{comp}(\text{CLS-A},t)=48\, \text{hours})\]

Finally, we can reason that:
\[\forall t \in T : (V_{\text{sys}}(\{\text{aEMT},\text{SMGW},\text{CLS-A}\},t) = \textsf{v} \; \land \; \delta^{\max}_{sys}(\{\text{aEMT},\text{SMGW},\text{CLS-A}\},t)=48\,\text{hours})\]

%% file: theory/clseedi.tex
\subsubsection*{General}
CLS.EEDI is an application-layer protocol that enables a backend system (e.g., the Issuer or an aEMT) to exchange information with a local system at the grid connection point, e.g. a CLS. It was specifically designed to integrate EEBUS-compatible CLS devices into the SMI via the CLS-A and the SMGW \cite{tech:standard:cls-eedi}.

The protocol supports the use cases LPC and LPP — also referred to as ad hoc limits — defined in VDE-AR 2829-6 \cite{tech:VDE:ARE_2829_6_1,tech:standard:cls-eedi}. Additionally, CLS.EEDI supports use cases for metering \cite{tech:standard:cls-eedi}, which we do not consider here.

CLS.EEDI has been implemented by BSI certified products \cite{tech:bsi:bszPpcCLSAdapterSteuerung,web:ppc:products} and is explicitly referenced in BSI TR-03109-5 \cite{tech:bsi:tr-03109-5}, underlining its practical relevance within the SMI ecosystem.

From a technical perspective, CLS.EEDI defines a JSON-based data model to represent the aforementioned use cases. Message exchange between the aEMT and the CLS-A is realized via the MQTT protocol \cite{tech:standard:cls-eedi}.

Consequently, $APPS_c$ consists of two layers, $CLS.EEDI_c$ and $MQTT_c$, yielding
\[L_{\text{aEMT}}=\{CLS.EEDI_{\text{aEMT}},MQTT_{\text{aEMT}},TCP_{\text{aEMT}},TLS_{\text{aEMT}}\},\]
and
\[L_{\text{CLS-A}}=\{CLS.EEDI_{\text{CLS-A}},MQTT_{\text{CLS-A}},TCP_{\text{CLS-A}},TLS_{\text{CLS-A}}\}.\] 

\subsubsection*{CLS.EEDI}

To analyze the vulnerability of CLS.EEDI (Version 1.2.0) to delay attacks, we examine the JSON schemata (data model) and the expected behavior defined in the specification. CLS.EEDI defines a single message type that encapsulates five different message payloads \cite{tech:standard:cls-eedi}. However, to demonstrate the general vulnerability, it is sufficient to consider only the payload types \textit{de.keo-connectivity.clseedi.control} and \textit{de.keo-connectivity.clseedi.ack}, as the remaining types are not related to active power control \cite{tech:standard:cls-eedi}.

Each message consists of a header and a payload \cite{tech:standard:cls-eedi}. The header is based on CloudEvents v1.0 and defines properties such as identifiers and relationships between messages \cite{tech:standard:cloudevents,tech:standard:cls-eedi}.

The payload type \textit{de.keo-connectivity.clseedi.control} contains two objects, \textit{limits} and \textit{failsafe}, which define production or consumption limits for the active power flow. The \textit{limit} object consists of the attributes \textit{value}, \textit{active}, and \textit{duration}. The attribute \textit{value} specifies the power limit in watts, while \textit{active} indicates whether the limit should be applied; in the following, we assume this attribute is set to true. The optional attribute \textit{duration} defines the time in seconds for which the limit remains active. CLS.EEDI does not define the behavior in the absence of this attribute \cite{tech:standard:cls-eedi}. For the purpose of this analysis, we assume a vulnerable case where the limit is applied for an infinite duration. %

The control signal does not include an absolute execution time or expiration timestamp \cite{tech:standard:cls-eedi}. Consequently, the duration is defined relative to the time of message reception at the CLS-A, or it is infinite if the optional \textit{duration} attribute is missing. A delayed message therefore remains semantically valid upon arrival, and its execution window is shifted. As a result, the CLS-A cannot detect whether a \textit{de.keo-connectivity.clseedi.control} message has been delayed based solely on its content.

The \textit{failsafe} object defines a fallback power limit that is applied in case of communication issues between the CLS-A and the CLS within the local network. Since the considered attack does not target this communication path, the \textit{limits} object is assumed to be effective.

Upon receiving a \textit{de.keo-connectivity.clseedi.control} message, the CLS-A must respond with a \textit{de.keo-conn\-ectivity.clseedi.ack} message indicating that the limit has been received and whether it can be applied by the CLS \cite{tech:standard:cls-eedi}. This acknowledgment is generated after the CLS-A has already forwarded the control signal to the CLS. While the aEMT may detect a missing acknowledgment, it cannot reliably distinguish between a delay attack and a system malfunction or unavailability if an attacker suppresses the \textit{de.keo-connectivity.clseedi.ack} message.

CLS.EEDI does not impose requirements on the $TCP_c$, $TLS_c$, or $MQTT_c$ layers for $c \in \{\text{CLS-A}, \text{SMGW}, \text{aEMT}\}$. 

Based on this analysis, we conclude 
\[\forall t \in T : (V_{layer}(\text{CLS.EEDI}_\text{CLS-A},t)=\textsf{v} \; \land \; \delta^{\max}_{layer}(\text{CLS.EEDI}_\text{CLS-A},t) = \infty)\]
as the CLS-A cannot distinguish between delayed and non-delayed control messages. Furthermore, 
\[V_{layer}(\text{CLS.EEDI}_\text{aEMT},\textsf{TM1})=\textsf{v} \text{ with } \delta^{\max}_{layer}(\text{CLS.EEDI}_\text{aEMT},\textsf{TM1}) = \infty\] and \[V_{layer}(\text{CLS.EEDI}_\textsf{aEMT},\text{TM2})=\textsf{lv} \text{ with } \delta^{\max}_{layer}(\text{CLS.EEDI}_\text{aEMT},\textsf{TM1}) < \Delta_{crit}\] 
following the same reasoning as for the generalized delay attack, since the \textit{de.keo-connectivity.clseedi.ack} message corresponds to the control response.
\subsubsection*{MQTT}

Since CLS.EEDI relies on MQTT for the transportation of CLS.EEDI messages, MQTT must also be analyzed. The CLS.EEDI specification does not mandate a specific MQTT version \cite{tech:standard:cls-eedi}. Therefore, MQTT version 5.0 (latest) \cite{tech:standard:mqtt-v5} is considered in the following.

In the considered system architecture, the MQTT broker is provided by the aEMT, while the Issuer and CLS-A act as MQTT clients. CLS.EEDI requires a dedicated MQTT topic per CLS and communication direction \cite{tech:standard:cls-eedi}. Messages addressed to the CLS-A use the topic \textit{clseedi/to-localdevice/A}, whereas messages addressed to the Issuer use \textit{clseedi/from-localdevice/A}, where \textit{A} denotes the device identifier.

The PUBLISH message is used to transmit CLS.EEDI control messages \cite{tech:standard:cls-eedi}. It consists of a fixed header and a variable header and payload \cite{tech:standard:mqtt-v5}. The fixed header includes, among others, the QoS level and flags for retransmission and retain. Since the CLS-A is an MQTT client the retain flag is irrelevant as this flag only applies to the MQTT broker \cite{tech:standard:mqtt-v5}.

The variable header includes the topic name, packet identifier, and additional properties. Among these properties, the Message Expiry Interval defines the maximum lifetime of an application message in seconds \cite{tech:standard:mqtt-v5}. The specification requires the broker to discard messages whose expiry interval has elapsed and to reduce the remaining expiry interval by the time the message is stored on the broker. However, the specification does not define constraints on transmission delays between sender and receiver. Therefore, delays introduced during transmission do not invalidate the message at the receiver, and the Message Expiry mechanism does not prevent delayed delivery once the message is in flight.

Additional MQTT control packets relevant for message delivery include PUBACK, PUBREC, PUBREL, and PUBCOMP \cite{tech:standard:mqtt-v5}. These packets contain a packet identifier, a reason code, and optional properties. However, neither the MQTT specification nor CLS.EEDI defines semantics for these fields that would allow a receiver to detect whether a message was delayed. Consequently, these packet types do not provide information enabling delay detection.

MQTT is a session-based protocol \cite{tech:standard:mqtt-v5}. The optional session expiry mechanism could, in principle, limit the maximum delay. However, the MQTT specification explicitly states that the session state must not be discarded if the network connection remains open \cite{tech:standard:mqtt-v5}. Consequently, delayed PUBLISH messages remain valid within an active session. Furthermore, MQTT sessions are independent of underlying transport connections, and the MQTT specification does not impose a strict upper bound on session lifetime \cite{tech:standard:mqtt-v5}. Therefore, MQTT does not impose a strict upper bound on the delay of messages within a session.

MQTT allows a client to request reauthentication after the initial authentication \cite{tech:standard:mqtt-v5}. In this context, an attacker may drop either the reauthentication request or the corresponding server response. In both cases, the client does not receive a response. Since the MQTT specification does not define a timeout or required reaction for this situation \cite{tech:standard:mqtt-v5}, the resulting behavior is implementation-specific. Furthermore, authentication and reauthentication are optional features according to the specification \cite{tech:standard:mqtt-v5}. Hence, no mandatory mechanism exists to bound message delay via reauthentication.

MQTT further specifies an optional keep-alive mechanism based on the PINGREQ and PINGRESP packet types \cite{tech:standard:mqtt-v5}. A client sends PINGREQ messages to the server, which responds with PINGRESP \cite{tech:standard:mqtt-v5}. The server expects to receive any MQTT control packet within 1.5 times the negotiated keep-alive interval, which is defined as a two-byte integer in seconds and therefore limited to $2^{16}-1 ~\text{seconds} \sim 18,2\, ~\text{hours}$. The specification states that the server may close the connection if this condition is violated, whereas the client is only recommended to close the connection if a PINGRESP is not received in time \cite{tech:standard:mqtt-v5}. Since no strict timeout behavior is mandated for the client, implementations may tolerate longer delays. Therefore, the keep-alive mechanism does not provide a guaranteed upper bound on message delay.

MQTT supports three QoS levels for message delivery \cite{tech:standard:mqtt-v5}. QoS 0 provides \textit{at most once} delivery, QoS 1 provides \textit{at least once} delivery, and QoS 2 provides \textit{exactly once} delivery. QoS 1 and QoS 2 guarantees are achieved through additional acknowledgment flows, while QoS 0 solely depends on a single PUBLISH message. However, the specification does not define timing constraints for these acknowledgment exchanges. Therefore, an attacker can delay or suppress acknowledgment packets without violating protocol requirements.

For QoS 1, the sender retransmits a PUBLISH message until a PUBACK is received. For QoS 2, a four-step acknowledgment flow involving PUBREC, PUBREL, and PUBCOMP is required. The specification does not define at which point the application message is processed, leaving this decision implementation-specific. Consequently, an attacker can influence the timing of message processing by delaying specific packets in the exchange while remaining compliant with the protocol behavior.

Furthermore, for QoS levels 1 and 2, reconnecting to an existing MQTT session causes retransmission of all PUBLISH messages for which the acknowledgment flow has not been completed, with the retransmission flag set \cite{tech:standard:mqtt-v5}. Since MQTT sessions can persist beyond individual TCP/TLS connections, and the MQTT specification does not impose a strict upper bound on session lifetime \cite{tech:standard:mqtt-v5}, retransmissions may occur after extended delays. This behavior increases the maximum achievable delay to the duration of the MQTT session, which may be unbounded.

MQTT does not impose requirements on the $TCP_c$ or $TLS_c$ layers for $c \in \{\text{CLS-A}, \text{SMGW}, \text{aEMT}\}$. 

In summary, MQTT specifies mechanisms to ensure reliable message delivery but does not define constraints on end-to-end message delay. As a result, the maximum delay $\Delta_{\max}$ is not bounded by the protocol specification.

Consequently,
\[\forall t \in T : (V_{layer}(MQTT_{\text{CLS-A}},t)=\textsf{lv} \; \land \; \delta^{\max}_{layer}(MQTT_{\text{CLS-A}},t)=\infty)\]
since $\Delta_{\max}$ is not bounded by the protocol but may be limited by specific configurations (e.g., keep-alive settings). Furthermore,
\[V_{layer}(MQTT_{\text{aEMT}},\textsf{TM1})=\textsf{v} \text{ with } \delta^{\max}_{layer}(MQTT_{\text{aEMT}},\textsf{TM1}) = \infty\]
and
\[V_{layer}(MQTT_{\text{aEMT}},\textsf{TM2})=\textsf{lv} \text{ with } \delta^{\max}_{layer}(MQTT_{\text{aEMT}},\textsf{TM2}) = \infty\]
following the same reasoning as for the generalized delay attack, since only QoS~1 or QoS~2 provides indicators of anomalies but cannot prevent a delayed control signal from being executed.

\subsubsection*{Sum Up}
\begin{figure}[h]
    \centering
    \begin{subfigure}{0.49\textwidth}
        \centering
        \includegraphics[width=0.99\textwidth]{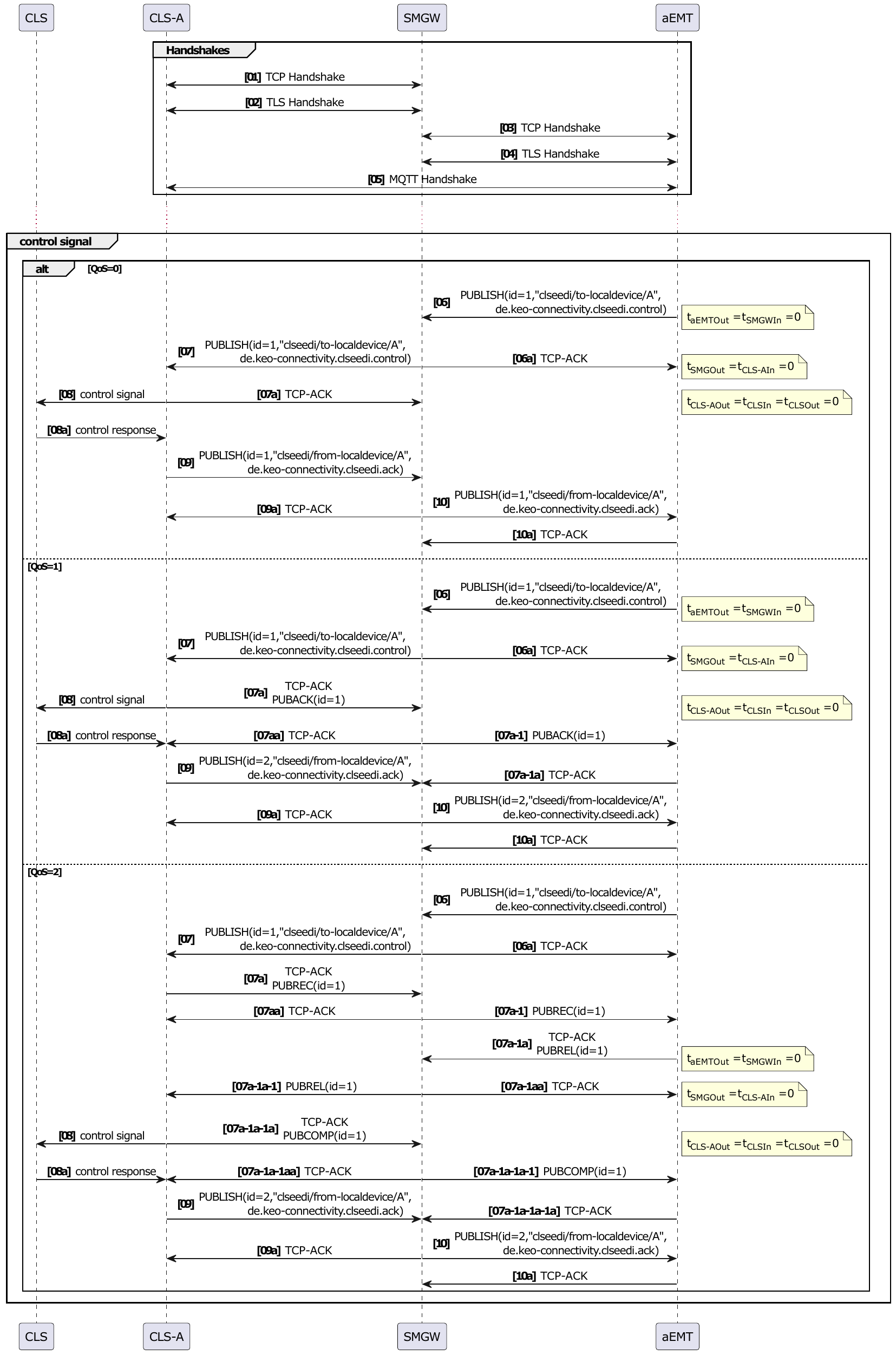}
        \caption{Intended}
        \label{fig:theory:clseedi:seqEEDI}
    \end{subfigure}
    \begin{subfigure}{0.49\textwidth}
        \centering
        \includegraphics[width=0.99\textwidth]{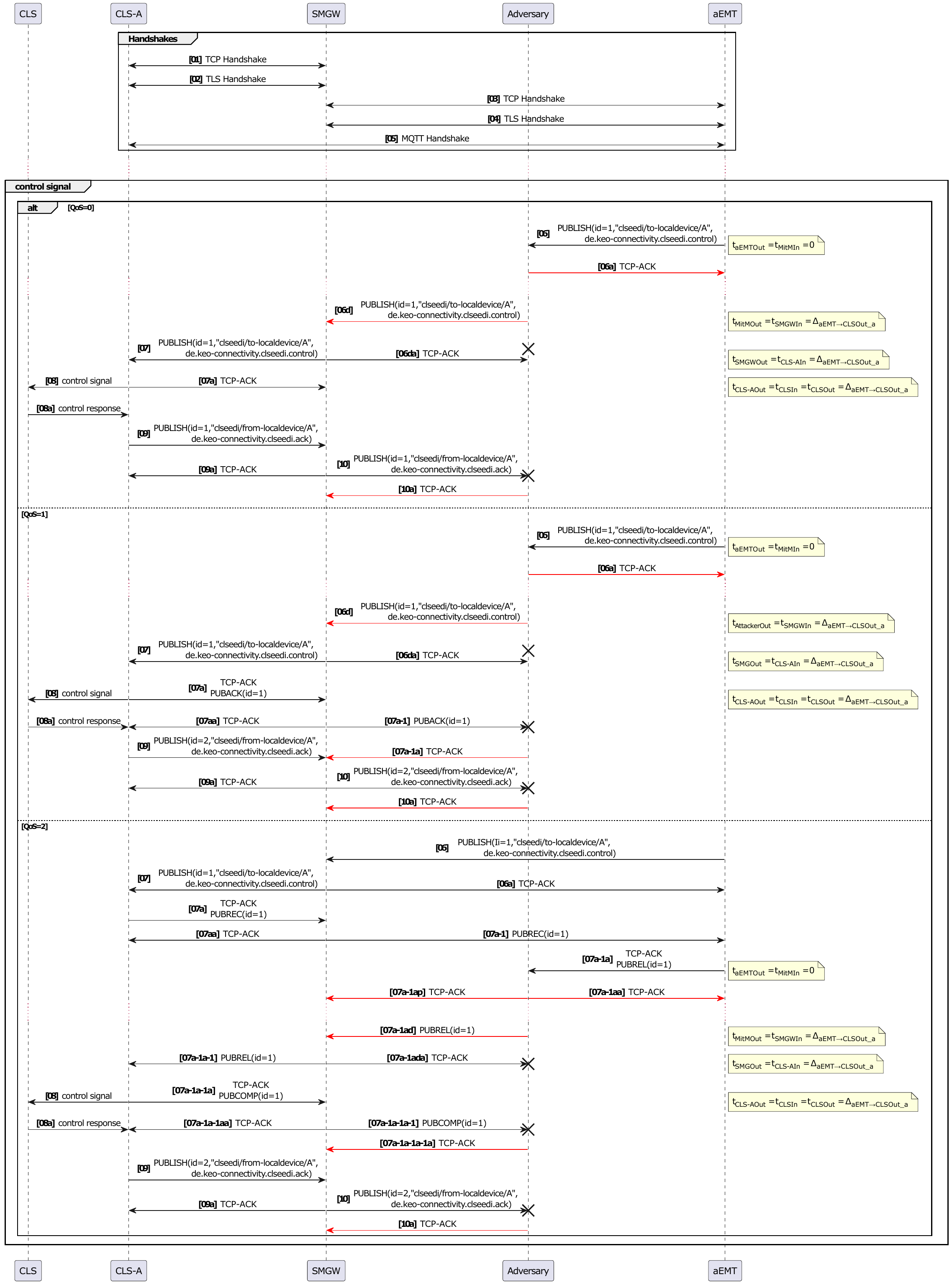}
        \caption{Attacked}
        \label{fig:theory:clseedi:seqEEDIAttacked}
    \end{subfigure}
    \caption{Sequence diagrams for CLS.EEDI}
    \label{fig:theory:clseedi}
\end{figure}

An example of a control signal transmitted using CLS.EEDI is shown in \cref{fig:theory:clseedi:seqEEDI}. For simplicity and comparability with other sequence diagrams, the Issuer is illustrated as part of the aEMT. For a more compact representation, TCP acknowledgments are consolidated with payloads in compliance with the TCP specification \cite{rfc9293}, given our assumption $\Delta_{\text{IssuerOut}\rightarrow \text{CLSOut},n}=0$.

Message 05 illustrates an abstracted MQTT handshake between the CLS-A and aEMT, transmitted over the CLS channel. To cover all three QoS levels, one case per QoS level is created. For the \textit{de.keo-connectivity.clseedi.ack} message, a QoS level of 0 is assumed, as this part is not relevant for the delay attack.

The QoS level 0 case is similar to the generalized sequence diagram in \cref{fig:theory:delay:normal}, with differences in the payload of messages 06, 07, 09, and 10. 

The QoS level 1 case requires additional message exchanges. Message 07a contains both the TCP acknowledgment for message 07 and the PUBACK corresponding to the PUBLISH message. Message 07a-1 contains the forwarded PUBACK, while messages 07aa and 07a-1a contain the respective TCP acknowledgments.

For the QoS level 2 case, it is assumed that the CLS-A forwards the control signal to the CLS after receiving the corresponding PUBREL. Messages 06 and 07 carry the actual control signal data, which is buffered at the CLS-A until the PUBREL message is received. To integrate this behavior into the timestamp scheme, the PUBREL is treated as $m$ since its reception causes the control signal to be forwarded to the CLS. Messages 07a, 07aa, and 07a-1 are analogous to the QoS level 1 case, but use type PUBREC instead of PUBACK. Message 07a-1a contains the TCP acknowledgment for message 07a-1 together with the PUBREL. The PUBREL is forwarded to the CLS-A in message 07a-1a-1, which triggers forwarding of the control signal to the CLS via message 08. Additionally, a PUBCOMP message is generated and acknowledged by message 07a-1a-1a. Message 07a-1a-1aa contains the TCP acknowledgment for PUBCOMP. Message 08a contains the control response. The PUBCOMP is finally forwarded to the aEMT in message 07a-1a-1a-1 and acknowledged using message 07a-1a-1a-1a. The remaining messages are analogous to the other QoS levels.

The previous analysis of CLS.EEDI and MQTT leads to \cref{fig:theory:clseedi:seqEEDIAttacked}, which illustrates a delay attack on a control signal for all three QoS levels. Attacks on QoS levels 0 and 1 follow the same general approach as described in the generalized delay attack in \cref{sec:theory:delay}.

For QoS level 2, the adversary must additionally drop message 07a-1 and send a corresponding TCP acknowledgment (message 07a-1a). Furthermore, the adversary must inject a TCP acknowledgment for message 07a-1, either by stripping the TCP payload of message 07a-1a or by injecting a new packet based on message 07a-1. In this case, instead of delaying the PUBLISH message, the adversary delays the PUBREL message, which controls when the control signal is forwarded to the CLS.

By combining the analysis of CLS.EEDI and MQTT, we derive 
\[\forall t \in T : (V_{\text{comp}}(\text{CLS-A}, t) = \textsf{lv} \; \land \; \delta^{\max}_{comp}(\text{CLS-A}, t) = \infty)\] since mitigation mechanisms (e.g., keep-alive) depend solely on optional MQTT features to which the aEMT must respond upon request. Furthermore, \[V_{\text{comp}}(\text{aEMT}, \textsf{TM2}) = \textsf{lv} \text{ with } \delta^{\max}_{comp}(\text{aEMT}, \textsf{TM2}) = \infty,\] and \[V_{\text{comp}}(\text{aEMT}, \textsf{TM1}) = \textsf{v} \text{ with } \delta^{\max}_{comp}(\text{aEMT}, \textsf{TM1}) = \infty,\] following the same argumentation as for the generalized delay attack.

Consequently, 
\[V_{\text{sys}}(\{\text{CLS-A}, \text{SMGW}, \text{aEMT}\}, TM1) = \textsf{v} \; \land \; \delta^{\max}_{sys}(\{\text{CLS-A}, \text{SMGW}, \text{aEMT}\}, TM1)=\infty \; \text{and}\]

\[V_{\text{sys}}(\{\text{CLS-A}, \text{SMGW}, \text{aEMT}\}, TM2) = \textsf{lv} \; \land \; \delta^{\max}_{sys}(\{\text{CLS-A}, \text{SMGW}, \text{aEMT}\}, TM2)=\infty \]
while $\Delta_{\max}$ may be limited in specific configurations, it is not bounded by the protocol specifications. In particular, $\Delta_{\max}$ may reach values on the order of typical session lifetimes (e.g., 48\,hours) and becomes infinite if QoS level 2 is used in combination with an infinite MQTT session lifetime.

%% file: theory/iec61850.tex
\subsubsection*{General}

IEC 61850 is a communication standard designed to control intelligent electronic devices in power utility automation systems \cite{tech:standard:iec61850-1}. Due to the complexity of the IEC 61850 standard and the numerous related standards it references, this work focuses on a specific application and implementation that is highly relevant in the context of the SMI: the FNN Steuerbox (FNN-STB).

The FNN-STB is a standardized CLS-A that is compatible with BSI TR-03109-5 and enables grid-friendly control via the SMGW's CLS channel \cite{tech:FNN:STB}. A CLS connected to a CLS-A may be interfaced either digitally (e.g., via EEBUS) or via analog relay contacts \cite{tech:FNN:STB}.

Within the system architecture, the FNN-STB takes the IEC 61850 server role, while the aEMT acts as the IEC 61850 client. The FNN-STB specification primarily defines a data model based on IEC 61850. To enable interoperable communication between devices, this data model and its associated services must be mapped to concrete communication protocols \cite{tech:standard:iec61850-8-1,tech:standard:iec61850-8-2}. The FNN-STB specification permits the use of Manufacturer Message Syntax (MMS) and Extensible Messaging and Presence Protocol (XMPP) as such communication protocols \cite{tech:FNN:STB}. In this work, XMPP is not considered.

Consequently, $APPS_c$ consists of two layers, denoted as $61850_c$ and $MMS_c$. This yields the layer sets \[L_{\text{aEMT}}=\{61850_{\text{aEMT}}, MMS_{\text{aEMT}}, TCP_{\text{aEMT}}, TLS_{\text{aEMT}}\} \; \text{and}\] 

\[L_{\text{CLS-A}}=\{61850_{\text{CLS-A}}, MMS_{\text{CLS-A}}, TCP_{\text{CLS-A}}, TLS_{\text{CLS-A}}\}.\]

\subsubsection*{IEC 61850 Data Model}

The FNN-STB specification defines a data model based on IEC 61850 that can be used to monitor, configure, and control the FNN-STB itself or one of the connected CLSs (monitoring and configuration are not possible in the case of an analog connection between the CLS and the FNN-STB). In this work, we solely consider the parts of the data model used to implement a control signal issued by an aEMT.

The FNN-STB uses IEC 61850-90-10 schedules to determine the limitation value for a CLS. For this purpose, multiple control functions are defined, of which we focus on the \textit{Notbefehl} (``emergency command'') control function, as it has the highest priority and cannot be interrupted by a loss of communication between the CLS-A and the aEMT. The schedules of this control function are identified by \textit{DoEc\_FSCH2500x} with $x \in \{1,2,3,4\}$ in the case of an analog connection between the CLS and the FNN-STB, which is the scenario considered in this work. The data model is based on the FSCH logical node defined in \cite{tech:standard:iec61850-90-10}.

A schedule of type FSCH mainly consists of a timestamp defining its start (\textit{StrTm01}) and a set of limitation values (\textit{ValASx}) with $x \in \mathbb{N}$, which are iterated using a configured interval (\textit{SchdIntv}) \cite{tech:standard:iec61850-90-10}. The object contains several read-only attributes (\textit{Beh}, \textit{NamePlt}, \textit{SchdSt}, \textit{SchedEntr}, \textit{SchedPrio}, \textit{ActStrTm}, \textit{NxtStrTm}, \textit{SchdEnaErr}, \textit{ValMV}, \textit{NumEntr}) \cite{tech:standard:iec61850-90-10}. Some of these attributes additionally contain a timestamp and a quality indicator \cite{tech:standard:iec61850-90-10}, which refer to the last update time and the reliability of the value \cite{tech:standard:iec61850-7-3}.

In the case of schedules of type \textit{DoEc\_FSCH2500x}, the set of limitation values has a cardinality of one \cite{tech:FNN:STB}. Hence, the interval defines the duration of the schedule, starting at the specified start timestamp. If the current time reaches the end of this duration, a schedule with a lower priority is applied instead. We do not consider this case to be vulnerable, because the duration restricts the control signal in an absolute manner. For example, if the duration is five minutes, a delay of more than five minutes does not change the outcome, as the schedule has already expired and will no longer be applied.

However, a schedule can additionally be configured to be cyclic (\textit{SchedReuse} set to true) \cite{tech:FNN:STB}, resulting in a gapless, effectively infinite duration (this behavior is required to be supported by STB\_0228 \cite{tech:FNN:STB}). If a schedule is defined as cyclic with a start time in the past, it is executed as if it had already been active before \cite{tech:FNN:STB}. Hence, the start time does not affect the execution.

A schedule can be enabled or disabled, with the limitation value being applied only if the schedule is enabled. We assume that the schedule has been configured in a previous or the current TLS session and remains disabled until explicitly enabled. A contrary assumption would contradict the IEC 61850 specification, since some attributes cannot be modified while a schedule is in the running state \cite{tech:standard:iec61850-90-10}. Hence, for a gapless cyclic \textit{DoEc\_FSCH2500x} schedule, the limitation value is applied only from the moment the \textit{EnaReq} arrives, provided that the start time (\textit{StrTm01}) lies in the past.

In contrast to the generalized attack from \cref{fig:theory:delay}, the information used to configure the schedule does not itself constitute the control signal. Therefore, we consider the request to enable a schedule as the control signal to be delayed.

A schedule can be enabled using an Enable Request (\textit{EnaReq}) \cite{tech:standard:iec61850-90-10}. The \textit{EnaReq} is a standard control operate request composed of several attributes. The attribute \textit{t} (of type timestamp) refers to the last time the value of the attribute \textit{stVal} was modified \cite{tech:standard:iec61850-7-3}. The value of \textit{stVal} must be set to true; otherwise, the \textit{EnaReq} must be ignored \cite{tech:standard:iec61850-90-10}.

The timestamp type consists of the attributes \textit{SecondSinceEpoch}, \textit{FractionOfSeconds}, and \textit{TimeQuality}, representing a millisecond-precise timestamp with an associated quality indicator. If an attacker delays packets containing an \textit{EnaReq}, the value of \textit{t} deviates from the time at which the FNN-STB receives the request by at least the introduced delay. Hence, a delay attack on the \textit{EnaReq} could be detected and mitigated if the FNN-STB compares \textit{t} against its local time using a threshold $< \Delta_{crit}$.

However, neither IEC 61850 nor the FNN-STB specification mandates such a check \cite{tech:standard:iec61850-90-10,tech:standard:iec61850-7-2,tech:standard:iec61850-5}. IEC 61850-5 mentions in Annex F (informative) an optional check to avoid ``issuing an old control that would have been stacked into the network''\cite{tech:standard:iec61850-5}. Hence, such a check may be implemented but is not required.

The FNN-STB specification requires the control mode \textit{direct-with-normal-security} \cite{tech:FNN:STB} for the \textit{EnaReq}. ``Direct'' means that control objects can be operated without a prior selection phase \cite{tech:standard:iec61850-7-2}. Otherwise, the object must first be selected and locked for a certain period, during which the control operation must occur \cite{tech:standard:iec61850-7-2}. ``Normal-security'' requires the STB to determine the result of the operation based on the validity of the control signal \cite{tech:standard:iec61850-7-2}. The result of the control operate command is sent back to the aEMT \cite{tech:standard:iec61850-7-2,tech:standard:iec61850-7-3}. This response can be dropped by an adversary to conceal the delayed execution.

The FNN-STB specification refers to IT security mechanisms for IEC 61850 described in IEC 62351-4 \cite{tech:FNN:STB}. IEC 62351-4 describes several Transport (T)-profiles and Application (A)-profiles that can be combined to provide different levels of IT security \cite{tech:standard:iec62351-4}. These mechanisms and profiles are neither mandatory to implement nor to use; only recommendations for combining T- and A-profiles to achieve specific security properties are provided \cite{tech:standard:iec62351-4}. A T-profile describes transport-layer security mechanisms, which are already required by the SMI and are therefore not considered here. A-profiles describe application-layer security mechanisms.

IEC 62351-4 specifies an end-to-end encryption mode at the application layer, which encapsulates the payload message into an additional PDU called \textit{SecPDU} \cite{tech:standard:iec62351-4}. The \textit{SecPDU} contains a object (``token'') that includes, among other attributes, a timestamp \cite{tech:standard:iec62351-4}. To comply with IEC 62351-4, the receiver (the FNN-STB) must verify that the timestamp does not deviate by more than 10 minutes from the local clock \cite{tech:standard:iec62351-100-4,tech:standard:iec62351-4}.

The FNN-STB specification mandates the implementation of IEC 62351-4 security mechanisms but does not require their enforcement \cite{tech:FNN:STB}. Note that this mode requires both the aEMT and the CLS-A to implement and use end-to-end encryption. Consequently, the use of this mechanism ultimately depends on the aEMT.

The FNN-STB specification requires to implement and use the TCP keep alive mechanism for the connection to the SMGW (STB\_0519 and STB\_0531) \cite{tech:FNN:STB}. This mechanism is further required by IEC 61850 if MMS is used for transportation (6.2.3.2.1) \cite{tech:standard:iec61850-8-1}. This keep alive mechanism is below the TLS layer therefore it can be spoofed by the adversary.

IEC 61850 and the FNN-STB do not impose any requirements on the $TCP_c$, $TLS_c$, or $MMS_c$ layers with $c \in \{\text{CLS-A}, \text{SMGW}, \text{aEMT}\}$ that are relevant to our analysis besides the TCP keep alive mechanism which can be spoofed by the adversary.

Based on this analysis, we conclude that 
\[\forall t \in T : (V_{layer}(61850_{\text{CLS-A}}, t) = \textsf{lv} \; \land \; \delta^{\max}_{\text{layer}}(61850_{\text{CLS-A}}, t) < \infty),\] 
since timestamp validation is not mandatory and IT security mechanisms are only active upon request by the aEMT.

Furthermore, \[V_{layer}(61850_{\text{aEMT}}, \textsf{TM1}) = \textsf{lv} \text{ with } \delta^{\max}_{\text{layer}}(61850_{\text{aEMT}}, \textsf{TM1}) = \infty,\] 
since the aEMT may request the use of end-to-end encryption, and 
\[V_{layer}(61850_{\text{aEMT}}, \textsf{TM2}) = \textsf{lv} \text{ with } \delta^{\max}_{\text{layer}}(61850_{\text{aEMT}}, \textsf{TM2}) = \infty,\] 
due to the IT security mechanisms stated in IEC 62351-4 and by following the same reasoning as in the generalized delay attack, as the control operate command result message corresponds to the control response message.

\subsubsection*{MMS}

The \textit{EnaReq} operation is transported as an MMS confirmed-RequestPDU \cite{tech:standard:iec61850-8-1}. MMS uses ASN.1 to structure the data and BER to encode a confirmed-RequestPDU, which only adds type and length information for each value to the raw data \cite{tech:standard:iec61850-8-1,tech:standard:ISO9506-1,tech:standard:ISO9506-2}.

A confirmed-RequestPDU consists of an \textit{invokeID} and a service of type \textit{ConfirmedServiceRequest}, which is of type \textit{write-Request} in our case. The write request contains the object \textit{variableAccessSpecification}, referring to the \textit{EnaReq} object of the IEC 61850 schedule discussed earlier (\textit{DoEc\_FSCH25001\$CO\$EnaReq\$Oper}), as well as a \textit{listOfData} object that carries the actual \textit{EnaReq} payload. \cite{tech:standard:ISO9506-1,tech:standard:ISO9506-2}

MMS does not impose any requirements on the $TCP_c$ and $TLS_c$ layers with $c \in \{\text{CLS-A}, \text{SMGW}, \text{aEMT}\}$.

Based on this analysis, we conclude that \[\forall t \in T : (V_{layer}(MMS_{\text{CLS-A}}, t) = \textsf{v} \; \land \;  \delta^{\max}_{\textit{layer}}(MMS_{\text{CLS-A}}, t) = \infty).\]  
Furthermore, \[V_{layer}(MMS_{\text{aEMT}}, \textsf{TM1}) = \textsf{v} \text{ with } \delta^{\max}_{\textit{layer}}(MMS_{\text{aEMT}}, \textsf{TM1}) = \infty,\] 
and 
\[V_{layer}(MMS_{\text{aEMT}}, \textsf{TM2}) = \textsf{lv} \text{ with } \delta^{\max}_{\textit{layer}}(MMS_{\text{aEMT}}, \textsf{TM2}) = \infty,\] 
following the same reasoning as in the generalized delay attack, as the confirmed-ResponsePDU corresponds to the control response message.

\subsubsection*{Sum Up}

\begin{figure}[h]
    \centering
    \begin{subfigure}{0.49\textwidth}
        \centering
        \includegraphics[width=0.99\textwidth]{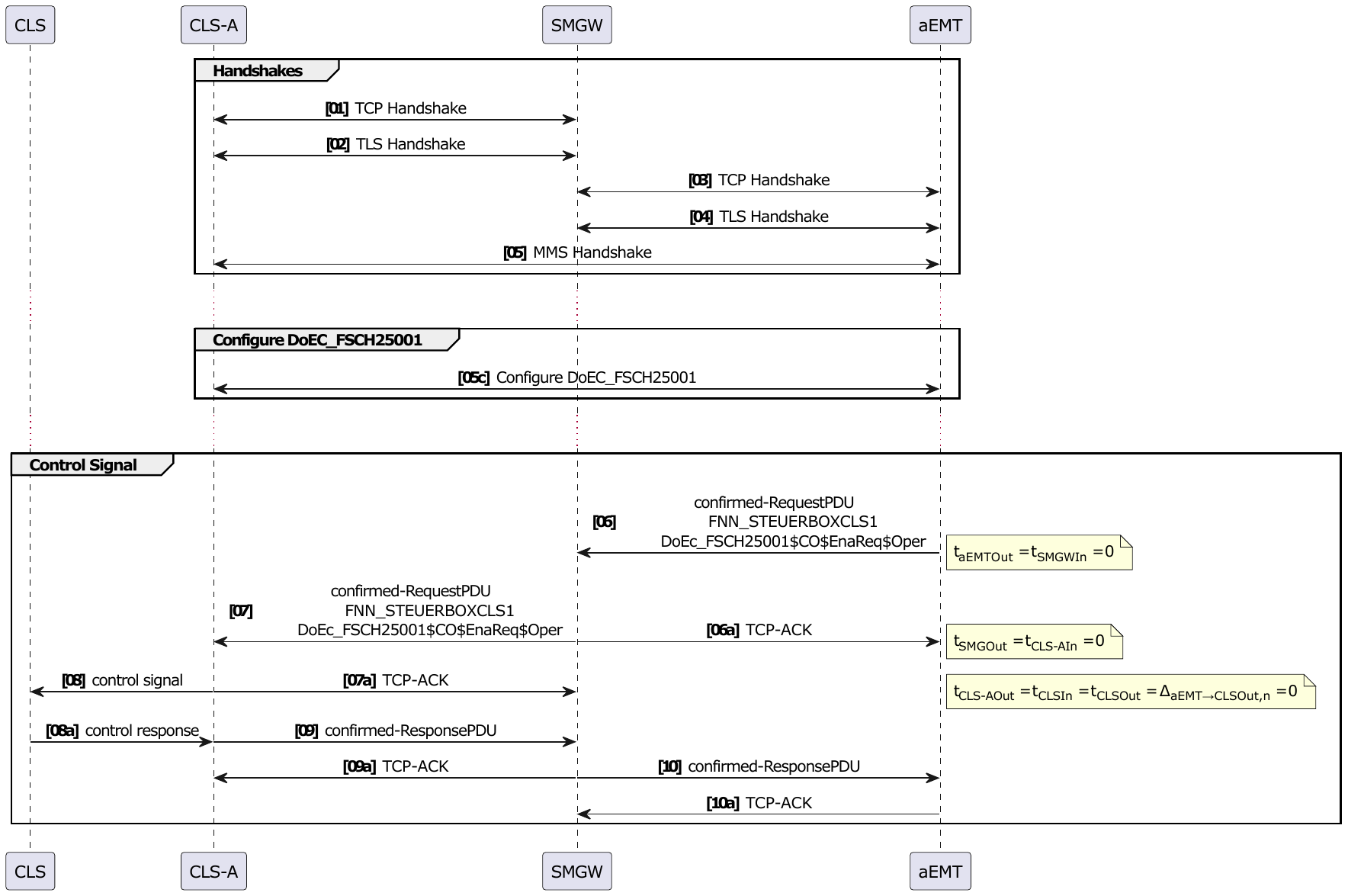}
        \caption{Intended}
        \label{fig:theory:iec61850:seq}
    \end{subfigure}
    \begin{subfigure}{0.49\textwidth}
        \centering
        \includegraphics[width=0.99\textwidth]{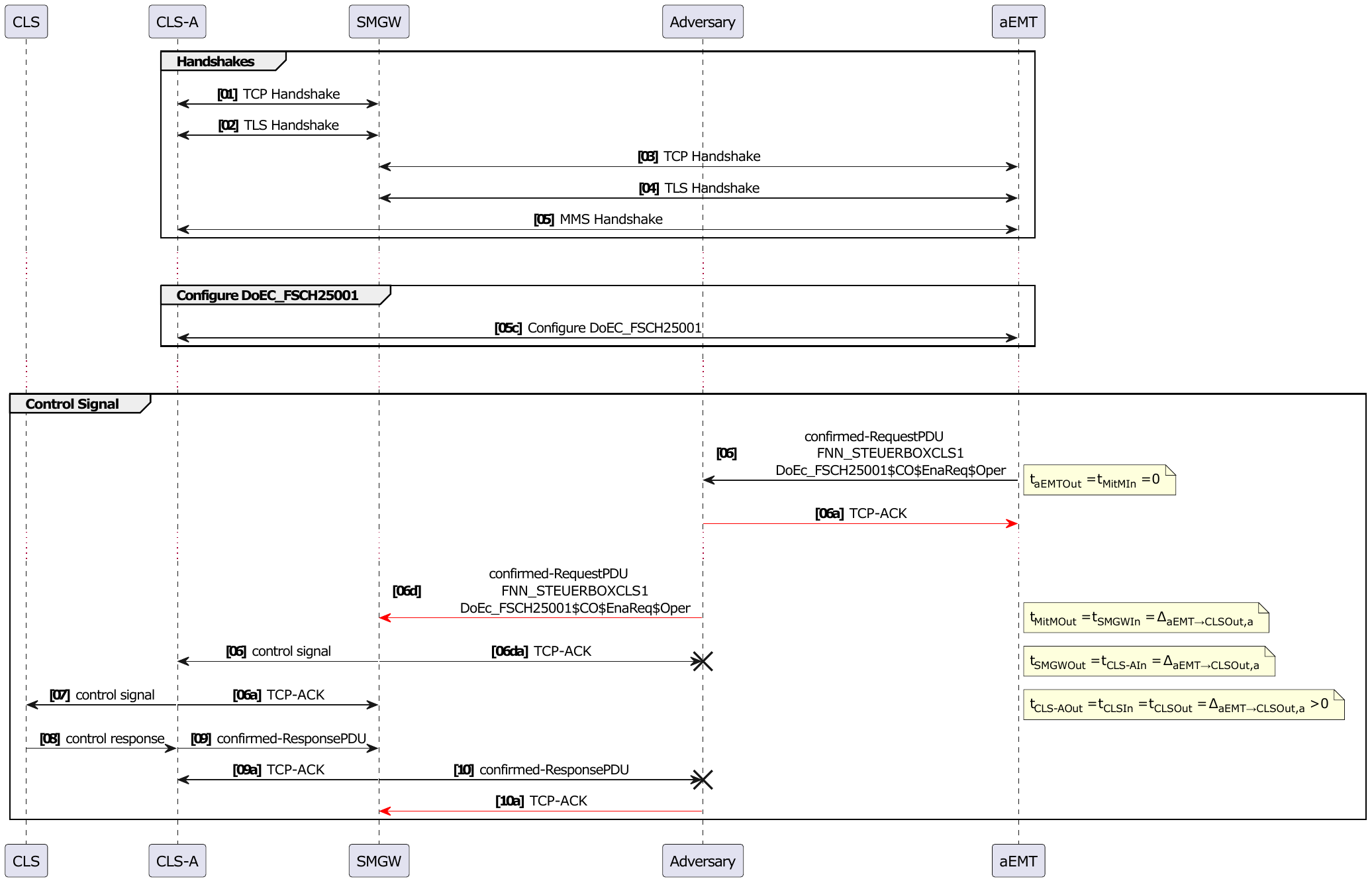}
        \caption{Attacked}
        \label{fig:theory:iec61850:seqAttacked}
    \end{subfigure}
    \caption{Sequence diagram of IEC 61850}
    \label{fig:theory:iec61850}
\end{figure}

\Cref{fig:theory:iec61850:seq} illustrates an example of a control signal transported using IEC 61850 with MMS used as the communication protocol. Message 05 represents the abstract MMS-Handshake. Message 05c refers to any communication related to the configuration of the schedule \textit{FSCH25001}. Messages 06 and 07 carry the confirmed-RequestPDU containing the \textit{EnaReq}. The response to the \textit{EnaReq} is carried in messages 09 and 10 as confirmed-ResponsePDU.

The attack is illustrated in \cref{fig:theory:iec61850:seqAttacked}. It aligns with the generalized delay attack, except that the adversary must await the completion of the configuration of \textit{DoEc\_FSCH25001}. Hence, the adversary delays message 06 and spoofs the TCP-ACK (message 06a). The delayed \textit{EnaReq} is contained in 06d. The TCP-ACK of it (message 06da) must be dropped. Further, message 10 must be dropped and the associated TCP-ACK (message 10a) must be spoofed.

By combining the previous analysis we can derive that 
\[\forall t \in T : (V_\text{comp}(\text{CLS-A},t)=\textsf{lv} \; \land \; \delta^{\max}_{\textit{comp}}(\textit{CLS-A},t) = \infty),\]
due to the timestamp included in an \textit{EnaReq}, which can be used to verify its age. Similarly, 
\[V_\text{comp}(\text{aEMT},\textsf{TM1})=\textsf{lv} \text{ with } \delta^{\max}_{\textit{comp}}(\textit{aEMT},\textsf{TM1})= \infty \; \text{and}\] 

\[V_\text{comp}(\text{aEMT},\textsf{TM2})=\textsf{lv} \text{ with } \delta^{\max}_{\textit{comp}}(\textit{aEMT},\textsf{TM2})= \infty.\] 
Consequently, 
\[V_\text{sys}(\{\text{SMGW},\text{aEMT},\text{CLS-A}\},\textsf{TM1})=\textsf{lv} \text{ with } \delta^{\max}_{\textit{sys}}(\{\text{SMGW},\text{aEMT},\text{CLS-A}\},\textsf{TM1})= \infty \; \text{and}\]

\[V_\text{sys}(\{\text{SMGW},\text{aEMT},\text{CLS-A}\},\textsf{TM2})=\textsf{lv} \text{ with } \delta^{\max}_{\textit{sys}}(\{\text{SMGW},\text{aEMT},\text{CLS-A}\},\textsf{TM2})= \infty.\]

%% file: evaluation/lab.tex
\subsubsection*{General}
This section describes the laboratory setup used to empirically validate the delay attack proposed in \cref{sec:theory}. This setup is designed as a controller hardware-in-the-loop (CHiL) environment, enabling controlled, repeatable experiments under realistic conditions of the SMI. 

The laboratory setup was separated into \textit{System under Test} (SuT) and the \textit{Test Environment} (TE) where the SuT consists of the system architecture outlined in \cref{sec:theory:scenario} and the TE of observation and test control systems.

\subsubsection*{Requirements}
The experimental setup was designed to fulfill the following requirements:
\begin{enumerate}
\item Emulate the system architecture outlined in \cref{sec:theory} while achieving a high Technology Readiness Level (TRL) to ensure realistic and generalizable results.
\item Provide full observability and control by granting the test control system access to the Issuer, CLS-A, and CLS.
\item Provide a suitable integration point for an adversary in accordance with the TM 1.
\item Ensure that the operation of the GWA and aEMT is logically segregated from other participants, as unintended interference with additional SMGW connections cannot be excluded.
\end{enumerate}

\subsubsection*{Function under Test}

The Function under Test (FuT) is defined as the ``process of sending a control signal''. This function is subdivided into the following stages:
\begin{itemize}
\item \textbf{FuT ``Connect'':} Establishment of the CLS channel using HKS3 between the CLS-A and the aEMT (transport layer handshake). 
\item \textbf{FuT ``Initialize'':} Initialization of protocol-specific communication between CLS-A and Issuer (application layer handshake).
\item \textbf{FuT ``Issue Command'':} Transmission of the control signal from the Issuer to the CLS via the SMGW and CLS-A.
\end{itemize}

The attack described in \cref{sec:theory} targets the function ``Issue Command'', while the preceding stages establish the necessary communication context.

\subsubsection*{Laboratory Setup Overview}

The SuT comprises the Issuer, aEMT, SMGW and CLS-A. In addition, the GWA is considered part of the SuT, as it provides essential services such as time synchronization. The CLS is not considered part of the SuT, since the interface to the CLS constitutes the system boundary between the SuT and the test environment.

The laboratory setup was implemented at the THU Smart Grid Laboratory. All certificates were issued by the Smart Metering Test Public Key Infrastructure (SM-Test-PKI) which is defined in \cite{tech:bsi:certPolicy}. The SM-Test-PKI is structurally equivalent to the production PKI but is not approved for general use due to relaxed organizational security requirements, making it suitable for controlled experiments. During the operating time of the setup, several certificates had to be manually renewed.

All physical network connections are realized via Ethernet. The adversary is positioned inline on the communication paths between SMGW $\rightleftharpoons$ aEMT and SMGW $\rightleftharpoons$ GWA. It is implemented using a system equipped with two network interfaces, acting as a Machine-in-the-Middle between communication endpoints. This setup ensures that all traffic for TM 1 traverses the adversary, enabling the capabilities required by both considered threat models.

Due to the co-location of the aEMT and GWA on the same system, it was not possible to restrict the adversary to a single communication path without significantly increasing the laboratory setup's complexity. Therefore, a strict experimental separation of the two threat models is not feasible in this setup. Such separation can only be achieved by selectively restricting the adversary's capabilities within the attack script.

The laboratory GWA and EMT are based on a product evaluation kit for on-site testing. With the manufacturer's permission and support, the otherwise isolated kit was integrated into the laboratory network. The system was isolated from the common network, but access was granted to the PTB time servers and the SM-Test-PKI servers. The laboratory network was also configured to allow the GWA to connect directly to the SMGW for the required wakeup packet.

Using the GWA, the SMGW was configured according to \cref{sec:theory}. Specifically, the required WAN configuration and proxy profiles were deployed on the SMGW, while corresponding HAN configuration profiles were deployed on the CLS-A to ensure compatibility. The SMGW was configured to synchronize its time from the GWA, which in turn synchronized from the PTB timeservers specified in \cite{tech:bsi:tr-03109-1}.

The aEMT implements a proxy between the CLS channel and the Issuer. It accepts incoming TLS connections on its WAN interface and subsequently establishes outbound connections to the Issuer. %
The chosen aEMT utilizes comprehensive management of CLS applications. Therefore, the aEMT application tunnels multiple data streams inside a single TLS stream with additional mandatory services such as heartbeat mechanisms which potentially mitigate the delay attack outlined in this work. The manufacturer calls this technique ``Multiplex Channel''. The system was configured not to use this manufacturer specific extension but to interconnect the aEMT directly with the CLS-A since the heartbeat mechanisms potentially poses a mitigation to delay attacks. The manufacturer calls this configuration ``Plain CLS''. The application data between the aEMT and SMGW (CLS channel) is encrypted using TLS while the connection between the aEMT and Issuer is in plain text (no TLS). 

The adversary is hosted on a Kali Linux system which synchronizes its time from the NTP Pool project's servers \cite{web:ntpPool}. The attacks were scripted in python using scapy version v2.6.1 \cite{web:scapy}. 

As multiple protocols are evaluated, the implementations of the Unit under Test (UuT) CLS-A, CLS, and Issuer vary depending on the protocol, particularly at the application layer.

The test control system and the adversary are operated remotely via a separate management network. This results in a CHiL-setup, which enables automated and repeatable execution of experiments over extended periods with minimal on-site intervention.

\subsubsection*{Test Case: DAP}

\begin{figure}[h]
    \centering
        \includegraphics[width=0.99\textwidth]{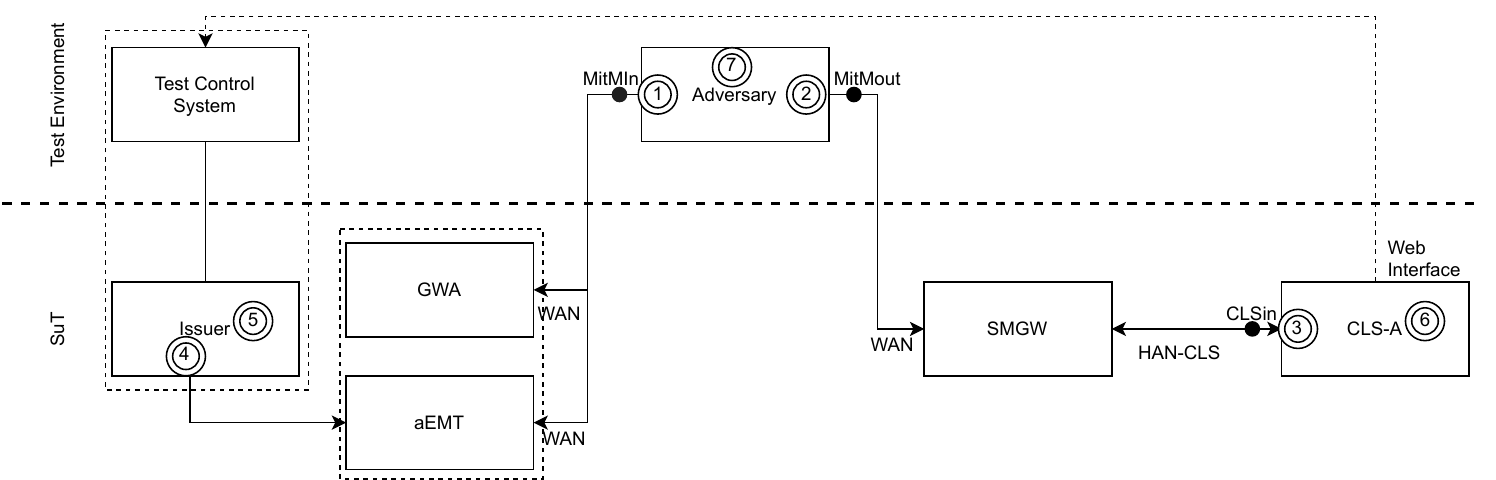}
    \caption{Experiment Setup - DAP Protocol}
    \label{fig:eval:lab:ExperimentSetupDAP}
\end{figure} 

An overview of the experiment setup for the protocol DAP can be found in \cref{fig:eval:lab:ExperimentSetupDAP}. The implementation details of each UuT are summarized in \cref{tab:eval:lab:overviewDAP}. 

At several positions of the laboratory setup application logs and network traffic were recorded (indicated by circled numbers in \cref{fig:eval:lab:ExperimentSetupDAP}). These recordings are used to form the basis for the delay calculation defined in \cref{sec:theory:scenario}. \Cref{tab:eval:lab:recordDAP} provides an overview over the recorded points mapped to the timestamps defined in \cref{tab:theory:delay:timestamps}.

To evaluate the DAP (see \cref{sec:theory:dap}), a custom implementation in Python was developed for the DAP client and DAP server.

The utilized CLS-A provides the capability to host Docker containers while managing communication with the SMGW in accordance with BSI TR-03109-5 \cite{tech:bsi:tr-03109-5}. This makes it a suitable platform for deploying custom protocol implementations. In this setup, a containerized DAP server was deployed on the CLS-A, logging each received command together with the current timestamp to persistent storage. The container accesses the time provided by the CLS-A which had no time synchronization active and was therefore configured manually. 

From a transport-layer perspective, both the Issuer and the CLS-A container act as servers. At the application layer, however, the Issuer implements the DAP client, while the CLS-A container implements the DAP server.

The Issuer (DAP client) is hosted on the same physical system as the test control system and logged each outgoing command together with the current timestamp to a log file. The test control system additionally establishes an SSH connection to the CLS-A to deploy the container and retrieve log data. The host synchronizes its time from the NTP Pool servers \cite{web:ntpPool}. 

\begin{table}[h]
\centering
\begin{tabular}{|l l l l|} 
\hline
\textbf{UuT} & \textbf{Manufacturer} & \textbf{Classification Rationale} & \textbf{TRL}\\
\hline
SMGW & A & certified hardware and firmware & TRL 9\\ %
CLS-A & A & final hardware, developer firmware & TRL 7\\ %
GWA & C & demonstrator tested at multiple customer sides & TRL 7 \\ %
aEMT & C & demonstrator close to the production version & TRL 8 \\ %
Issuer & HAS & Custom implementation & TRL 5 \\
Receiver & HAS & Custom implementation hosted on CLS-A & TRL 5 \\ 
\hline
\end{tabular}
\caption{Utilized Components within the DAP Laboratory Setup}
\label{tab:eval:lab:overviewDAP}
\end{table}

\begin{table}[h]
\centering
\begin{tabular}{|l l l|} 
\hline
\textbf{Point} & \textbf{Timestamp} & \textbf{Description}\\
\hline
1 & $t_{\text{MitMIn}}$ & Network dump on the Adversary interface towards the aEMT/GWA\\
2 & $t_{\text{MitMOut}}$ & Network dump on the Adversary interface towards the SMGW\\
3 & $t_{\text{CLS-AIn}}$ & Network dump on the CLS-A interface towards the SMGWs\\
4 & $t_{\text{aEMTIn}}$ & Network dump on the Issuer interface towards the aEMT\\
5 & $t_{\text{IssuerOut}}$& Application Log of the DAP client\\
6 & $t_{\text{CLSIn}}$ & Application Log of the DAP server\\
7 & - & Log of the attack script\\
\hline
\end{tabular}
\caption{Recordings Created within the DAP Laboratory Setup}
\label{tab:eval:lab:recordDAP}
\end{table}

\subsubsection*{Test Case: IEC 61850}

\begin{figure}[h]
    \centering
    \includegraphics[width=0.99\textwidth]{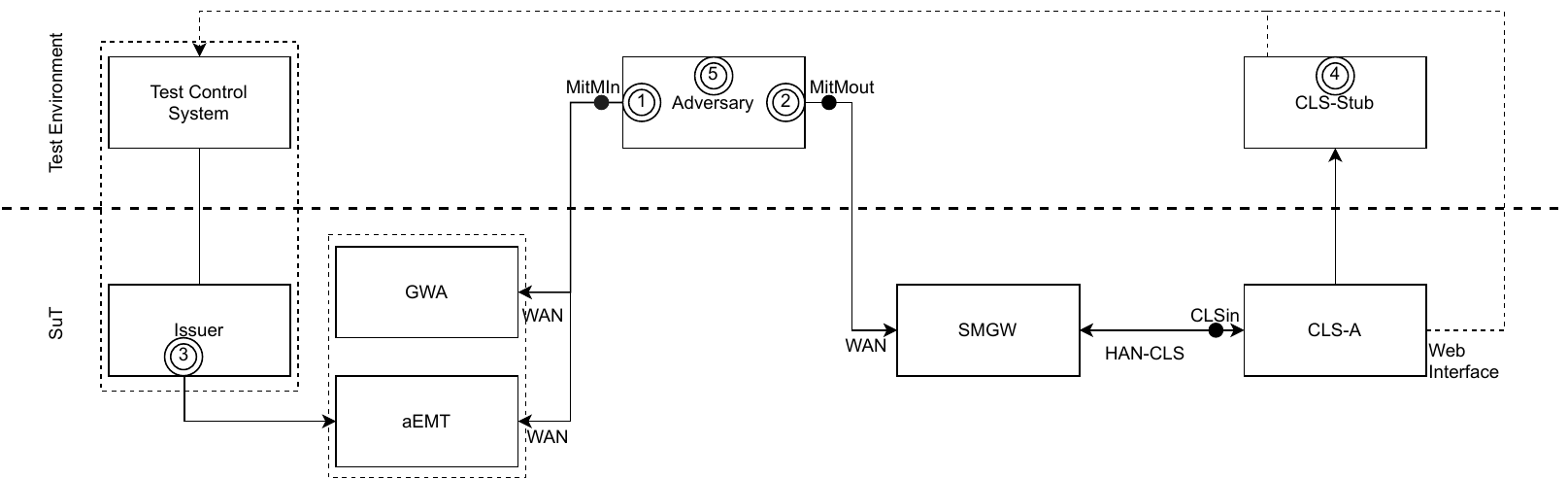}
    \caption{Experiment Setup - IEC 61850}
    \label{fig:eval:lab:ExperimentSetupIEC}
\end{figure} 

An overview of the IEC 61850 experiment setup using a FFN-STB as CLS-A is shown in \cref{fig:eval:lab:ExperimentSetupIEC}. The implementation details of each UuT are summarized in \cref{tab:eval:lab:overviewIEC}.

Application logs and network traffic were recorded at several points in the laboratory setup (indicated by circled numbers in \cref{fig:eval:lab:ExperimentSetupIEC}). These recordings serve as the basis for the delay calculation defined in \cref{sec:theory:scenario}. \cref{tab:eval:lab:recordIEC} provides an overview of the recording points with a mapping to the timestamps defined in \cref{tab:theory:delay:timestamps}.

The FNN-STB was configured with all schedules disabled except for SyRe\_FSCH01001, which cannot be disabled \cite{tech:FNN:STB}. SyRe\_FSCH01001 was configured to continuously enforce a power limit of 11 kW. The schedule DoEc\_FSCH25001 was configured to limit the power consumption to 0 kW. 

Since the FNN-STB implementation used does not provide a reliable method to detect the currently applied power limitation using onboard tools, a Raspberry Pi (CLS-Stub) was added to directly observe changes in the relay states. A script running on the CLS-Stub monitors relay state changes and logs them together with timestamps. The CLS-Stub synchronizes its time from NTP Pool project's servers \cite{web:ntpPool}. 

All four relay contacts (S1, S2, W3, W4) were connected to the GPIO pins of the CLS-Stub. S1 and S2 are normally open contacts \cite{tech:FNN:STB}, meaning they have two states: open (no current flows) or closed (current flows). W3 and W4 are changeover contacts \cite{tech:FNN:STB}, which always connect a common contact to one of two outputs. Thus, they switch between two positions (e.g., operating mode A or B) and are never completely disconnected.

The relays were configured to switch at specific power limits. This setup allows the CLS-Stub to continuously determine whether each monitored relay output is active or inactive. When a relay output is active, the CLS-Stub detects a voltage signal and records the state as HIGH; otherwise, it records LOW.

For the IEC 61850 client, a custom implementation based on libiec61850 \cite{web:libiec} was used. Since both the client and the aEMT software expect to connect to a server, a TCP proxy based on socat was implemented. This proxy provides a TCP server interface towards the aEMT while simultaneously establishing a corresponding server endpoint towards the Issuer upon connection.

Internally, socat accepts one connection on each port and links the corresponding sockets, forwarding data bidirectionally. This effectively creates a transparent proxy between two TCP endpoints. Since the FNN-STB closes network connection if no IEC 61850 session is established, the IEC 61850 client periodically polls if the respective port is open to ensure that the SMGW connection is not missed once the CLS channel is established.

The test control system accesses the CLS-Stub via SSH to retrieve the GPIO log file. This enables independent verification that IEC 61850 control signals issued by the aEMT control script result in the expected physical switching behavior of the CLS-A.

The Issuer and the socat proxy were hosted on the same physical system as the test control system. Additionally, the test control system had access to the web interface of the FNN-STB, allowing it to restart the device when necessary. The host synchronizes it's time from the NTP Pool servers \cite{web:ntpPool}. 

\begin{table}[h]
\centering
\begin{tabular}{|l l l l|}
\hline
\textbf{UuT} & \textbf{Manufacturer} & \textbf{Classification Rationale} & \textbf{TRL}\\
\hline
SMGW & A & certified hardware and firmware & TRL 9\\ %
CLS-A & B & final hardware, near-production firmware & TRL 7\\ %
GWA & C & demonstrator tested at multiple customer sides & TRL 7 \\ %
aEMT & C & demonstrator close to the production version & TRL 8 \\ %
Issuer & THU, HAS & Custom implementation & TRL 6\\ %
\hline
\end{tabular}
\caption{Utilized Components within the IEC 61850 Laboratory Setup}
\label{tab:eval:lab:overviewIEC}
\end{table}

\begin{table}[h]
\centering
\begin{tabular}{|l l l|} 
\hline
\textbf{Point} & \textbf{Timestamp} & \textbf{Description}\\
\hline
1 & $t_{\text{MitMIn}}$ & Network dump on the adversary interface towards the aEMT/GWA\\
2 & $t_{\text{MitMOut}}$ & Network dump on the adversary interface towards the SMGW\\
3 & $t_{\text{IssuerOut}}$ & Network dump on the Issuer interface towards the aEMT\\
4 & $t_{\text{CLSIn}}$ & Log of the relay observer\\
5 & - & Log of the attack script \\
\hline
\end{tabular}
\caption{Recordings Created within the IEC 61850 Laboratory Setup}
\label{tab:eval:lab:recordIEC}
\end{table}

%% file: evaluation/meth.tex
\subsubsection*{General}
To validate the theoretical analysis, a multi-trial experiment was conducted to ensure reproducibility. The experiment stop criterion was defined as three consecutive successful trials, excluding trials that failed due to issues traceable to the test environment, such as kernel panics on the adversary system or bugs in the attack script.

To maintain TCP-layer validity throughout the attack, the attacker script was designed to perform the following actions in addition to delaying packets containing the control signal:
\begin{enumerate}
    \item Spoof TCP keep-alive responses expected by the SMGW.
    \item Spoof TCP acknowledgments for delayed packets.
    \item Update timestamps included in the TCP timestamp option.
\end{enumerate}

To delay the control signal, we used Python's sleep function (\texttt{time.sleep()}) in a dedicated thread to avoid affecting the main thread responsible for handling the TCP keep-alive and the traffic on the path $\text{GWA}\rightleftharpoons \text{SMGW}$.

Verifying $\Delta_{\max}$ in practice is impossible, as it was derived under the assumption that no natural delay exists. Therefore, an offset of $\epsilon = 60\,\text{seconds}$ was introduced to compensate for natural delays. This offset is negligible compared to the overall delay of more than $172{,}740$ seconds. 

To verify that the delay was applied correctly, the following condition must hold $\Delta_{\max} - \epsilon \leq \Delta_{\text{MitMIn}\rightarrow\text{MitMOut}} \leq \Delta_{\text{IssuerOut}\rightarrow\text{CLSIn}}$.

\subsubsection*{DAP}
To identify the packet carrying the control signal, a context-based approach was used. This approach selects the first TCP packet based on the elapsed time after the attack script has started. The attack script was configured to wait 15 seconds after startup before delaying any TCP packet with a non-empty payload. Using this approach, the FuT ``Connect'' phase is not targeted. Note that the FuT ``Initialize'' does not exist for DAP due to its specification (see \cref{sec:theory:dap}).

The DAP server was configured to accept any incoming connection, wait until data is received, and write the received data to a file. In contrast, the DAP client was configured to wait 30 seconds after establishing a connection before sending the control signal to ensure the attack script is in the correct state.

To calculate $\Delta_{\text{IssuerOut}\rightarrow\text{CLSIn}}$, the timestamps $t_{\text{IssuerOut}}$ and $t_{\text{CLSIn}}$ were used, as the latter only exists if the control signal has been accepted. During the experiments, a negative drift of the CLS-A's clock used to log the arrival of the control signal was observed. This drift was manually compensated by adding $\theta = -\Delta_{\text{MitMOut}\rightarrow\text{CLSIn}}$ seconds to $t_{\text{CLSIn}}$. The calculation of $\theta$ has the downside that $\Delta_{\text{MitMOut}\rightarrow\text{CLSIn},n}$ is canceled out. Consequently, $\Delta_{\text{IssuerOut}\rightarrow\text{CLSIn}} = t_{\text{CLSIn}} - t_{\text{IssuerOut}} + \theta$.

\subsubsection*{IEC 61850}
The identification of the packet carrying the control packet was performed based on its size. Due to the behavior of the FNN-STB, which terminates the CLS channel and attempts to re-establish it with increasing delay if the application-layer session is not established within a predefined timeframe, it was necessary to restart the FNN-STB before each trial.

As a consequence of this behavior, parts of the attack procedure required manual intervention. In particular, the transition between the ``bypass'' and ``attack'' states of the attack script was performed manually after completion of the FuT ``Initialize''.

To ensure that the relay states were correctly observed by the CLS-Stub and that the IEC 61850 session was successfully established, the following control sequence was issued manually prior to the delayed control signal: \textit{DsaReq, EnaReq, DsaReq}. After executing this sequence, the attack script was switched to the attack state, and the control signal subject to delay was issued manually.

%% file: evaluation/res.tex
\subsubsection*{DAP}
\Cref{tab:eval:res:dap} summarizes the experimental results for DAP. All three trials demonstrate that a delay attack is feasible in practice. In none of the trials was the control signal rejected by the DAP server or dropped by the SMGW or CLS-A. However, the adversary system's network stack crashed unexpectedly during the third trial; therefore, the trial was repeated.

\begin{table}[h]
\centering
\begin{tabular}{|c c c c c c|}
\hline
$t_{\text{IssuerOut}}$ & $t_{\text{MitMIn}}$ & $t_{\text{MitMOut}}$ & $t_{\text{CLSIn}}$ & $\theta$ & $\Delta_{\text{IssuerOut}\rightarrow\text{CLSIn}} + \theta$ \\
\hline
1760937735.082388 & 1760937735.081188 & 1761110475.126328 & 1761110469.313180 & 5.813148 & 172740.043940 \\
\hline
1761112495.171656 & 1761112495.178219 & 1761285235.198342 & 1761285226.890479 & 8.307862 & 172740.026685 \\
\hline
1761655992.740740 & 1761655992.743485 & 1761828732.774333 & 1761828716.713694 & 16.060638 & 172740.033592 \\
\hline
\end{tabular}
\caption{Experimental results for DAP}
\label{tab:eval:res:dap}
\end{table}

\begin{figure}[h]
    \centering
    \begin{subfigure}{0.49\textwidth}
        \centering
        \includegraphics[width=\textwidth]{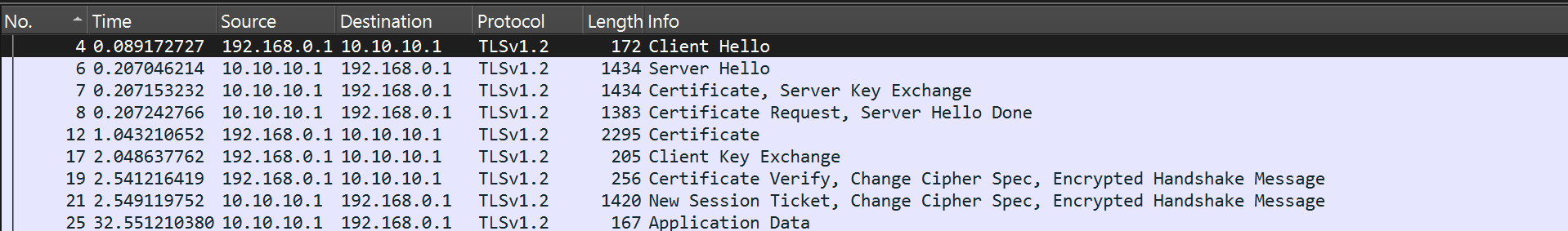}
        \caption{MitMIn TLS traces}
        \label{fig:eval:res:MitMDAP:In}
    \end{subfigure}
    \begin{subfigure}{0.49\textwidth}
        \centering
        \includegraphics[width=\textwidth]{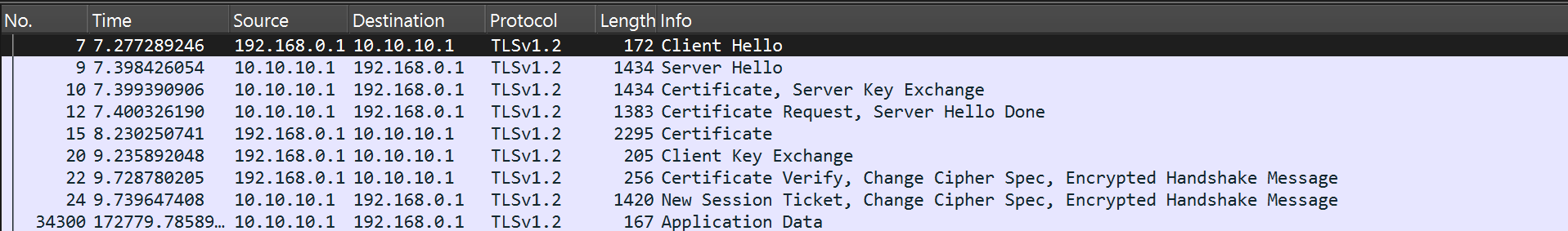}
        \caption{MitMOut TLS traces}
        \label{fig:eval:res:MitMDAP:Out}
    \end{subfigure}
    \caption{Network traces at the adversary: DAP}
    \label{fig:eval:MitMDAP}
\end{figure}

\Cref{fig:eval:MitMDAP} shows the network traffic captured on both interfaces of the adversary using Wireshark for the first trial. The traces confirm that the TCP session between the aEMT and the SMGW remained intact throughout the attack. Furthermore, they confirm that the control signal was delayed by approximately $\Delta_{\max}-\epsilon$ after initial interception. All Layer~2 and Layer~3 addresses were randomized using TraceWrangler Version 0.6.9 \cite{web:tracewrangler}.

\begin{figure}[h]
    \centering
    \begin{subfigure}{0.49\textwidth}
        \centering
            \includegraphics[width=\textwidth]{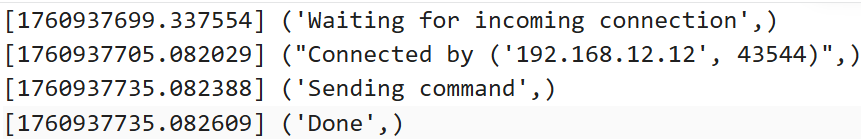}
        \caption{DAP client logs}
        \label{fig:eval:res:AppDAP:In}
    \end{subfigure}
    \begin{subfigure}{0.49\textwidth}
        \centering
        \includegraphics[width=\textwidth]{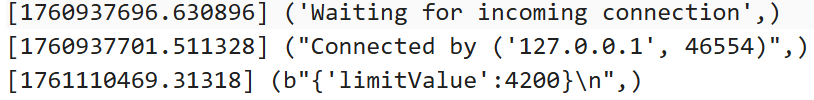}
        \caption{DAP server logs}
        \label{fig:eval:res:AppDAP:Out}
    \end{subfigure}
    \caption{Application-layer logs: DAP}
    \label{fig:eval:AppDAP}
\end{figure}

\Cref{fig:eval:AppDAP} presents a comparison of the application-layer logs of the DAP client and server for the first trial. These logs clearly show that the control signal was received approximately $\Delta_{\max}-\epsilon$ after being sent.

\subsubsection*{IEC 61850}
\Cref{tab:eval:res:iec} summarizes the experimental results for IEC 61850. All three trials demonstrate that a delay attack is feasible in practice. In none of the trials was the control signal rejected by the FNN-STB or dropped by the SMGW or CLS-A. However, the second trial had to be repeated twice. In one instance, the CLS-Stub observed unexpected changes in relay states. The manufacturer confirmed that this behavior is known for the installed firmware version if \textit{SyRe\_FSCH01001} reaches the end of its current definition period. In another instance, the attack script crashed causing the delayed control signal not to be sent, requiring a repetition of the trial.

\begin{table}[h!]
\centering
\begin{tabular}{|c c c c c|}
\hline
$t_{\text{IssuerOut}}$ & $t_{\text{MitMIn}}$ & $t_{\text{MitMOut}}$ & $t_{\text{CLSIn}}$ & $\Delta_{\text{IssuerOut}\rightarrow\text{CLSIn}}$ \\
\hline
1771594699.649062 & 1771594399.860618 & 1771767439.740247 & 1771767440.479987 & 172740.830925 \\
\hline
1772204576.810141 & 1772204576.810141 & 1772377317.643659 & 1772377317.643659 & 172740.833518 \\
\hline
1773327504.576281 & 1773327504.578931 & 1773500244.700064 & 1773500245.458447 & 172740.882166 \\
\hline
\end{tabular}
\caption{Experimental results for IEC 61850}
\label{tab:eval:res:iec}
\end{table}

\begin{figure}[h]
    \centering
    \begin{subfigure}{0.49\textwidth}
        \centering
        \includegraphics[width=\textwidth]{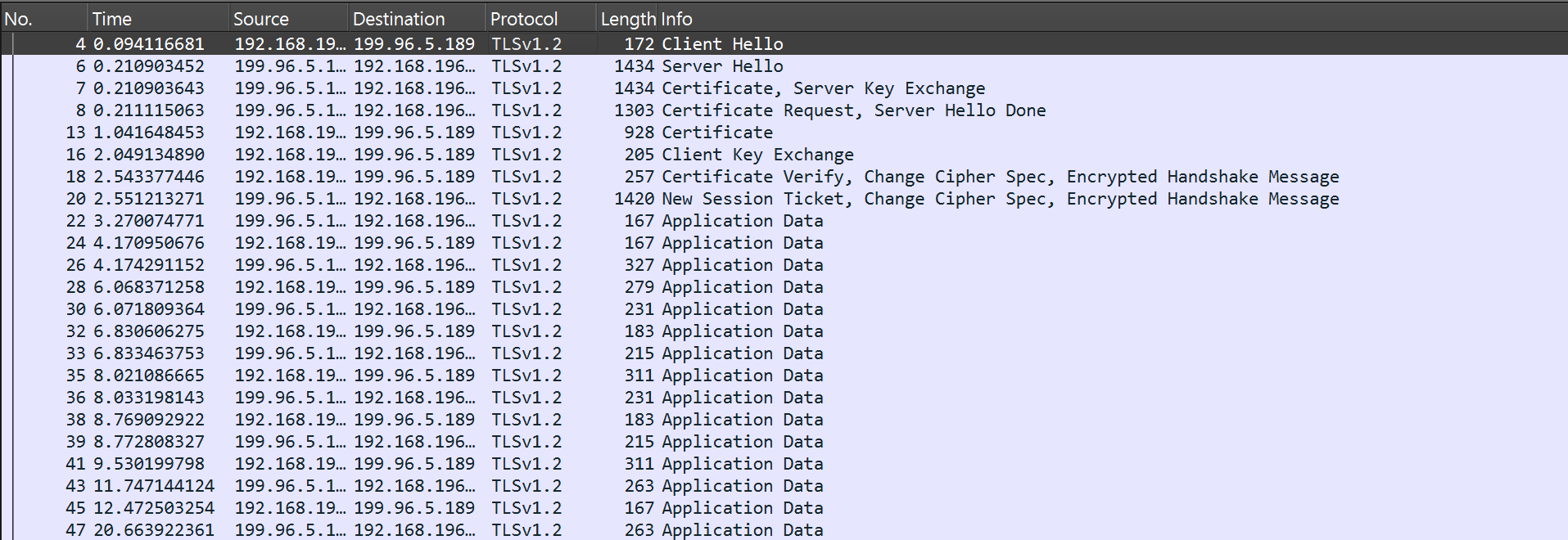}
        \caption{MitMIn TLS traces}
        \label{fig:eval:res:MitMIEC:In}
    \end{subfigure}
    \begin{subfigure}{0.49\textwidth}
        \centering
        \includegraphics[width=\textwidth]{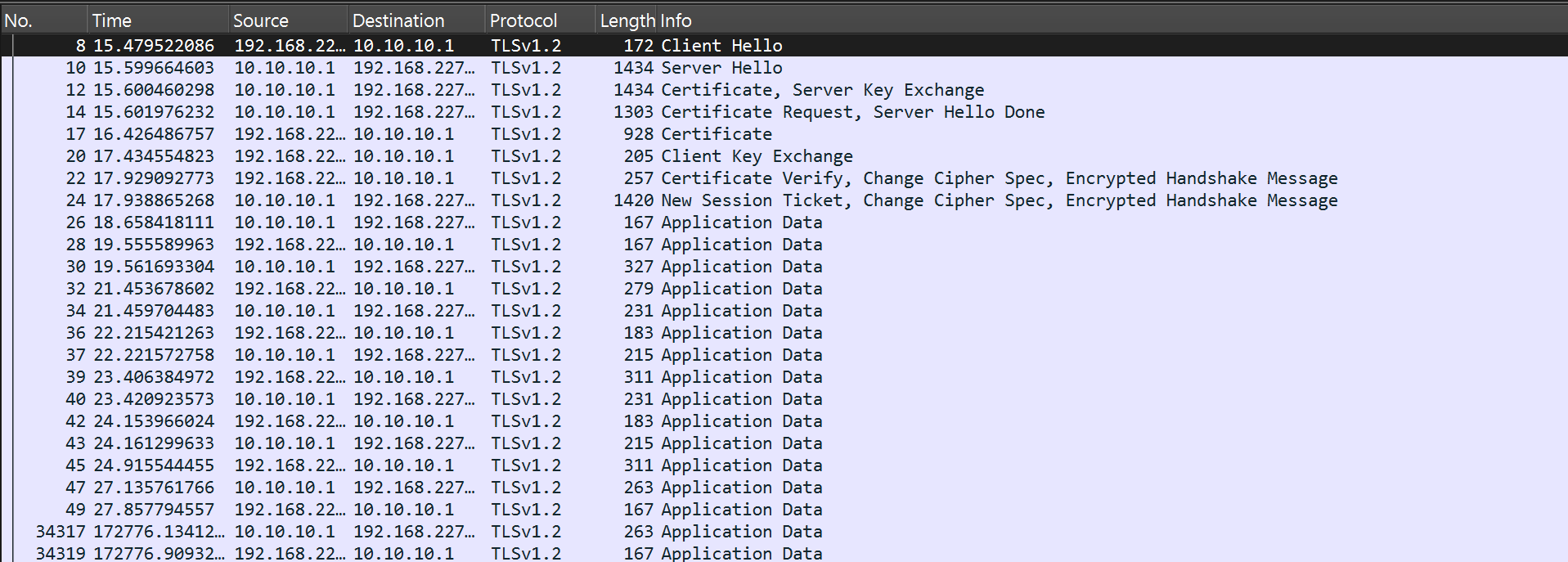}
        \caption{MitMOut TLS traces}
        \label{fig:eval:res:MitMIEC:Out}
    \end{subfigure}
    \caption{Network traces at the adversary: IEC~61850}
    \label{fig:eval:MitMIEC}
\end{figure}

\Cref{fig:eval:MitMIEC} shows the network traffic captured on both interfaces of the adversary using Wireshark for the first trial. The trace confirms that the control signal was delayed by approximately $\Delta_{\max}-\epsilon$ after initial interception. The traces were modified similar to the DAP traces.

\begin{figure}[h]
    \centering
    \begin{subfigure}{0.49\textwidth}
        \centering
            \includegraphics[width=\textwidth]{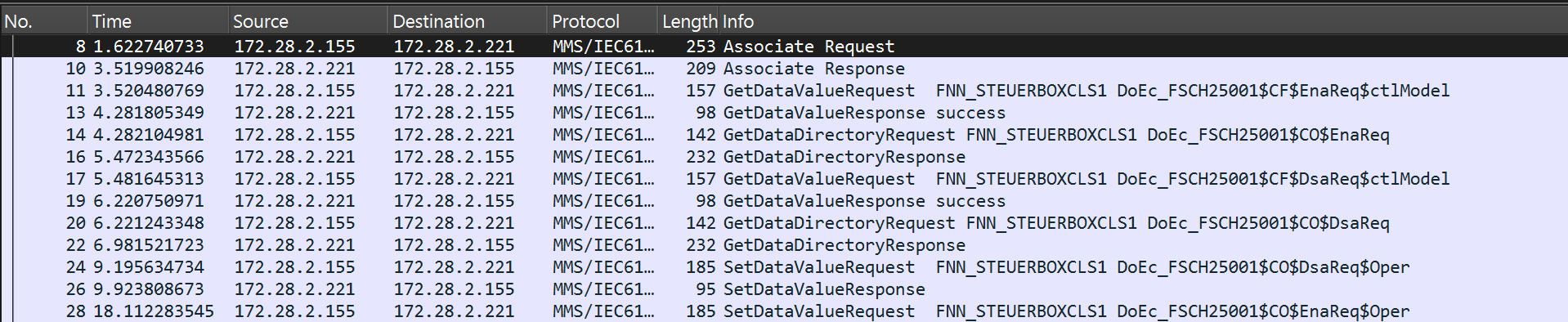}
        \caption{Issuer MMS traces}
        \label{fig:eval:res:AppIEC:In}
    \end{subfigure}
    \begin{subfigure}{0.49\textwidth}
        \centering
        \includegraphics[width=\textwidth]{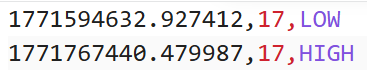}
        \caption{CLS-Stub Logs}
        \label{fig:eval:res:AppIEC:Out}
    \end{subfigure}
    \caption{Application-layer observations: IEC 61850}
    \label{fig:eval:AppIEC}
\end{figure}

\Cref{fig:eval:AppIEC} presents a comparison between the network traffic observed at the Issuer (towards the aEMT) and the relay observer logs for the first trial. Comparing the \textit{EnaReq} message with the relay state transitions shows that the control signal was executed approximately $\Delta_{\max}-\epsilon$ after being issued. The observed relay behavior matches the expected behavior based on its configuration described in \cref{sec:evaluation:lab}.

%% file: impact/usecases.tex
According to §19 (2) MsbG, energy-industry-relevant metering and control operations must be transmitted via the SMGW; these use cases are primarily enumerated in §34 MsbG \cite{web:bnetza:erd}. They include congestion-control operations by a DSO under §14a EnWG (§34 (1) 6. MsbG), by a TSO under §13a EnWG (§34 (1) 7. MsbG), and control signals for direct marketing of EEG- and KWKG-producers (§34 (1) 8. a. MsbG).

To calculate the potential impact of a delayed control signal for a use case $u$ on the power grid, we define $p_{\max}^u$ as the maximum load of a CLS and $p_{\min}^u$ as the minimum load of a CLS in use case $u$. Consequently, the impact of a delayed control signal is calculated as $p_{\text{impact}}^u = p_{\max}^u - p_{\min}^u$. $p_{\max}$ denotes the maximum nameplate load (in kW) at which a CLS can operate, independent of any use case. Note that $p_{\text{impact}}^u$ is a static quantity: it captures the magnitude and sign of the load deviation a delayed control signal can ultimately produce, but not how this deviation depends on the length of the delay or on the state of the grid and CLS at the moment of delayed execution. Such a function is out of scope for this work.

The use case §14a EnWG ($u = \text{14a}$) allows a DSO to limit the load of consumers at grid level 6 and 7 \cite{law:bnetza:bk6_22_300}, guaranteeing a minimum load of 4.2\,kW if their combined maximum load exceeds 4.2\,kWp, in order to prevent congestion within the DSO's grid. If the maximum load of a heat pump or air conditioner exceeds 11\,kWp, the guaranteed minimum load is instead 40\% of its maximum load \cite{law:bnetza:bk6_22_300}. Consumers affected by 14a include heat pumps, wallbox, air conditioners, and batteries \cite{law:deu:enwg,law:bnetza:bk6_22_300}. Consequently,
\[p_{\min}^\text{14a} =
\begin{cases}
4.2\,\mathrm{kW}, & \text{if } p_{\max} \leq 11\,\mathrm{kWp}, \\
0.4 \cdot{} p_{\max}^\text{14a}, & \text{if (heat pump or air conditioner)} \land p_{\max} > 11\,\mathrm{kWp}.
\end{cases}\]

The use case §13a EnWG ($u = \text{13a}$) allows a TSO to increase or limit the production of a producer or battery with a load of more than 100\,kWp in the context of the redispatch / redispatch 2.0 process. Consequently, $p_{\min}^\text{13a}=0$ and $p_{\max}^\text{13a}=p_{\max}$.

Producers participating in direct marketing with more than 25\,kWp are required to be controllable by the direct marketer ($u = \text{DV}$). Producers with more than 100\,kWp are mandated to participate in the direct marketing mechanism, while smaller producers may optionally participate. The direct marketer is able to control the producer's production and consumption entirely. Consequently, $p_{\min}^\text{DV}=0$ and $p_{\max}^\text{DV}=p_{\max}$.

§9 EEG ($u = \text{9}$) further defines a use case where the SMGW, if installed at the property, must be used by the DSO to control a producer if the installed load exceeds 2\,kWp. According to §29 (1) 2. MsbG, an SMGW is mandatory if the installed load exceeds 7\,kWp. For simplicity, we consider only producers exceeding 7\,kWp as relevant under §9 EEG here. Consequently, $p_{\min}^\text{9}=0$ and $p_{\max}^\text{9}=p_{\max}$.

%% file: impact/laa.tex
Our previous theoretical analysis showed that $\Delta_{\max} = 48\,\text{hours}$. If a control signal is delayed for a sufficiently long period (e.g., multiple hours), the state of the power grid (e.g., power flows and frequency) can differ from the state at the time the control signal was originally issued, for example because of changes in generation and consumption conditions \cite{smard:ConsumptionVsProduction:HourlyJuly2025}. Consequently, a delayed control signal may have a fundamentally different impact on the power grid (i.e., an effect that destabilizes rather than stabilizes it).

The attack described in \cref{sec:theory} corresponds to what is commonly referred to as a static load-altering attack (LAA) \cite{paper:laa:firstPaper,paper:laa:survey,paper:laa:gridshock}. An adversary controlling a sufficiently large amount of active power could induce frequency deviations, potentially causing blackouts, brownouts, or damage to grid infrastructure \cite{paper:laa:firstPaper,paper:laa:survey,paper:laa:gridshock}. Even if the controllable active power is insufficient to destabilize the grid in a relevant manner, an adversary could still impose economic costs by increasing the need for curative measures and ancillary services (e.g., forcing TSOs to activate additional balancing reserve capacity). The monetary impacts of delay attacks on energy pricing in low-voltage nano grids have been discussed in \cite{Paper:SmartMetersCyberAttacks}.

Dabrowski et al. analyzed the impact of LAAs in the Central European (CE) synchronous area and defined an attack as successful if the frequency reaches 49\,Hz, since load-shedding measures are triggered at that point \cite{paper:laa:gridshock}. They used the frequency containment reserve (FCR) capacity in combination with the power grid's total load to estimate the amount of active power an adversary would need to control to cause such a frequency deviation \cite{paper:laa:gridshock}. The FCR capacity is required to be at least $\pm 3\,\text{GW}$ according to regulatory requirements \cite{law:eu:sogl}. For 2026, the FCR capacity has been increased to $\pm 3.45\,\text{GW}$ \cite{web:entsoe:fcrAmount} and is likely to increase in the future \cite{law:eu:sogl}. Dabrowski et al. showed that, depending on the actual grid load at the time the LAA is carried out, between 2 and 3.5 times the FCR capacity in active power must be controllable by an adversary to cause a frequency deviation of 1 Hz. Considering the FCR capacity of 2026, this corresponds to $6.9\,\text{GW}$ to $12.0755\,\text{GW}$ of active power required to cause the frequency deviation of 1 Hz. The multipliers were derived by Dabrowski et al. for the grid conditions (load, inertia, and primary-control configuration) prevailing in their study; we adopt it as-is and do not re-derive it for the 2026 CE synchronous area, so the resulting thresholds should be read as indicative rather than precisely calibrated to current and future conditions. We further include the bare FCR capacity (i.e., the $1\times$ case) as a conservative lower reference point. If an adversary can observe the latency, it might be able to piggyback on a naturally occurring deviation, which would lower the effective threshold. The current frequency and load of the CE synchronous area are obtainable using public sources such as \cite{web:energychart:frequency} and \cite{web:entsoe:actualLoad}.

\begin{figure}[h]
    \centering
    \begin{subfigure}{0.49\textwidth}
        \centering
        \includegraphics[width=0.75\textwidth]{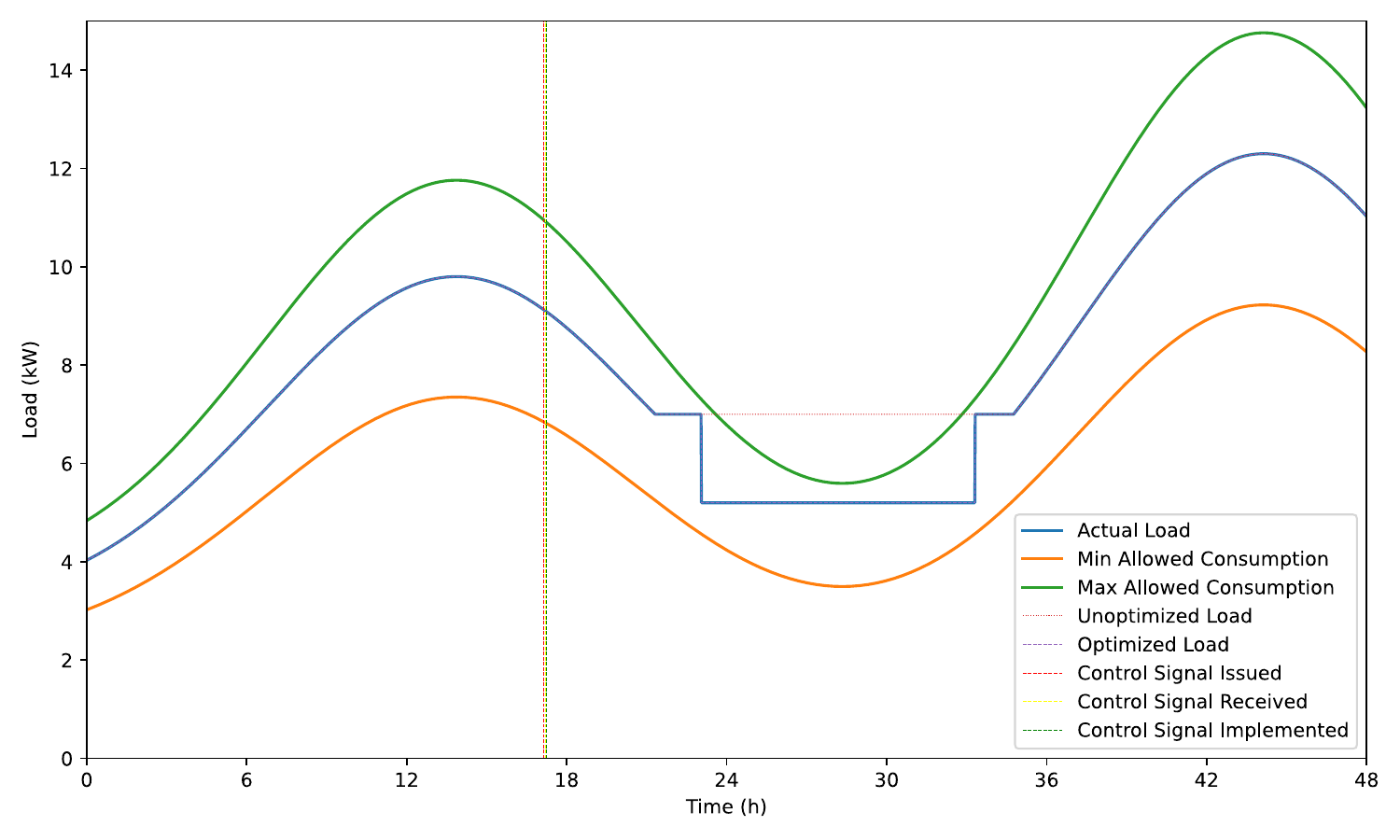}
        \caption{Intended}
        \label{fig:impact:general:normal}
    \end{subfigure}
    \begin{subfigure}{0.49\textwidth}
        \centering
        \includegraphics[width=0.75\textwidth]{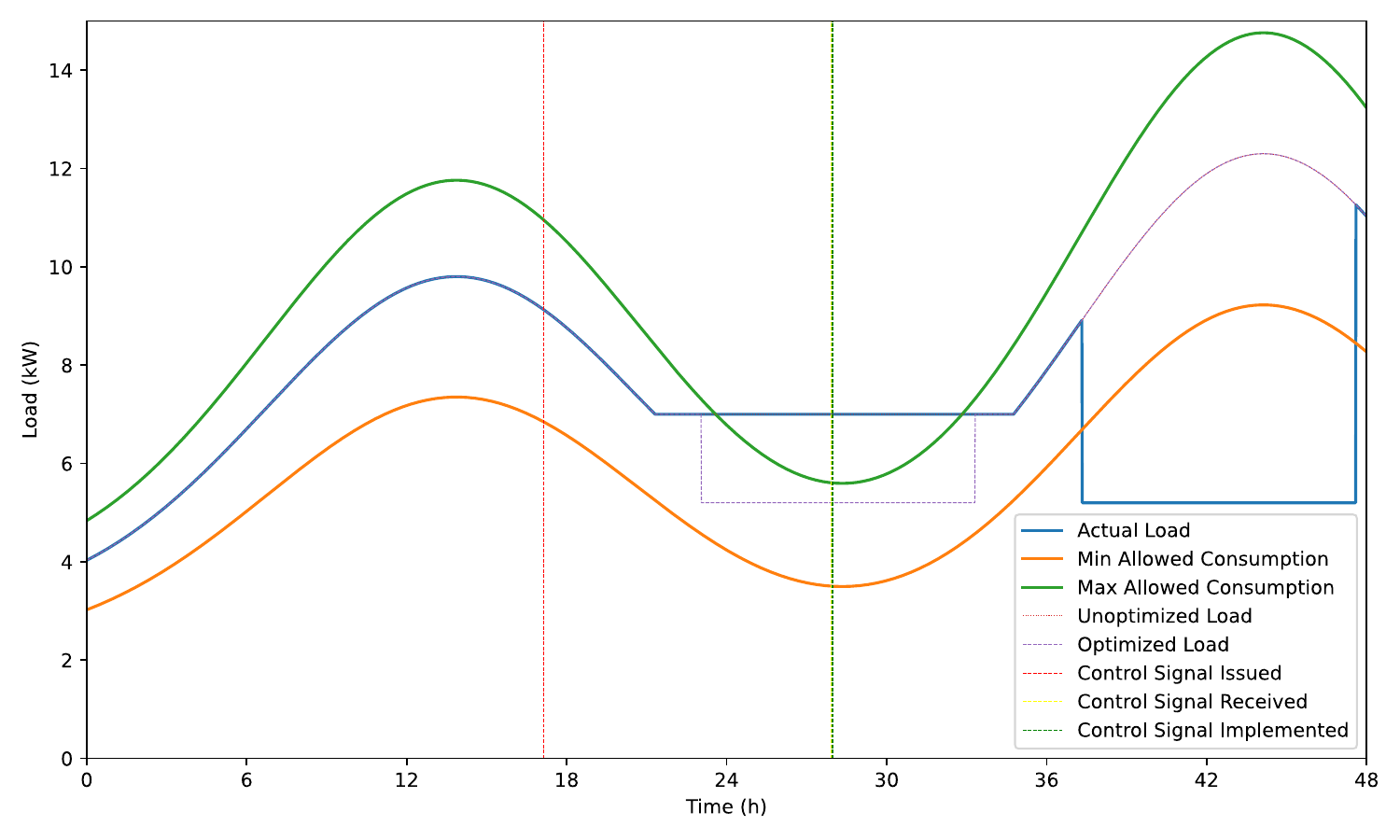}
        \caption{Delay-Attack}
        \label{fig:impact:general:delay}
    \end{subfigure}
    \begin{subfigure}{0.49\textwidth}
        \centering
        \includegraphics[width=0.75\textwidth]{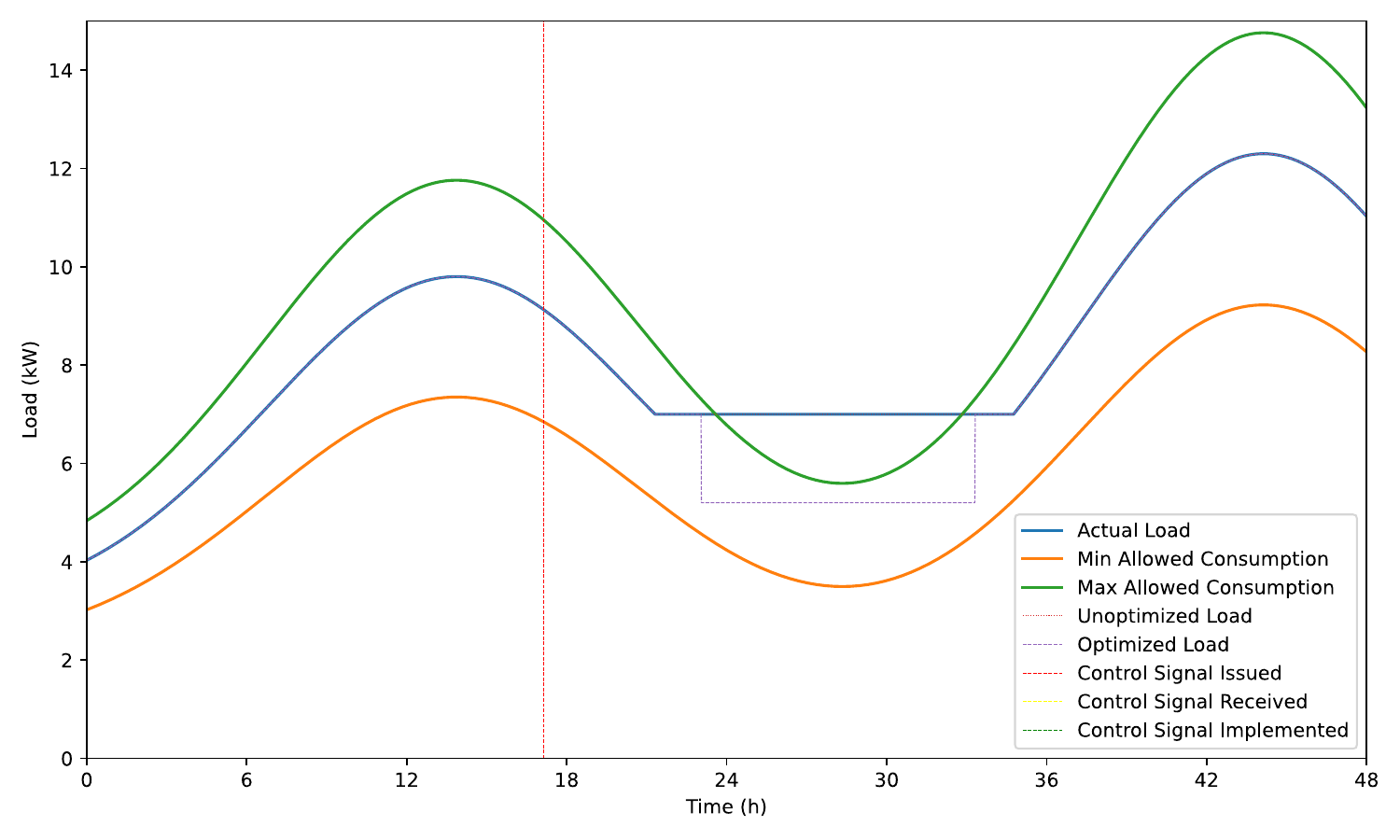}
        \caption{DoS-Attack}
        \label{fig:impact:general:dos}
    \end{subfigure}
    \begin{subfigure}{0.49\textwidth}
        \centering
        \includegraphics[width=0.75\textwidth]{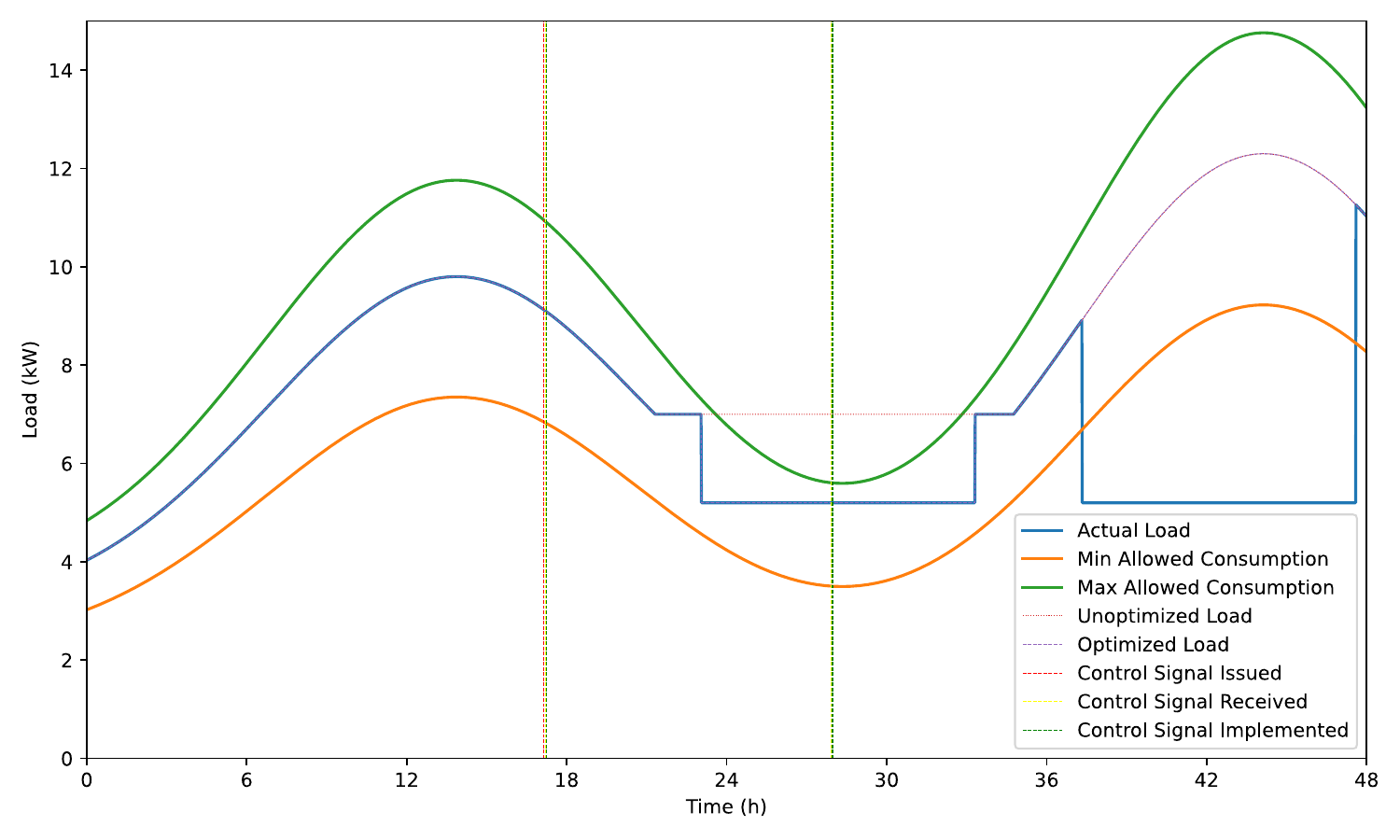}
        \caption{Replay-Attack}
        \label{fig:impact:general:replay}
    \end{subfigure}
    \caption{Impact of different attacks to load curves}
    \label{fig:impact:general}
\end{figure}

\Cref{fig:impact:general} compares the power curve intended by a DSO or TSO with possible power curves resulting from a successful delay, replay, or DoS attack. The data underlying these power curves are purely illustrative. The unoptimized power curve represents the consumption behavior of a consumer, for example as driven by dynamic energy prices. The maximum and minimum consumption limits define the range within which a consumer should operate to avoid negatively impacting the power grid.

\Cref{fig:impact:general:normal} illustrates the restriction of a consumer's power consumption (e.g., a heat pump) through control signals, for example in the context of §14a EnWG. It can be observed that the unoptimized curve would exceed the maximum permitted consumption if no control signal were issued by the DSO. However, because of the control signal, the actual consumption remains within the defined limits at all times.

\Cref{fig:impact:general:delay} demonstrates that, under a successful delay attack, the actual power consumption initially exceeds the maximum consumption limit and subsequently falls below the minimum consumption limit, resulting in an adverse impact on the power grid (e.g., frequency deviations). In comparison to a DoS attack (\cref{fig:impact:general:dos}), a delay attack affects the power grid over a longer period and results in a larger absolute load deviation. Compared to a replay attack (\cref{fig:impact:general:replay}), in which the control signal is executed twice, a delay attack merely shifts its execution in time. 

Using the impact definition from \cref{sec:impact:cases}, we can estimate the impact of a CLS on a per-use-case basis. In the case of the §14a use case, we can estimate the impact of an 11\,kW ($p_{\max}=11$) wallbox as $p_\text{impact}^\text{14a} = p_{\max}^\text{14a} - p_{\min}^\text{14a} = 11\,\text{kW}-4.2\,\text{kW} = 6.8\,\text{kW}$. Since the §14a mechanism is relevant for grid levels 6 and 7 without any kWp restriction, the worst-case impact per device can be estimated to be on the order of several hundred kW, e.g., for a large heat pump. Unfortunately, no detailed data on consumers currently part of the §14a mechanism is publicly available.

Producers that are mainly targeted by the remaining use cases discussed in \cref{sec:impact:cases} can have their impact estimated using the data available in the Marktstammdatenregister (MaStR) \cite{web:bnetza:MaStR}. The maximum load among such use cases is that of a coal-fired power plant, with $p_{\max}=1{,}110{,}000\,\text{kW} = 1.11\,\text{GWp}$ \cite{web:bnetza:MaStR:Large}. Hence, the impact is $p_\text{impact}^\text{u} = p_{\max}^\text{u} - p_{\min}^\text{u} = 1.11\,\text{GW}-0\,\text{kW} = 1.11\,\text{GW}$ for $u \in \{13a, DV, 9\}$. This figure uses $p_{\max}$ as nameplate capacity; for a coal-fired plant this is a reasonable proxy for dispatchable capacity, but this is not true for weather-dependent producers such as PV.

Large producers, especially those located at grid level 1 to 3, must additionally have voice-based communication as a fallback in case the digital communication path is not available \cite{tech:VDE:ARN_4130}. Therefore, an additional communication path not part of TM~1 and TM~2 exists, which the adversary must also be able to control.

A single CLS is unlikely to cause a significant impact on the power grid, since neither the FCR capacity nor the thresholds estimated based on Dabrowski et al.'s work can be exceeded by a single device.

%% file: impact/CoLAA.tex
As outlined previously, it is unlikely that an adversary can cause significant or non-compensable harm to the power grid by delaying a single control signal. If the power grid is in a state in which control signals for the §14a use case are required, it is also unlikely that a DSO would issue such a signal for only a single CLS, because §14a measures are typically applied only after other flexibilities have been exhausted \cite{law:bnetza:bk6_22_300}. The same reasoning applies to the §13a and §9 use cases, since their control signals are used to stabilize the grid. By contrast, control signals for the DV use case may be sent more frequently, for example when a direct marketer optimizes its fleet. If an issuer assumes a control signal failed (e.g. due to a missing application-layer response), the issuer might resend a similar control signal to another CLS which gives the adversary additional control signal.

If an adversary collects multiple control signals and releases them simultaneously to synchronize their execution (batch execution), the resulting load will suddenly increase or decrease, depending on the content of the signals, by the sum of all control signals. This type of attack is sometimes referred to as a coordinated static load altering attack \cite{paper:laa:gridshock}.

To estimate the critical mass of control signals on a per-use-case basis, we use $n_{crit}^{th,u}=\lceil\frac{th}{p_{impact}^{u}}\rceil$ with $th \in \{\text{FCR},\text{MIN},\text{MAX}\}$ and $\text{FCR}=3.45\,\text{GW}$, $\text{MIN}=6.9\,\text{GW}$, $\text{MAX}=12.0755\,\text{GW}$. Comparing $n_{crit}^{th,u}$ to the current number of CLS and projecting this number to 2045, we can estimate whether enough CLS exist in a use case to exceed the threshold. As a baseline for the state of the power grid in 2045, we use the scenarios from the Grid Development Plan Electricity \cite{law:bentza:netzentwicklungsplanSzenarien}.

For the remainder of this section, we make three assumptions. First, the adversary must collect a sufficient number of control signals until an exploitable grid state occurs with respect to $\Delta_{\max}$, thereby enabling control of the required amount of active power. Second, each CLS for which the adversary has collected a control signal must operate at its maximum (or minimum) when the delayed signals are released. Third, the non-execution of the delayed control signals at their intended time must not trigger a power cycle of the SMGW (or CLS-A), for example by overloading a local disruption station.

Only a Metering Point Operator (MPO) is allowed to operate a SMGW \cite{law:deu:msbg}. If the metering locations of a sufficiently large number of CLS are operated by the same MPO, an adversary positioned between the MPO's aEMT system and the connected SMGWs could execute the attack centrally from a single network location and affect all relevant control signals simultaneously while still operating within TM~1 and TM~2. This could be achieved, for example, by compromising the MPO's ISP or the ISP of a SaaS provider serving multiple MPOs. For metering locations operated by other MPOs, an adversary may need to compromise multiple network locations. Note, for SaaS-MPO-Systems like \cite{web:arvato:gwa} multiple MPOs may be attack via the same compromised network location. According to MaStR, 1,119 MPOs were registered in Germany as of 13 July 2026 \cite{web:bnetza:MaStR:MSB}. If a congestion event is limited to a specific region of the grid, however, the adversary would only need to target the subset of MPO operating in that region. The impact of such an attack would likewise be confined to the targeted region. Compromising this number of network location might be challenging but not impossible which contributes to the first assumption. 

To achieve a delay of $\Delta_{\max}$ for many control signals and execute them approximately simultaneously, those signals would need to be transmitted immediately after the CLS channel is established. Consequently, multiple CLS channels would need to be created at the same time, which is unlikely to occur naturally. However, an adversary in our threat model could inject TCP RST packets to force CLS channels to be re-established, thereby synchronizing their lifetimes. Because an exploitable grid state can occur at any time (i.e., after $\Delta_{\text{crit}}$ has elapsed), it may not be necessary to delay every control signal by the full $\Delta_{\max}$.

Considering solely 11\,kW chargers in the §14a use case, $n_{crit}^{FCR,\text{14a}} = \lceil\frac{3.45\,\text{GW}}{6.8\,\text{kW}}\rceil = 507{,}353$, $n_{crit}^{MIN,\text{14a}} = \lceil\frac{6.9\,\text{GW}}{6.8\,\text{kW}}\rceil = 1{,}014{,}706$, and $n_{crit}^{MAX,\text{14a}} = \lceil\frac{12.0755\,\text{GW}}{6.8\,\text{kW}}\rceil = 1{,}775{,}809$ chargers are required to cause a significant impact. As of 1 April 2026, 2,169,717 battery electric vehicles (BEVs) existed in Germany \cite{data:kba:FZ27_04_2026}. The scenarios used for the Grid Development Plan Electricity (German: Netzentwicklungsplan Strom, NEP) indicate that by 2045 between 32.4 and 39.8 million BEVs will exist in Germany \cite{law:bentza:netzentwicklungsplanSzenarien}. A 2025 market analysis \cite{web:uscale:studyPrivateCharging} states that 80\% of electric vehicle owners charge at home and that 85\% of these own a wallbox, implying approximately 0.68 wallboxes per BEV. To satisfy the second assumption, the coincidence factor must be considered. In extreme cases, it can be as high as 80\% if all wallboxes respond to price incentives, based on a simulation of 150 wallboxes in a study by the VDE that is used in German grid planning \cite{web:vde:coincidenceWallbox}. Applying these factors, we estimate that approximately 1,180,320 wallboxes existed in Germany as of 1 April 2026, which is reasonable for satisfying the second assumption. Projecting the same factor to 2045, between 17.6 and 21.6 million private wallboxes are expected to exist while satisfying the second assumption. This current estimate already exceeds both the FCR and MIN thresholds computed above, though not the MAX threshold; by 2045, all three thresholds are exceeded several times over. Note that a separate ADAC survey reports a higher, directly stated wallbox-ownership rate of 80\% among EV owners \cite{web:adac:eCar} rather than the compound 68\% used above; applying this alternative figure instead would yield approximately 1,735,774 wallboxes today and between 25.9 and 31.8 million by 2045. This indicates that, for the §14a use case, the current impact depends on the grid situation in CE, while in the future enough wallboxes are expected to exist to cause a significant impact largely independent of grid conditions they are released. Note, the NEP estimates that between 30\% and 80\% of the consumers are operated market-driven in 2045 \cite{law:bentza:netzentwicklungsplanSzenarien}. Therefore, the coincidence factor can be less than the worst case of 80\%. Since, 14a is related to congestion management, the missing execution can result in a violation of the third assumption. However, since wallboxes alone might exceed the threshold, an attack combining use cases require only a sub set of control signal. Hence, the third assumption can hold true.  

There are 142,235 producers relevant to the §13a use case, with an aggregated load of 242,385,685\,kW \cite{web:bnetza:MaStR:ezaGT100}. Consequently, we assume that the average impact of a CLS in this use case is $p_\text{impact}^\text{13a} \approx \frac{242{,}385{,}685\,\text{kW}}{142{,}235} \approx 1{,}704\,\text{kW}$. Therefore, $n_{crit}^{FCR,\text{13a}} = \lceil\frac{3.45\,\text{GW}}{1{,}704\,\text{kW}}\rceil = 2{,}025$, $n_{crit}^{MIN,\text{13a}} = \lceil\frac{6.9\,\text{GW}}{1{,}704\,\text{kW}}\rceil = 4{,}050$, and $n_{crit}^{MAX,\text{13a}} = \lceil\frac{12.0755\,\text{GW}}{1{,}704\,\text{kW}}\rceil = 7{,}087$. Considering only producers at grid levels 5--7 that are subject to the §13a use case, the number of producers is reduced to 121,578, with an aggregated load of 97,289,849\,kW \cite{web:bnetza:MaStR:ezaGT100GL57}. Consequently, the average impact of a CLS is reduced to $p_\text{impact}^\text{13a} \approx \frac{97{,}289{,}849\,\text{kW}}{121{,}578} \approx 800\,\text{kW}$. Therefore, $n_{crit}^{FCR,\text{13a}} = \lceil\frac{3.45\,\text{GW}}{800\,\text{kW}}\rceil = 4{,}313$, $n_{crit}^{MIN,\text{13a}} = \lceil\frac{6.9\,\text{GW}}{800\,\text{kW}}\rceil = 8{,}625$, and $n_{crit}^{MAX,\text{13a}} = \lceil\frac{12.0755\,\text{GW}}{800\,\text{kW}}\rceil = 15{,}095$. This indicates that there are enough producers, even solely at grid levels 5--7, to exceed our thresholds. As noted in \cref{sec:impact:cases}, these figures rest on nameplate rather than dispatchable capacity.

By 2045, the NEP estimates that production capacity will increase by between 220.5\% and 289.6\% relative to 2024 \cite{law:bentza:netzentwicklungsplanSzenarien}. However, there is no indication of the grid level or nameplate load at which these producers operate. Therefore, no estimate of the number of available producers can be deduced.

In the first half of 2026, there were a total of 9,456 redispatch operations, with a maximum of 126 operations and 36.935\,GW on a single calendar day, carried out by the four TSOs (data exclude exchange-based redispatch operations) \cite{web:uenb:rdp}. In addition, DSO-initiated redispatch must be considered; however, no consolidated data are publicly available. The number of redispatch operations shown here does not reflect the actual number of control signals sent to CLS, since a TSO can delegate the selection and control of the actual CLS to a DSO. Since it exceeds the threshold the first assumption can be satisfied for the 13a use case. 

For the DV and §9 use cases, we can reuse the calculation for the §13a use case, since all producers included in that calculation are part of the direct-marketing mechanism and must be controllable under §9 EEG, and this population already exceeds the thresholds defined above. The number of producers that fall under the DV and §9 use cases is, in fact, higher than the number under §13a; thus, the corresponding thresholds are met with an even larger margin. In the context of control signals issued by direct marketers, these signals are sent more frequently and more predictably than those used for stability and congestion management, particularly during periods of negative day-ahead market prices, since a direct marketer has an incentive to reduce production to 0 kW. This satisfies the first assumption for the DV use case. Since DV related control signals are not related to congestion management, it is likely that the third assumption holds true. 

To summarize, the estimates in this section indicate that, from a pure device-count perspective, sufficient controllable load or generation capacity plausibly exists to reach the thresholds derived from Dabrowski et al.'s work, both today (for §13a, DV, and §9) and, prospectively, for §14a as the wallbox population grows toward 2045. Note that an adversary can combine different use cases and control directions to obtain the required amount of controllable load. However, the three assumptions effectively limit such an attack to a specific window of opportunity, the quantification of which is beyond the scope of this work.

%% file: countermeasure/shortTem.tex
\subsubsection*{General}
Due to the strict certification requirements and overall complexity of the SMI, short-term mitigations avoid modifications to protocols or implementations.

A viable short-term approach is therefore to enforce restrictive configuration parameters on the SMGW that prevent vulnerable operational modes of the CLS channel.

\subsubsection*{SMGW Parameter Restriction}

Using our formal model introduced in \cref{sec:theory:delay}, we can show that it is sufficient to mitigate the vulnerability at the TLS layer of the SMGW ($TLS_{SMGW}$):
\\Assuming
\[\forall t \in T : \big(V_{layer}(TLS_{SMGW},t)=\textsf{i} \;\land\; \delta^{\max}_{layer}(TLS_{SMGW},t)<\Delta_{\text{crit}}\big),\]
it follows that
\[\forall t \in T : \big(V_{sys}(\{\text{SMGW},\text{CLS-A},\text{aEMT}\},t)=\textsf{i} \;\land\;  \delta^{\max}_{sys}(\{\text{SMGW},\text{CLS-A},\text{aEMT}\},t)<\Delta_{\text{crit}}\big).\]

To achieve such invulnerable configurations, the maximum TLS session lifetime must be restricted to $< \frac{\Delta_{\text{crit}}}{s}$ with $s\in\mathbb{N}$ as the security margin. Choosing $s=1$ mitigates the attacks proposed in this work. However, choosing $s>1$ may further mitigate delay attacks targeting the retry behavior of the aEMT, e.g., if the aEMT tries to retransmit the same control signal in subsequent TLS sessions for a certain amount of time.

In case of \textit{HKS.TLSPROXY.SOCKS} and \textit{HKS.TLSPROXY.SRV}, this mitigation would lead to a significant increase in WAN traffic and resource consumption at the SMGW and aEMT, since the TCP and TLS handshake process must be executed frequently (e.g., at least every $\frac{ \Delta_\text{crit}}{s}$). This mitigation results in
\[\forall t \in T : \big(V_{layer}(TLS_{SMGW},t)=\textsf{i} \; \land \; \delta^{\max}_{layer}(TLS_{SMGW},t)< \tfrac{\Delta_{\text{crit}}}{s}\big) \text{ with } s > 0\]
which satisfies our initial assumption.

To reduce the communication overhead during idle times (e.g., when no CLS control is required), the communication scenario \textit{HKS.TLSPROXY.CLI} can be used in combination with a maximum session lifetime of $< \frac{\Delta_{\text{crit}}}{s}$ for the CLS channel. In \textit{HKS.TLSPROXY.CLI}, the aEMT requests the GWA to instruct the SMGW to establish the CLS channel \cite{tech:bsi:tr-03109-1}. Upon receiving this instruction, the SMGW establishes two TLS sessions: one for the HAN-CLS side and one for the WAN side \cite{tech:bsi:tr-03109-1}. This process is referred to as \textit{FA.DoRequestProxyCh} in the SMI specification \cite{tech:bsi:tr-03109-1}.

Under TM~1, an adversary could target the CLS channel request issued by the GWA; this scenario must additionally be considered when choosing $s$. This mitigation results in
\[\forall t \in T : \big(V_{layer}(TLS_{SMGW},t)=\textsf{i} \; \land \; \delta^{\max}_{layer}(TLS_{SMGW},t)< \tfrac{\Delta_{\text{crit}}}{s}\big) \text{ with } s > 0\]
which satisfies our initial assumption.

The downside of using \textit{HKS.TLSPROXY.CLI} is that additional communication overhead is introduced for requesting a CLS channel, which increases the RTTs and traffic.

%% file: countermeasure/midTerm.tex
\subsubsection*{General}
While, as shown in \cref{sec:countermeasure:short}, mitigation against delay attacks can be achieved quickly at the TLS layer by restricting TLS session length, this comes at the cost of additional overhead. As an alternative, the mitigation could be achieved at the application layer without modifying the protocol itself, by enforcing previously optional security mechanisms or checks on a per-protocol basis.

\subsubsection*{CLS.EEDI}

As our analysis in \cref{sec:theory:clseedi} shows, the keep-alive feature of MQTT can be used to mitigate delay attacks.

MQTT does not mandate that a client disconnects if it does not ``[receive] a \textit{PINGRESP} packet within a reasonable amount of time after sending a \textit{PINGREQ}'' \cite{tech:standard:mqtt-v5}. By enforcing heartbeats and a disconnect within a time bound of $< \frac{\Delta_{\text{crit}}}{s}$, delay attacks can be mitigated. In this case, our formal model evaluates to
\[\forall t \in T : \big(V_{layer}(MQTT_{CLS-A},t)=\textsf{i} \;\land\; \delta^{\max}_{layer}(MQTT_{CLS-A},t)< \tfrac{\Delta_{\text{crit}}}{s}\big) \text{ with } s>0.\] 
Consequently,
\[V_{sys}(\{\text{SMGW},\text{CLS-A},\text{aEMT}\},\textsf{TM2})=\textsf{i} \text{ with } \delta^{\max}_{sys}(\{\text{SMGW},\text{CLS-A},\text{aEMT}\}, \textsf{TM2})<\tfrac{\Delta_{\text{crit}}}{s} \text{ with } s>0.\] 

This mitigation aligns with the one outlined in \cite{dis:ChenglongFu}.

\subsubsection*{IEC 61850}

The previous analysis in \cref{sec:theory:iec61850} showed that an \textit{EnaReq} already includes a timestamp $t$. If the FNN-STB specification mandates that this timestamp must be validated according to $lt-t < \tfrac{\Delta_{\text{crit}}}{s}$ where $lt$ denotes the current local time of the CLS-A a delay attack can be mitigated, since the formal model evaluates to 
\[\forall t \in T : \big(V_{layer}(61850_{CLS-A},t)=\textsf{i} \;\land\; \delta^{\max}_{layer}(61850_{CLS-A},t)< \tfrac{\Delta_{\text{crit}}}{s}\big) \text{ with } s > 0.\] 
Consequently,
\[V_{sys}(\{\text{SMGW},\text{CLS-A},\text{aEMT}\},\textsf{TM2})=\textsf{i} \text{ with } \delta^{\max}_{sys}(\{\text{SMGW},\text{CLS-A},\text{aEMT}\},\textsf{TM2})<\tfrac{\Delta_{\text{crit}}}{s} \text{ with } s>0.\] 

Another mitigation approach is to mandate the use of end-to-end encryption according to IEC 62351-4 \cite{tech:standard:iec62351-4}. This mechanism includes timestamp validation with a constraint such as $lt-t < 10\,\text{minutes}$ with $lt$ as the current local time of the CLS-A. Assuming that this validation is enforced, the formal model evaluates to
\[\forall t \in T : \big(V_{layer}(61850_{aEMT},t)=\textsf{i} \;\land\; \delta^{\max}_{layer}(61850_{aEMT},t) < 10\,\text{minutes} < \Delta_{\text{crit}}\big).\]
Consequently,
\[V_{sys}(\{\text{SMGW},\text{CLS-A},\text{aEMT}\},\textsf{TM2})=\textsf{i} \text{ with } \delta^{\max}_{sys}(\{\text{SMGW},\text{CLS-A},\text{aEMT}\},\textsf{TM2})\leq 10\,\text{minutes} < \Delta_{\text{crit}}.\]

The first mitigation introduces only negligible processing overhead for timestamp validation. In contrast, the latter mitigation introduces additional overhead during the handshake phase as well as per control signal, e.g., through additional cryptographic operations.

Another mitigation strategy is to enforce Select-Before-Operate (SBO) control instead of direct control (e.g., \textit{sbo-with-normal-security}) for critical operations such as \textit{EnaReq} with a constraint \textit{sboTimeout} $<\frac{\Delta_{\text{crit}}}{s}$. 

SBO control requires the client to first select and lock the control object for the duration defined in \textit{sboTimeout}. Only within this time window can the client successfully request the corresponding control operation. Once \textit{sboTimeout} expires, the object must be selected again; otherwise, the operate command fails.

Using our formal model we obtain
\[\forall t \in T : \big(V_{layer}(61850_{CLS-A},t)=\textsf{i} \;\land\;  \delta^{\max}_{layer}(61850_{CLS-A},t)=\text{sboTimeout} < \frac{\Delta_{\text{crit}}}{s}\big).\]
Consequently,
\[\forall t \in T : \big(V_{sys}(\{\text{SMGW},\text{CLS-A},\text{aEMT}\},t)=\textsf{i} \;\land\; \delta^{\max}_{sys}=\text{sboTimeout} < \frac{\Delta_{\text{crit}}}{s}\big).\]

The SBO-based mitigation introduces additional communication overhead due to the extra RTT required to select and lock the control object.

%% file: countermeasure/longTerm.tex
\subsubsection*{General}
To fix a protocol that previously evaluated to $\textsf{v}$ (e.g. TLS), an extension which is mandatory to implement by relevant components in the system architecture can mitigate the attack.

In the following, we focus on optional protocol modifications and extensions that preserve backward compatibility. Our formal model evaluates to $\textsf{i}$ only if such modifications are mandatory.

\subsubsection*{CLS.EEDI}

CLS.EEDI can be mitigated using two approaches. The first approach introduces an additional timestamp within the CLS.EEDI data model. The second approach requires CLS.EEDI to specify additional requirements for the MQTT layer to include a timestamp.

For the first approach, an optional timestamp $t$ is included in the message header and validated using $lt - t < \frac{\Delta_{\text{crit}}}{s}$, where $lt$ denotes the current time of the CLS-A. Alternatively, the timestamp can be embedded in the user properties of an MQTT PUBLISH message and processed equivalently. A plugin for MQTT implementations exists which adds a timestamp \cite{web:mqqt:timestampPlugin}. This mitigation introduces only minimal overhead in terms of message size and processing time.

Using our formal model, we can show that
\[\forall t \in T : \big(V_{layer}(CLS.EEDI_{CLS-A}, t) = \textsf{i} \;\land\; \delta^{\max}_{layer}(CLS.EEDI_{CLS-A}, t) < \frac{\Delta_{\text{crit}}}{s}\big),\]
since $\Delta_{\text{aEMTOut}\rightarrow {CLS-AIn}} = lt-t < \frac{\Delta_{\text{crit}}}{s}$ must be fulfilled. Consequently,
\[\forall t \in T : \big(V_{sys}(\{\text{SMGW},\text{CLS-A},\text{aEMT}\}, t) = \textsf{i} \;\land\;  \delta^{\max}_{sys}(\{\text{SMGW},\text{CLS-A},\text{aEMT}\}, t) < \frac{\Delta_{\text{crit}}}{s}\big).\]

An experimental version of CLS.EEDI (1.2.0-exp) introduces an additional message type that enables the specification of a power curve with a duration defined relative to an absolute point in time \cite{tech:standard:cls-eedi:draft}. If such a message is delayed sufficiently, its intended start time may already be in the past upon arrival. In this situation, the behavior of the CLS-A is not clearly specified \cite{tech:standard:cls-eedi:draft}. Nevertheless, this additional message type could provide a mitigation for delay attacks if the CLS-A behaviour is specified unambiguously.

\subsubsection*{TLS 1.3}

Mitigating delay attacks at the TLS layer may be the most robust solution, because it mitigates Delay Attack independently of the application-layer protocol.

We define an additional TLS~1.3 content type, \textit{application\_data\_timestamped}, which includes a timestamp $t$ in addition to all properties of the \textit{application\_data} content type. Support for this extension is negotiated via TLS~1.3 extensions during the handshake. A TLS~1.3 peer that receives a message of type \textit{application\_data\_timestamped} verifies that $lt - t < \frac{\Delta_{\text{crit}}}{s}$, where $lt$ denotes the local clock time of the receiving peer. This allows either a TLS implementation or the application to decide for which application-data messages a maximum transmission time should be enforced. This mitigation introduces communication overhead comparable to adding a timestamp at the application layer.

From an SMI perspective, it must be mandatory for the SMGW and aEMT to implement this extension. Furthermore, application-data messages must be sent using the \textit{application\_data\_timestamped} content type only.

Using our formal model, we can show that
\[\forall t \in T : \big(V_{layer}(TLS_{SMGW},t)=\textsf{i} \;\land\; \delta^{\max}_{layer}(TLS_{SMGW},t)<\frac{\Delta_{\text{crit}}}{s}\big),\]
since $\Delta_{aEMTOut\rightarrow SMGWIn} = lt-t < \frac{\Delta_{\text{crit}}}{s}$ with $s > 1$ can be fulfilled. Consequently,
\[\forall t \in T : \big(V_{sys}(\{\text{SMGW},\text{CLS-A},\text{aEMT}\},t)=\textsf{i} \;\land\; \delta^{\max}_{sys}(\{\text{SMGW},\text{CLS-A},\text{aEMT}\},t)<\frac{\Delta_{\text{crit}}}{s}\big).\]

\subsubsection*{DR-TL}
Another approach to provide resistance against delay attacks independently of the application-layer protocol can be the implementation of a protocol on top of TLS, referred to as Delay-Resistant Transport Layer (DR-TL). This protocol explicitly enforces resistance to delay attacks. Below we outline three variants: a timestamp-based, a keep-alive-based, and a delivery-process-based mitigation.

The timestamp-based variant adds a fixed-length header of 8 bytes to the payload, containing a timestamp $t$. The receiver verifies that $lt - t < \frac{\Delta_{\text{crit}}}{s}$. If this condition is violated, the connection is terminated. This mitigation introduces communication overhead comparable to adding a timestamp at the application layer.

Using our formal model, we can show that
\[\forall t \in T : \big(V_{layer}(\mathrm{DR\text{-}TLS}_{SMGW},t)=\textsf{i} \;\land\; \delta^{\max}_{layer}(\mathrm{DR\text{-}TLS}_{SMGW},t) = \frac{\Delta_{\text{crit}}}{s}\big),\]
since $\Delta_{aEMT\rightarrow CLS-A} = lt-t < \frac{\Delta_{\text{crit}}}{s}$. Consequently,
\[\forall t \in T : \big(V_{sys}(\{\text{SMGW},\text{CLS-A},\text{aEMT}\},t)=\textsf{i} \;\land\; \delta^{\max}_{sys}<\frac{\Delta_{\text{crit}}}{s}\big).\]

Another approach is to use a keep-alive mechanism, similar to the heartbeat mechanism of TLS~1.2, to ensure that both peers can verify liveness. The protocol defines two message types with a header size of 1 byte: 0x00 denotes a \textit{PAYLOAD} message and 0x01 denotes a \textit{PING} message.

Each participant sends a \textit{PING} message every $\frac{\Delta_{\text{crit}}}{2s}$ seconds. If a \textit{PING} message is not received within $\frac{\Delta_{\text{crit}}}{2s}$, the connection is silently closed. Since TLS does not allow message reordering, this protocol ensures $\Delta_{\max} = \frac{\Delta_{\text{crit}}}{s}$. A drawback of this approach is the continuous generation of additional traffic. This mitigation is in principle similar to the vendor-specific multiplex protocol provided by the aEMT used in \cref{sec:evaluation}, which includes a heartbeat within the CLS channel.

Using our formal model, we can show
\[\forall t \in T : \big(V_{layer}(\mathrm{DR\text{-}TLS}_{SMGW},t)=\textsf{i} \;\land\; \delta^{\max}_{layer}(\mathrm{DR\text{-}TLS}_{SMGW},t)=\Delta_{\text{crit}}\big),\]
since a \textit{PAYLOAD} message can only be delayed by $\frac{\Delta_{\text{crit}}}{s}$ before the connection is closed due to a missing \textit{PING} message. Consequently,
\[\forall t \in T : \big(V_{sys}(\{\text{SMGW},\text{CLS-A},\text{aEMT}\},t)=\textsf{i} \;\land\; \delta^{\max}_{sys}(\{\text{SMGW},\text{CLS-A},\text{aEMT}\},t)<\Delta_{\text{crit}}\big).\]

A third approach is to define a delivery process that ensures delay resistance. This protocol uses three message types with a fixed-length header of 3 bytes. The most significant byte (MSB) encodes the message type: 0x00 denotes a \textit{PAYLOAD} message, 0x01 a \textit{REQUEST} message, and 0x02 a \textit{RELEASE} message. The remaining bytes contain a unique identifier for each payload.

Upon receiving a \textit{PAYLOAD} message, the receiver sends a corresponding \textit{REQUEST} message. The sender responds with a \textit{RELEASE} message if the time between sending the \textit{PAYLOAD} and receiving the \textit{REQUEST} is less than $\frac{\Delta_{\text{crit}}}{2s}$. Upon receiving the \textit{RELEASE} message, the receiver verifies that the time between receiving the \textit{PAYLOAD} and the \textit{RELEASE} is less than $\frac{\Delta_{\text{crit}}}{2}$.

This protocol ensures that a single DR-TL message cannot be delayed by more than $\Delta_{max}<\frac{\Delta_{\text{crit}}}{2s}$. An adversary delaying two messages by $<\frac{\Delta_{\text{crit}}}{2s}$ each results in $\Delta_{max}<\frac{\Delta_{\text{crit}}}{s}$. Delaying all three messages beyond $\frac{\Delta_{\text{crit}}}{4s}$ each is not possible, as this would violate the timing constraint for the \textit{REQUEST} message.

This mitigation increases both latency and communication overhead, as an additional RTT is required. In its core it is similar to the QoS 2 processes of MQTT.

Using our formal model, we can show
\[\forall t \in T : \big(V_{layer}(\mathrm{DR\text{-}TLS}_{SMGW},t)=\textsf{i} \;\land\; \delta^{\max}_{layer}(\mathrm{DR\text{-}TLS}_{SMGW},t) < 2 \cdot \frac{\Delta_{\text{crit}}}{2s}<\frac{\Delta_{\text{crit}}}{s}\big).\]
Consequently,
\[\forall t \in T : \big(V_{sys}(\{\text{SMGW},\text{CLS-A},\text{aEMT}\},t)=\textsf{i} \;\land\; \delta^{\max}_{sys}(\{\text{SMGW},\text{CLS-A},\text{aEMT}\},t)<\frac{\Delta_{\text{crit}}}{s}\big).\]

%% file: discussion/gen.tex
The argumentation in this work is based on control signals and data sent from the aEMT to a CLS-A. However, the proposed attack is theoretically not limited to this type of data and direction. It is reasonable to assume that measurements (e.g., using RLM over the CLS channel \cite{web:vde:irlmsys}) transmitted via the CLS channel can be delayed in the same way, provided that no application-layer countermeasures are implemented. Note, the TLS session length of the aEMT is not directly specified by SMI requirements \cite{tech:bsi:certPolicy}. Hence, the maximum achievable delay may exceed 48 hours.

This work focuses on delay attacks targeting the CLS channel provided by the SMGW. However, the underlying vulnerability is neither specific to the SMGW nor to the SMI and may exist in other technologies used within power grids. In particular, systems may be vulnerable if vendors assume that TLS inherently mitigates all threats affecting the integrity of transmitted data. The protocol analyses conducted for CLS.EEDI and IEC 61850 can therefore be extended to any component in which a TLS session remains idle for longer than $\Delta_{\max}$. It is worth noting that this could be the case for communication between a manufacturer's cloud and the CLS on the premises, especially in the context of the power to issue a statutory instrument under \S94(2) EEG, which permits the German Federal Ministry for Economic Affairs and Energy to allow the use of manufacturer clouds for grid-friendly control \cite{law:deu:eeg}.

%% file: discussion/limit.tex
\subsubsection*{Evaluation}

For the empirical evaluation, we used components and software that are either development versions or self-developed (TRL < 9). Therefore, we cannot make definitive statements about the vulnerability of production systems. However, our theoretical analysis is based on protocol specifications rather than specific implementations. Since we did not implement the protocols ourselves (with the exception of DAP), but instead relied on existing implementations to validate our theoretical analysis, this does not affect the validity of our results.

As we used a certified version of the SMGW firmware, we can infer that the SMGW is vulnerable in production environments. However, this vulnerability is not caused by a vendor-specific implementation as we discussed earlier in \cref{sec:theory:delay}.

We did not empirically evaluate CLS.EEDI; therefore, conclusions regarding this protocol are limited to theoretical analysis.

\subsubsection*{Vulnerability Model}

The proposed vulnerability model was specifically designed for the considered system architecture and is not intended to be directly generalizable to other system architectures.

In this work, we exclusively considered attacks targeting a single TLS and application-layer session. If either layer implements retransmission mechanisms for control signals or their intended effects, our formal model is no longer directly applicable, as it does not account for application-layer retransmissions, which are typically implementation-specific.

For protocols that expect an application-layer response to a control signal, the absence of such a response may cause the aEMT to retransmit the control signal, either within the same TLS session or in a new one. An adversary aware of such behavior could drop all but the final retransmission, which can then be delayed according to the proposed attack. Additionally, an adversary could delay the establishment of new TLS sessions to further increase $\Delta_{\max}$. Attacks involving multiple sessions therefore contribute to a greater $\Delta_{\max}$.

\subsubsection*{Countermeasures}

\Cref{sec:countermeasure} provides several countermeasures each with its own pros and cons to mitigate the attacks we proposed. Since we do not consider retransmission mechanisms on the application layer in this work, such countermeasures might not be sufficient to mitigate all kinds of delay attacks.

We further did not evaluate the proposed countermeasures within the laboratory setup. Hence, no empirical indication regarding their effectiveness exists yet.

%% file: discussion/impa.tex
Producers with active power capacity $\geq 135\,\mathrm{kWp}$ are also subject to real-time requirements for reactive power control \cite{law:deu:enwg,tech:VDE:ARN_4110,tech:VDE:ARN_4120,tech:VDE:ARN_4130}. While these requirements could, in principle, be exploited to affect grid stability, reactive power control has not been considered in this works.

Many producers are required to prioritise control signals and to ignore them if the grid frequency deviates by more than approximately $\pm$200\,mHz to $\pm$500\,mHz (depending on DSO/TSO configuration); they must then increase or reduce production according to the direction of the deviation \cite{tech:VDE:ARN_4105,tech:VDE:ARN_4110,tech:VDE:ARN_4120,tech:VDE:ARN_4130}. Consequently, attempting to manipulate producers in critical grid situation via delayed control signals may have limited or no effect in some configurations.

The threshold criteria derived from \cite{paper:laa:gridshock} do not consider current or future grid load, nor scenarios where the entire load change occurs within the German portion of the Central European synchronization area. A system split that islands Germany would reduce available FCR capacity and change load patterns, which could substantially affect thresholds. Additionally, transfer capacity between Germany and neighbouring TSOs and events such as (distribution) transformer overloads are not considered in our estimates.

The estimates in \cref{sec:impact:CoLAA} for a coordinated static load-altering attack establish only whether a sufficient number of CLS exist in principle for $n_{crit}^{th,u}$ to be reached; it does not quantify the cost or likelihood of an adversary mounting such an attack at scale. In particular, the contextual-knowledge requirement in our threat models — identifying which packets on a communication path carry a relevant control signal — depends on implementation- and configuration-specific details. This work does not analyse the generalizability of the packet-identification mechanism or whether parameters derived from a single control signal transfer to many other signals. §34 MsbG defines additional use cases beyond those discussed in \cref{sec:impact:cases} for which communication requirements are not yet fully covered by existing standards \cite{tech:FNN:STB,tech:standard:cls-eedi,tech:standard:cls-eedi:draft,tech:uenb:rr-itAnf,tech:bsi:tr-03109-1,tech:bsi:tr-03109-5}. Increased standardization may, over time, make generalization easier.

%% file: conclusion/summary.tex
This work demonstrated that the current set of countermeasures mandated in the Protection Profile and TRs \cite{tech:bsi:tr-03109-1,tech:bsi:tr-03109-5,tech:bsi:tr-03116-3,tech:bsi:pp-smgw} are not sufficient to prevent an adversary from manipulating the behavior of a CLS located at the customer's premises (T.DataModificationWAN). Furthermore, this work estimates that an adversary exploiting the identified vulnerability at scale (coordinated load-altering attack) could cause a significant frequency deviation potentially resulting in load shedding, since the adversary is theoretically able to control several times the current FCR capacity in the Central European synchronous area. To address this, we propose several countermeasures, each with different implementation efforts and trade-offs.

In detail, this work answers the initial research questions posed in \cref{sec:intro:motivation} as follows:
\begin{itemize}
\item \textit{What is the theoretical maximum delay that an attacker could introduce for control signals between an aEMT and a CLS while keeping the control signal successfully implemented?}
\item \textit{Do application-layer protocols such as CLS.EEDI or IEC 61850, when used within the CLS channel, sufficiently mitigate delay attack vulnerabilities?}
\end{itemize}

We were able to demonstrate a delay attack on control signals transmitted via the CLS channel of the SMGW on a theoretical basis and validated these findings in the Smart Grid Laboratory at THU. Our results show that an adversary operating within a threat model comparable to that defined in the SMGW protection profile \cite{tech:bsi:pp-smgw} can delay the implementation of control signals by up to 48 hours. Whether an attack can be successfully executed depends on the implementation of the protocols.

\begin{itemize}
\item \textit{What impact could delayed control signals have on the German power grid?}
\end{itemize}

If a control signal is sufficiently delayed, it may be executed in a grid state that is fundamentally different from the one in which it was issued or be synchronized with other delayed control signals. When executed at scale (e.g., targeting several thousand CLS devices in grid level 7), the cumulative effect could reach several gigawatts, potentially exceeding available Frequency Containment Reserve and higher thresholds derived from Dabrowski et al. \cite{paper:laa:gridshock}, possibly causing load shedding due to a frequency deviation of 1 Hz. However, this estimation is based on projection and does not consider system interdependencies or the likelihood of the required grid states. To exploit this vulnerability at scale, contextual knowledge to identify the correct packet across many CLS channels would need to be generalizable or efficiently obtainable which is still an open research question.

\begin{itemize}
\item \textit{Which countermeasures can be implemented at the SMGW, CLS-A or aEMT to mitigate delay attacks?}
\end{itemize}

We proposed countermeasures ranging from enforcing secure SMGW configurations to strengthening application-layer protocols and enhancing the underlying communication stack. Specifically, reducing the idle timeout of the CLS channel can limit the feasibility of delay attacks. In IEC 61850, validating timestamps in \textit{EnaReq} messages enables the detection of delayed control signals. For CLS.EEDI, the MQTT keep-alive mechanism can restrict the maximum delay of transmitted control signals. Additionally, extending the CLS.EEDI data model with timestamps, similar to IEC 61850, can further mitigate the attack.

To support long-lasting TLS sessions while maintaining protection against delay attacks, we propose an extension to TLS that introduces timestamped records. Additionally, we propose a protocol layer between TLS as an alternative to a TLS extension. Since this work does not consider retransmission and retry mechanisms at the application layer, the proposed mitigations may not be sufficient to address all types of delay attacks.

%% file: conclusion/future.tex
At the time of writing, conclusions regarding the vulnerability of CLS.EEDI to delay attacks are based on theoretical analysis and limited laboratory validation. Conducting broader empirical evaluations in real-world or more diverse laboratory settings is an important next step to assess feasibility.

As outlined in \cref{sec:discussion:limit}, this work does not consider delay attacks that specifically target the retry behaviour of application-layer protocols. Extending the formal model to include such attack vectors and developing corresponding mitigations remain open research directions.

\Cref{sec:impact} projects the potential impact a successful delay attack could have on the power grid. As discussed in \cref{sec:discussion:impa}, these projections may over- or underestimate the actual impact. A more realistic assessment requires incorporating additional factors such as power band and curved which must be followed in most use cases \cite{tech:VDE:ARN_4100,tech:VDE:ARN_4105,tech:VDE:ARN_4110,tech:VDE:ARN_4120,tech:VDE:ARN_4130}, the behaviour of producers and consumers, and the probability of exploitable grid states. Future work should therefore incorporate power-flow analysis or simulation-based approaches and update baseline simulations (e.g., Dabrowski et al.) to current and projected future grid loads for more precise thresholds.

As discussed in \cref{sec:discussion:impa} this work does not analyze the generalizability of the contextual-knowledge and therefore the resources required to execute the proposed coordinated static load altering attack at scale. This is required to estimate the risk associated with delay attacks on a whole grid 

This work defines 15 minutes as the threshold for a successful delay attack. This threshold may not be appropriate for all use cases and may, in some scenarios, be too strict (e.g., network congestion). For example, the Federal Network Agency (German: Bundesnetzagentur, BNetzA) states that control signals which are part of the §14a-mechanism must arrive at the premise within 6 seconds of issuance \cite{tech:bnetza:gpke}. Even stricter timing constraints apply to control signals associated with balancing reserve, which must be received within 5 seconds \cite{tech:uenb:pq-bedingungen,tech:uenb:rr-itAnf}. In cases where a CLS channel is used for multiple use cases, the most restrictive threshold must be applied. However, instead of relying on a static configuration, the sender of a message could determine such thresholds dynamically, either on a per-control-signal or per-session basis. This approach may also provide a mitigation against delay attacks targeting retransmission mechanisms, as the sender could reduce the acceptable delay threshold with each retry.

As outlined in \autoref{sec:discussion:gen}, the attack vector of delay attacks might not be limited to control signals between an aEMT and CLS-A. Therefore, other possible communication paths that might be used to control CLS on the premise should be analyzed regarding their vulnerability.

%% file: appendix/contri.tex
\textbf{Fabio Stoll}: Conceptualization, Methodology, Software, Investigation, Formal analysis, Validation, Writing -- original draft, Writing -- review \& editing.\\
\textbf{Heiko Lorenz \& Shalaka Kale}: Resources (laboratory setup), Methodology (laboratory setup), Software, Writing -- original draft (description of the laboratory setup), Writing -- review \& editing.\\
\textbf{Benjamin Pottkamp}: Validation, Writing -- review \& editing. \\
\textbf{Joachim Gerlach, Jessica Rövekamp}: Funding acquisition, Supervision, Writing -- review \& editing.\\
All authors have read and approved the final manuscript. Unless explicitly stated otherwise, all contributions were carried out solely by Fabio Stoll.

%% file: appendix/ack.tex
\begin{tabular}{@{}m{0.8\textwidth} m{0.2\textwidth}@{}}
Parts of the presented research has been carried out in the research projects ``MeGA - Measuring systems for large-scale generation plant'' (FKZ 03EI6108C, FKZ 03EI6108E) and ``eMpowerSYS - Development of a value-added platform for the digital energy transition'' (FKZ 03EI6074F, FKZ 03EI6074G). The German Federal Ministry for Economic Affairs and Energy (BMWE) funded both projects. &

\includegraphics[width=\linewidth]{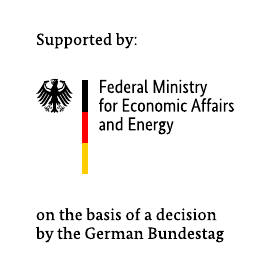}
\end{tabular}

%% file: appendix/ai.tex
This paper was reviewed and edited with the assistance of AI tools to improve clarity, grammar, and style. All ideas, research, and conclusions presented remain those of the authors. The use of AI did not alter the original content, analysis, or interpretation of the work.

%% file: appendix/conf.tex
The authors declare that they have no competing interests. Furthermore, the funding sponsor had no involvement in the study design, data collection, data analysis, interpretation of the results, preparation of the manuscript, or the decision to publish the findings.

The manufacturers of the SMGW, FNN-STB, aEMT, GWA and CLS-A used to host the DAP server requested that the authors refrain from disclosing their identities. The authors decided to honor this request to not disclose the manufacturers' names and revised the screenshots accordingly since the vulnerability derives from the specification itself rather than from specific vendor implementations.

%% file: appendix/disc.tex
On 7 January 2026, the vulnerability was disclosed to the BSI.\\
On 16 April 2026, the vulnerability was disclosed to project partners, the BSI and the BMWE.\\
On 24 April 2026, an early draft of this paper was sent to the BSI, BMWE, Webolution, FNN and project partners.

The authors discussed the issue with project partners, which acknowledged the vulnerability and decided to contribute countermeasures to the standardization processes.